

\input harvmac
\input tables
%
\def\np#1#2#3{Nucl. Phys. B{#1} (#2) #3}
\def\pl#1#2#3{Phys. Lett. {#1}B (#2) #3}

\def\ev#1{\langle#1\rangle}

\def\tilde{\widetilde}

\def\CC{{\cal C}}
\def\Tr{{\rm Tr ~}}
\def\tr{{\rm Tr ~}}
\font\zfont = cmss10 
\def\ZZ{\hbox{\zfont Z\kern-.4emZ}}
\def\bigone{{\bf 1}}
\def\vl{p}\def\vlt{\tilde{\vl}}
\def\nc{{N_{c}}}\def\nf{{N_{f}}}\def\ncd{{\tilde\ncl}}
\def\lk#1{\lambda_{#1}}

\def\nfl{{N_{f}}}\def\nfr{{N'_{f}}}
\def\ncl{{\nc}}\def\ncr{{N'_{c}}}
\def\ncld{{\tilde{N}_c}}\def\ncrd{{\tilde{N}'_c}}
\def\nncl{N}\def\nncr{{N'}}
\def\nncld{\tilde{N}}\def\nncrd{{\tilde{N}'}}
\def\mcl{{M}}\def\mcld{{\tilde{M}}}
\def\mfl{{m_{f}}}\def\mfr{{\tilde m_{f}}}
\def\Sl{S}\def\Sr{S'}\def\sll{s}\def\sr{s'}

\def\om{\eta}
\def\nfrp{n'_f}

\def\nflp{n_f}
%
%
\def\ts{X}
\def\Ql{Q}\def\Qr{Q'}
\def\Pl{M}\def\Pr{M'}
\def\Mm{P}\def\Mmt{\tilde{\Mm}}
\def\Qlt{\tilde{\Ql}}\def\Qrt{\tilde{\Qr}}
\def\tst{\tilde{\ts}}
%
%
\def\tsD{Y}\def\tstD{\tilde\tsD}
\def\ql{q}\def\qr{\ql'}
\def\qlt{\tilde\ql}\def\qrt{\tilde\qr}

\Title{hep-th/9506148, RU-95-38}
{\vbox{\centerline{New Examples of Duality in Chiral and}
\centerline{}
\centerline{Non-Chiral Supersymmetric Gauge Theories}}}
\bigskip
\centerline{K. Intriligator, R.G. Leigh, and M.J. Strassler}
\vglue .5cm
{\it \centerline{Department of Physics and Astronomy}
\centerline{Rutgers University}
\centerline{Piscataway, NJ 08855-0849, USA}}

\bigskip\bigskip

\noindent

We present evidence for renormalization group fixed points
with dual magnetic descriptions in fourteen new classes of
four-dimensional $N=1$ supersymmetric models. Nine of these
classes are chiral and many involve two or three gauge groups.
These theories are generalizations of models presented earlier
by Seiberg, by Kutasov and Schwimmer, and by the present authors.
The different classes are interrelated; one can flow from one
class to another using confinement or symmetry breaking.

\Date{6/95}

\lref\sw{N. Seiberg and E. Witten, \np{426}{1994}{19}, hep-th/9407087;
\np{431}{1994}{484}, hep-th/9408099.}%

\lref\sem{N. Seiberg, \np{435}{1995}{129}, hep-th/9411149.}%

\lref\isson{K. Intriligator and N. Seiberg, {\it Duality,
Monopoles,  Dyons, Confinement and Oblique Confinement in
Supersymmetric $SO(\nc)$ Gauge Theories}, RU--95--3,
hep-th/9503179, to appear in Nucl. Phys. B.}

\lref\intpou{K. Intriligator and P. Pouliot, {\it Exact
Superpotentials, Quantum Vacua and Duality in Supersymmetric
$Sp(\nc)$ Gauge Theories}, RU--95--23, hep-th/9505006, to appear
in Phys. Lett. B.}

\lref\kut{D. Kutasov, {\it A Comment on Duality in N=1
Supersymmetric Non-Abelian Gauge Theories}, EFI--95--11,
hep-th/9503086.}

\lref\kutsch{ D. Kutasov and A. Schwimmer, {\it On Duality in
Supersymmetric Yang-Mills Theory}, EFI--95--20, WIS/4/95,
hep-th/9505004.}

\lref\Ahsonyank{O. Aharony, J. Sonnenschein, and S. Yankielowicz,
{\it Flows and Duality Symmetries in N=1 Supersymmetric Gauge
Theories}, TAUP--2246--95, CERN-TH/95--91, hep-th/9504113.}

\lref\berk{M. Berkooz, {\it The Dual of Supersymmetric SU(2k)
with an Antisymmetric Tensor and Composite Dualities}, RU-95-20,
hep-th/9505088.}

\def\kref{\refs{\kut , \kutsch}}

\lref\privcom{N. Seiberg, private communication.}

\lref\rlmsspso{R.G. Leigh and M.J. Strassler, {\it Duality of
$Sp(2\nc)$ and $SO(\nc)$ Supersymmetric Gauge Theories with
Adjoint Matter}, RU--95--30, hep-th/9505088.}

\lref\emop{R.G. Leigh and M.J. Strassler,
{\it Exactly Marginal Operators and Duality in
Four Dimensional N=1 Supersymmetric Gauge Theory},
RU--95--2, hep-th/9503121, to appear in Nucl. Phys. B.}

\lref\kispso{K. Intriligator, {\it New RG Fixed Points and
Duality in Supersymmetric $Sp(\nc)$ and $SO(\nc)$ Gauge
Theories}, RU--95--27, hep-th/9505051, to appear in Nucl. Phys.
B.}

\lref\powerh{N. Seiberg, {\it The Power of Holomorphy -- Exact
Results in 4D SUSY Field Theories}  In the Proc. of PASCOS 94,
 RU-94-64, IASSNS-HEP-94/57, hep-th/9408013.}

\lref\powerd{N. Seiberg, {\it The Power of Duality
 -- Exact Results in 4D SUSY Field Theories},
In the Proc. of PASCOS 95 and the Proc. of the
Oskar Klein Lectures, RU-95-37, IASSNS-HEP-95/46, hep-th/9506077.}

\lref\apsnii{P.C. Argyres, M.R. Plesser and A. Shapere,
{\it The Coulomb Phase of N=2 Supersymmetric QCD},
IASSNS-HEP-95/32, UK-HEP/95-06, hep-th/9505100.}

\lref\hanoz{A. Hanany and Y. Oz, {\it On the Quantum Moduli
Space of Vacua of $N=2$ Supersymmetric $SU(N_c)$ Gauge Theories},
TAUP-2248-95, WIS-95/19, hep-th/9505075.}

\def\Zsuasym{3}
\def\Zsusym{4}
\def\Zsusa{5}
\def\Zspsp{6}
\def\Zsoso{7}
\def\Zsusu{8}
\def\Zsosp{9}
\def\Zsusos{10}
\def\Zsusoso{11}
\def\Zsuspa{12}
\def\Zsuspsp{13}
\def\Zsusps{14}
\def\Zsusoa{15}
\def\Zsuspso{16}

\newsec{Introduction}\seclab{\intro}

Recent work has shown that $N=1$ supersymmetry is a fruitful
arena for studying strongly coupled dynamics of gauge theories.
See \refs{\powerh, \powerd} for
recent reviews and references therein for earlier work.
One of the most intriguing of the recent
results is the discovery by Seiberg
that $N=1$ theories can have interacting superconformal fixed
points, in a non-Abelian Coulomb phase, which
have dual descriptions in terms of different,
``magnetic'', gauge theories \sem.   This duality is a
generalization  \refs{\sem, \emop, \isson} of the Montonen-Olive
electric-magnetic duality \ref\omd{C. Montonen and D. Olive, \pl
{72}{1977}{117}.} of $N=4$ theories \ref\dualnf{H. Osborn,
\pl{83}{1979}{321}; A. Sen, \pl{329}{1994}{217}, hep-th/9402032.}
and finite $N=2$ theories \refs{\sw, \hanoz, \apsnii} to
$N=1$ supersymmetric theories.

The original examples of $N=1$ duality
include $SU(N_c)$ with matter in the fundamental representation
\sem, $SO(N_c)$ with matter in the vector
representation \refs{\sem, \isson}, and $Sp(N_c)$ with matter
in the fundamental representation\refs{\sem , \intpou}.  New
examples of non-trivial infrared fixed points with dual
descriptions have recently been discovered by Kutasov \kut,
Kutasov and Schwimmer \kutsch, and the present authors
\refs{\kispso, \rlmsspso}; related models appear in
\refs{\Ahsonyank, \berk}.  The new theories are all based on the
original theories discussed in \sem\ along with an additional
field $X$ and a superpotential $W(X)$.
The superpotential plays an important role, controlling
the infrared fixed point.  The need in these models for the
superpotential is a disappointment --- it would be preferable
to have the dual for the theory without the superpotential, since
perturbing it with a superpotential and flowing to the infrared
would give the dualities for all of the theories with
superpotentials. Unfortunately, the dual of the theory without
any superpotential for the field $X$ is not yet known.  In
addition, there are limits to our understanding of the
physics even in the theories with the added superpotential
\refs{\kut,\kutsch}; many dynamical questions remain.
Given the gaps in our present understanding of duality, it
is important to have as many examples as possible. In this paper
we will present fourteen new classes of theories which are
generalizations of the models of \refs{\kut-\rlmsspso}.

In \refs{\sem,\emop} it was suggested that the duality of
$N=1$ models is closely related to that of finite $N=2$ theories.
The explicit relation between the $N=1$ theories studied in \sem\
and certain finite $N=2$ theories was presented in \emop. Finite
$N=2$ models have a marginal coupling constant (the gauge
coupling) with an associated continuous space of fixed points.
The duality transformation involves a reflection on this space,
along with a transformation of the flavor representations.
An analogous situation occurs for the $N=1$ models of \sem\ when
the number of flavors is such that the gauge group is self dual.
In all classes studied in \refs{\kut-\rlmsspso} and in this
paper, it has been found that the models with self dual gauge
group always have exactly marginal composite meson operators.
The full significance of this observation is still under
investigation. As in \emop, it appears to play an important role
in the duality transformation \refs{\kutsch,\rlmsspso}.

Because a theory in one class often flows to a theory in another
class under a relevant perturbation, new examples lead to
non-trivial consistency checks of duality.  Many of our examples
have flat directions in which they flow to a previously known
model.  Additional terms in the superpotential can have a similar
effect.  Perhaps the most interesting phenomenon that we observe
is associated with confinement; in cases with more than one
gauge group we find that the duality is consistent even when one
gauge coupling is much larger than the other.  This is essential
for the renormalization group flow to be sensible at
all values of the gauge couplings.  A detailed illustration of
how this occurs is presented in one class of models (sect.
\Zspsp.3).

 Nine of our new classes of models are chiral. Previous
studies of duality in chiral models have been carried out by
Berkooz \berk\ and by Seiberg \privcom. In principle,
chiral models can be very different from non-chiral ones; for
example, they can dynamically break supersymmetry.  In our
examples, however, the duality in the chiral
theories appears to be completely analogous to that in the
non-chiral theories.  We do not know if this is an indication of
a general similarity --- it could be that some unsolved chiral
models are completely unlike those that we have discovered.   One
feature of our chiral models
is that the $R$-charges, and therefore the dimensions at the
fixed point, of some chiral operators are not uniquely specified;
only dimensions of operators which are uncharged under all other
$U(1)$ symmetries are determined.

Because we will present so many models, we will discuss each
example only briefly.  In each case we will provide a few
pieces of evidence for the duality, mention interesting features,
and show how it is connected to other models when subjected to
a relevant perturbation.  We hope that our extensive
collection of interrelated theories will serve as a compelling
demonstration of the validity and generality of the phenomenon
of $N=1$ duality.

The structure of this paper is as follows.  In the next section
we provide an overview, reviewing the models of
\refs{\kut-\rlmsspso}
and summarizing the duality in our new examples.  Readers
interested only in our main results should read this section.
In sects. 3--16, each of our new classes of models is discussed
in more detail.

\newsec{Overview of the old and new theories}
\seclab{\squibs}

In this section we list the models of \refs{\kut-\rlmsspso},
and then present a short summary of each of our new models.  For
every new model, a later section provides additional details.

Each summary follows roughly the same pattern.  First,
we will introduce the electric theory.  Each model consists
of one or two 2-index tensors and some fields in the defining
representation (fundamental representation for $SU(\nc)$ and
$Sp(\nc)$, vector representation for $SO(\nc)$) and has a
superpotential for the two-index tensor(s).  (When the
superpotential is
a mass term, the duality always reduces to that of
\refs{\sem,\isson,\intpou}.)   Next, we will present its magnetic
dual, which is of the same type but with a different number of
colors and with singlet fields which
enter in the superpotential.  We then will discuss the effect of
a perturbation of the theory by a certain superpotential which
causes both the electric and magnetic theories to flow to a
product of simpler theories; in all cases the low energy theories
are dual to one another.  Lastly, any additional special features
of the model, including interesting flat directions and behavior
under confinement, will be outlined.

Before beginning we discuss a few notational issues.  We refer
to the symplectic group whose fundamental representation is
$2\nc$-dimensional as $Sp(\nc)$.  A ``flavor'' of
$Sp(\nc)$ represents two fields in the fundamental
representation; we will use $\nf$ when referring to the number
of flavors and $\nflp$ when referring to the number of fields.
Similarly a ``flavor'' of $SU(\nc)$ represents a field in the
fundamental representation and another in the antifundamental
representation; we will use $\nf$ when referring to the number
of flavors and $(\tilde m_f) m_f$ when referring to the number
of (anti)fundamental representations.  A ``defining model''
is a theory which only contains charged fields in the
defining representation of the gauge group, as in
\refs{\sem,\isson,\intpou}.

\subsec{$SU(N_c)$ with an adjoint field and $N_f$ fundamental
flavors}
\subseclab{\SQsuadj}

These models were presented by Kutasov and Schwimmer \kutsch.
The field $X$ is in the adjoint
representation of $SU(\nc)$
and the $\nf$ fields $Q^f$ ($\tilde Q^{\dot g}$) are in the
(anti)fundamental representation.  The superpotential $W=\tr
X^{k+1}$ truncates the chiral ring; the chiral mesons are
$(M_j)^{f\dot g}\equiv
Q^fX^{j} \tilde Q^{\dot g}$, $j=0\dots k-1$.

The dual theory has gauge group $SU(\ncd)$, with $\ncd =
k\nf-\nc$. It has an adjoint field $Y$, $\nf$ flavors $q_f$,
$\tilde q_{\dot g}$, and singlet fields $(M_j)^{f\dot g}$. The
superpotential is
\eqn\WDsuadj{
W=\Tr Y^{k+1}+\sum _{j=0}^{k-1}M_{k-j-1}qY^{j}\tilde q \ .}

Under perturbation by a mass term, $W = \tr X^{k+1} + m \tr X^2$,
the gauge group breaks to a product of decoupled defining models
\eqn\MPsuadj{
SU(n_1)\times U(n_2) \cdots \times U(n_k)}
where $\sum n_j = \nc$; each group has $\nf$ flavors in the
fundamental representation.  The mass term is mapped under
duality
to a mass term for $Y$; the magnetic theory flows to a similar
product
of defining models, which are dual to those above
\refs{\sem,\kutsch}.

If $\nc$ is divisible by $k$, then the model also has a
flat direction under which it breaks to $[SU(\nc/k)]^{k}$, with
each factor a defining model with $\nf$ flavors.  The
dual theory breaks to $[SU(\ncd/k)]^{k}=[SU(\nf-\nc/k)]^{k}$,
which satisfies the duality of \sem.

\subsec{$SO(N_c)$ with a traceless symmetric tensor and
$N_f$ vectors}
\subseclab{\SQsosym}

These models were presented in \kispso.
The field $X$ is in the traceless
symmetric tensor representation of
$SO(\nc)$ and the $\nf$ fields $Q^f$ are in the
vector representation.  The superpotential $W=\tr X^{k+1}$
truncates the chiral ring; the chiral mesons are $(M_j)^{f
g}\equiv Q^fX^{j} Q^{g}$, $j=0\dots k-1$, which are symmetric
tensors of the flavor group.

The dual theory has gauge group $SO(\ncd)$, with $\ncd =
k(\nf+4)-\nc$. It has a traceless symmetric tensor field $Y$,
$\nf$ vectors $q_f$, and singlet fields $(M_j)^{fg}$. The
superpotential is
\eqn\WDsosym{
W=\Tr Y^{k+1}+\sum _{j=0}^{k-1}M_{k-j-1}qY^{j}q \ .}

Under perturbation by a mass term, $W = \tr X^{k+1} + m \tr X^2$,
the gauge group breaks to a product of decoupled defining models
\eqn\MPsosym{
SO(n_1)\times SO(n_2) \cdots \times SO(n_k)}
where $\sum n_j = \nc$; each group has $\nf$ fields in the
vector representation. The magnetic theory flows to the dual of
this product.

If $\nc$ is divisible by $k$, then the model also has a
flat direction under which it breaks to $[SO(\nc/k)]^{k}$, with
each factor a defining model with $\nf$ fields in the vector
representation.  The
dual theory breaks to $[SU(\ncd/k)]^{k}=[SU(\nf+4-\nc/k)]^{k}$,
which satisfies the duality of \sem.

\subsec{$Sp(N_c)$ with a traceless antisymmetric tensor and
$N_f$ flavors}
\subseclab{\SQspasym}

These models were also presented in \kispso.
The field $X$ is in the traceless
antisymmetric tensor representation of $Sp(\nc)$ and the $\nf$
flavors ($2\nf$ fields) $Q^f$ are in the $2N_c$ dimensional
fundamental representation.  The superpotential $W=\tr X^{k+1}$
truncates the chiral ring; the chiral mesons are $(M_j)^{f
g}\equiv
Q^fX^{j} Q^{g}$, $j=0\dots k-1$, which are antisymmetric
tensors of the flavor group.

The dual theory has gauge group $Sp(\ncd)$, with $\ncd =
k(\nf-2)-\nc$. It has a traceless
antisymmetric tensor field $Y$,
$\nf$ flavors ($2\nf$ fields) $q_f$, and singlet fields
$(M_j)^{fg}$. The superpotential is
\eqn\WDspasym{
W=\Tr Y^{k+1}+\sum _{j=0}^{k-1}M_{k-j-1}qY^{j} q \ .}

Under perturbation by a mass term, $W = \tr X^{k+1} + m \tr X^2$,
the gauge group breaks to a product of decoupled defining models
\eqn\MPspasym{
Sp(n_1)\times Sp(n_2) \cdots \times Sp(n_k)
}
where $\sum n_j = \nc$; each group has $\nf$ flavors in the
fundamental representation.
The magnetic theory flows to the dual of this product.

If $\nc$ is divisible by $k$, then the model also has a
flat direction under which it breaks to $[Sp(\nc/k)]^{k}$, with
each factor a defining model with $\nf$ flavors.  The
dual theory breaks to $[Sp(\ncd/k)]^{k}=[Sp(\nf-2-\nc/k)]^{k}$,
which satisfies the duality of \sem.

\subsec{$Sp(N_c)$ with an adjoint field and $N_f$ flavors}
\subseclab{\SQspadj}

These models were presented in \rlmsspso.
The field $X$ is in the adjoint
representation of $Sp(\nc)$ and the $\nf$ flavors $Q^f$ are in
the fundamental representation.  The superpotential $W=\tr
X^{2(k+1)}$ truncates the chiral ring; the chiral mesons are
$(M_j)^{f g}\equiv
Q^f cX^{j} Q^{g}$, $j=0\dots 2k$, which are (anti)symmetric
tensors of the flavor group for odd (even) $j$.

The dual theory has gauge group $Sp(\ncd)$, with
$\ncd = (2k+1)\nf-\nc-2$.
It has a field $Y$ in the adjoint representation,
$\nf$ flavors  $q_f$, and singlet fields
$(M_j)^{fg}$. The superpotential is
\eqn\WDspadj{
W=\Tr Y^{2(k+1)}+\sum _{j=0}^{2k}M_{2k-j}qY^{j} q \ .}

Under perturbation by a mass term,
$W = \tr X^{2(k+1)} + m \tr X^2$,
the gauge group breaks to a product of decoupled defining models
\eqn\MPspadj{
Sp(n_0)\times U(n_1) \cdots \times U(n_k)
}
where $\sum_0^k n_j = \nc$; the symplectic group factor has $\nf$
flavors ($2\nf$ fields) while each unitary group has
$2\nf$ flavors in the fundamental representation.
The magnetic theory flows to the dual of this product.

\subsec{$SO(N_c)$ with an adjoint field and
$N_f$ vectors}
\subseclab{\SQsoadj}

These models were also presented in \rlmsspso.
The field $X$ is in the adjoint
representation of $SO(\nc)$
and the $\nf$ flavors $Q^f$ are in the vector
representation.  The superpotential $W=\tr X^{2(k+1)}$
truncates the chiral ring; the chiral mesons are $(M_j)^{f
g}\equiv
Q^fX^{j} Q^{g}$, $j=0\dots 2k$, which are (anti)symmetric
tensors of the flavor group for even (odd) $j$.

The dual theory has gauge group $SO(\ncd)$, with
$\ncd = (2k+1)\nf-\nc+4$.
It has a field $Y$ in the adjoint representation,
$\nf$ vectors  $q_f$, and singlet fields
$(M_j)^{fg}$. The superpotential is
\eqn\WDsoadj{
W=\Tr Y^{2(k+1)}+\sum _{j=0}^{2k}M_{2k-j}qY^{j} q \ .
}

Under perturbation by a mass term,
$W = \tr X^{2(k+1)} + m \tr X^2$,
the gauge group breaks to a product of decoupled defining models
\eqn\MPsoadj{
SO(n_0)\times U(n_1) \cdots \times U(n_k)
}
where $n_0+2\sum_1^k n_j = \nc$; the orthogonal group factor has
$\nf$ vectors while each unitary group has $\nf$ flavors in the
fundamental representation.
The magnetic theory flows to the dual of this product.

\subsec{$SU(\ncl)$ with an antisymmetric flavor and $\nfl$
fundamental flavors}
\subseclab{\SQsuasym}

The fields $X$ and $\tilde X$ are a flavor
of antisymmetric tensor representations of $SU(\nc)$ and
the $\nf$ fields $Q^f$ ($\tilde Q^{\dot g}$) are in the
(anti)fundamental representation.  The superpotential
\eqn\Wsuasym{W=\tr(X\tilde X)^{k+1}}
truncates the chiral ring; the chiral mesons
are $(M_j)^{f\dot g}\equiv Q^f(\tilde X X)^{j} \tilde Q^{\dot
g}$, $j=0\dots k$, $(P_r)^{fg}\equiv Q^f (\tilde X X)^{r}\tilde
X Q^g$, and $(\tilde P_r)^{\dot f\dot g}\equiv
\tilde Q^{\dot f}X(\tilde X X)^{r}\tilde Q^{\dot g}$, $r=0\dots
k-1$.
The $P$ and $\tilde P$ operators are antisymmetric in their
flavor indices.

The dual theory has gauge group $SU(\ncd)$, with
$\ncd = (2k+1)\nf-4k-\nc$.
It has fields $Y$ and $\tilde Y$ which form a flavor
of antisymmetric tensor representations,
$\nf$ fundamental flavors $q_f$, $\tilde q_{\dot g}$, and singlet
fields $(M_j)^{f\dot g}$, $(P_r)^{fg}$, and
$(\tilde P_r)^{\dot f\dot g}$.  The superpotential is
\eqn\WDsuasym{W=\Tr (Y\tilde Y)^{k+1}
+\sum _{j=0}^{k}M_{k-j} q(\tilde Y Y)^{j}\tilde q
+\sum _{r=0}^{k-1} \Big[P_{k-r-1} q(\tilde Y Y)^{r}\tilde Y q
+\tilde P_{k-r-1}\tilde qY(\tilde Y Y)^{r}\tilde q\Big]\ .}

The duality exhibits a feature not previously observed.  There
are two vector-like $U(1)$ symmetries, one counting $X$-number
and one counting $Q$-number.  Under duality these are mixed;
$Y$-number and $q$-number are linear combinations of $X$-number
and $Q$-number.

Under perturbation by a mass term,
$W = \tr (X\tilde X)^{k+1} + m \tr (X\tilde X)$,
the gauge group breaks to a product of decoupled defining models
\eqn\MPsuasym{
SU(n_0)\times Sp(n_1) \times \cdots \times Sp(n_k)}
where $n_0+2\sum n_j = \nc$; each group has $\nf$ flavors of
fundamental representations.
The magnetic theory flows to the dual of this product.

The model has a set of flat directions in which an operator
$B_n\equiv X^n Q^{\nc-2n}$, with gauge indices
contracted using an epsilon tensor,
gets an expectation value.  The gauge group is broken to
$Sp(n)$ with an antisymmetric tensor $\tilde X$ and $\nf$
flavors; the superpotential is $W=\tr\tilde X^{k+1}$.  This
is the theory considered in \kispso\ and discussed in
sect. \SQspasym.  Under duality the operator $B_n$ is mapped to
$Y^{k(\nf-2)-n} q^{\nf+2n-\nc}$.  Its expectation value causes
the magnetic theory to flow to the dual expected from
sect. \SQspasym\ \kispso.

Additional discussion of this model is presented in sect.
\Zsuasym.

\subsec{$SU(\ncl)$ with a symmetric flavor and $\nfl$
fundamental flavors}
\subseclab{\SQsusym}

The fields $X$ and $\tilde X$ are a flavor
of symmetric tensor representations of $SU(\nc)$  and
the $\nf$ fields $Q^f$ ($\tilde Q^{\dot g}$) are in the
(anti)fundamental representation.  The superpotential
\eqn\Wsusym{W=\tr(X\tilde X)^{k+1}}
truncates the chiral ring; the chiral mesons
are $(M_j)^{f\dot g}\equiv Q^f(\tilde X X)^{j} \tilde Q^{\dot
g}$, $j=0\dots k$, $(P_r)^{fg}\equiv Q^f(\tilde X X)^{r}\tilde
X Q^g$, and $(\tilde P_r)^{\dot f\dot g}\equiv
\tilde Q^{\dot f}X(\tilde X X)^{r}\tilde Q^{\dot g}$, $r=0\dots
k-1$. The $P$ and $\tilde P$ operators are symmetric in their
flavor indices.

The dual theory has gauge group $SU(\ncd)$, with
$\ncd = (2k+1)\nf+4k-\nc$. It has fields $Y$ and $\tilde Y$ which
form a flavor of symmetric tensor representations,
$\nf$ fundamental flavors $q_f$, $\tilde q_{\dot g}$, and singlet
fields $(M_j)^{f\dot g}$, $(P_r)^{fg}$, and
$(\tilde P_r)^{\dot f\dot g}$.  The superpotential is
\eqn\WDsusym{
W=\Tr (Y\tilde Y)^{k+1}
+\sum _{j=0}^{k}M_{k-j} q(\tilde Y Y)^{j}\tilde q
+\sum _{r=0}^{k-1}\Big[P_{k-r-1} q(\tilde Y Y)^{r}\tilde Y q
+\tilde P_{k-r-1}\tilde qY(\tilde Y Y)^{r}\tilde q\Big]\ .}
As in the previous case there are
two vector-like $U(1)$ symmetries, one counting $X$-number and
one counting $Q$-number, which are mixed under duality.

Under perturbation by a mass term,
$W = \tr (X\tilde X)^{k+1} + m \tr (X\tilde X)$,
the gauge group breaks to a product of decoupled defining models
\eqn\MPsusym{
SU(n_0)\times SO(n_1) \times \cdots \times SO(n_k)}
where $n_0+\sum n_j = \nc$; the unitary factor has $\nf$ flavors
of fundamental representations while each orthogonal group
has $2\nf$ fields in the vector representation.
The magnetic theory flows to the dual of this product.

The model has a set of flat directions in which an operator
$B_n\equiv X^n Q^{\nc-n}Q^{\nc-n}$, with gauge indices
contracted using two epsilon tensors, gets an expectation value.
The gauge group is broken to
$SO(n)$ with a symmetric tensor $\tilde X$ and $2\nf$ vectors;
the superpotential is $W=\tr\tilde X^{k+1}$.  This
is the theory considered in \kispso\ and discussed in
sect.  \SQsosym.  Under duality the operator $B_n$ is mapped to
$Y^{k(2\nf+4)-n} q^{\nf+n-\nc} q^{\nf+n-\nc}$.
Its expectation value causes
the magnetic theory to flow to the dual expected from sect.
\SQsosym\ \kispso.

Additional discussion of this model is presented in sect.
\Zsusym.

\subsec{$SU(\nc)$ with an antisymmetric tensor and a symmetric
tensor; a chiral theory}
\subseclab{\SQsusa}

The gauge group is $SU(\nc)$; the field $X$ is in the
${\bf \half \nc (\nc-1)}$ representation, the field $\tilde X$
is in the ${\bf \overline{\half \nc(\nc+1)}}$ representation, and
the $m_f (\tilde m_f)$ fields $\Ql^f$ $(\Qlt^{\dot g})$ are in
the (anti)fundamental representation of $SU(\nc)$.  This model
is chiral; anomaly cancellation requires that
$m_f=\tilde m_f+8$.   The superpotential
\eqn\Wsusa{W= \tr(\ts\tst)^{2(k+1)}}
truncates the chiral ring; the chiral mesons
are $(M_j)^{f\dot g}\equiv Q^f(\tst\ts)^{j} \tilde Q^{\dot g}$,
$j=0\dots 2k+1$, $(P_r)^{fg}\equiv Q^f (\tst\ts)^{r}\tst Q^g$ and
$(\tilde P_r)^{\dot f\dot g}\equiv
\tilde Q^{\dot f}\ts(\tst\ts)^{r}\tilde Q^{\dot g}$,
$r=0\dots 2k$.  The $P_r$ ($\tilde P_r$) operators are
symmetric in their flavor
indices for even (odd) $r$ and antisymmetric in their flavor
indices for odd (even) $r$.

The dual theory has gauge group $SU(\tilde \nc)$, with
$\tilde \nc=\half (4k+3)(m_f+\tilde m_f)-\nc$.
It has a field $Y$ in the
${\bf \half \ncd(\ncd-1)}$ representation, a field $\tilde Y$
in the
${\bf \overline{\half \ncd(\ncd+1)}}$ representation,
$m_f$ ($\tilde m_f$) fields $q_f$ ($\tilde q_{\dot g}$) in the
(anti)fundamental representation of $SU(\tilde \nc)$, and singlet
fields $(M_j)^{f\dot g},(P_r)^{fg}$ and $(\tilde P_r)^{\dot f\dot
g}$. The gauge anomaly vanishes since we still have $m_f=\tilde
m_f+8$. The superpotential is
\eqn\WDsusa{W=\Tr (Y\tilde Y)^{2(k+1)}
+\sum _{j=0}^{2k+1}M_{2k-j+1} q(\tilde Y Y)^{j}\tilde q
+\sum _{r=0}^{2k} \Big[P_{2k-r} q(\tilde Y Y)^{r}\tilde Y q
+\tilde P_{2k-r}\tilde qY(\tilde Y Y)^{r}\tilde q\Big]\ .}
As in earlier cases there are two $U(1)$ global symmetries, in
addition to the $R$ symmetry, which are rotated under duality.

Under the perturbation by a term
$W = \tr (\ts\tst)^{2(k+1)} + \lambda \tr(\ts\tst)^2$
the gauge group breaks to a product of decoupled models
\eqn\MPsusa{
SU(n_0)\times U(n_1) \times \cdots \times U(n_k)}
where $n_0+2\sum n_j = \nc$.  The first factor is a model of the
type discussed in this section, with $k=0$; it has an
antisymmetric tensor $\hat X$, a symmetric tensor $\hat{\tilde
X}$, a superpotential $\hat W = \tr(\hat X\hat{\tilde X})^2$,
and $m_f (\tilde m_f)$ fields in the (anti)fundamental
representation.  The other factors
are defining models with $m_f+\tilde m_f$ flavors.
The magnetic theory flows to the dual of this product.

The model has a set of flat directions in which an operator
$B_n\equiv X^n Q^{\nc-2n}$, with gauge indices
contracted using an epsilon tensor,
gets an expectation value.  The gauge group is broken to
$Sp(n)$ with a symmetric (adjoint) tensor $\tilde X$ and
$m_f+\tilde m_f$ fundamental fields; the superpotential is
$W=\tr\tilde X^{2(k+1)}$.  This is the theory considered in
\rlmsspso\ and discussed in sect. \SQspadj.  Under duality the
operator $B_n$ is mapped to $Y^{(k+\half)(m_f+\tilde m_f)-n-2}
q^{m_f+2n-\nc}$. Its expectation value causes the magnetic theory
to flow to the dual expected from sect. \SQspadj\ \rlmsspso.

The model also has a set of flat directions in which an operator
$\tilde B_n\equiv \tilde X^n \tilde Q^{\nc-n}\tilde Q^{\nc-n}$,
with gauge indices contracted using two epsilon tensors,
gets an expectation value.  The gauge group is broken to $SO(n)$
with an antisymmetric (adjoint) tensor $X$ and $m_f+\tilde m_f$
vectors;  the superpotential is $W=\tr X^{2(k+1)}$.  This
is the theory considered in \rlmsspso\ and discussed in
sect. \SQsoadj.  Under duality the operator $\tilde B_n$ is
mapped to $\tilde Y^{(2k+1)(m_f+\tilde m_f)-n+4}
\tilde q^{\tilde m_f+n-\nc}\tilde q^{\tilde m_f+n-\nc}$.
Its expectation value causes the magnetic theory to flow to the
dual expected from sect. \SQsoadj\ \rlmsspso.

Additional discussion of this model is presented in sect. \Zsusa.

\subsec{$Sp(\ncl)\times Sp(\ncr)$}
\subseclab{\SQspsp}

The gauge group is $Sp(\ncl)\times Sp(\ncr)$; the field $X$ is
in the ${\bf (2\ncl,2\ncr)}$, and
the $\nfl$ ($\nfr$) flavors $Q^f$ ($ Q'^{g'}$) are in the
fundamental representation of $Sp(\ncl)\ ( Sp(\ncr))$ .  The
superpotential
\eqn\Wspsp{W=\tr \ts^{2(k+1)}}
truncates the chiral ring; the chiral mesons
are $(\Mm_{r})^{fg'}= Q^f\ts^{(2r+1)} Q'^{g'}$, $r=0\dots k-1$
$(\Pl_{j})^{fg}=Q^f\ts^{2j} Q^g$, and
$(\Pr_{j})^{f'g'}=Q'^{f'}\ts^{2j} Q'^{g'}$, $j=0\dots k$.
The $M$ and $M'$ operators are antisymmetric in their flavor
indices.

The dual theory has gauge group $Sp(\ncld)\times Sp(\ncrd)$, with
$\ncld = (k+1)(\nfl+\nfr-2)-\nfl-\ncr$ and
$\ncrd = (k+1)(\nfl+\nfr-2)-\nfr-\ncl$.
It has a field $Y$ in the ${\bf (2\ncld,2\ncrd)}$,
$\nfr$ ($\nfl$) flavors $q_{g'}$ ($q'_f$) in the fundamental
representation of $Sp(\ncld)\ (Sp(\ncrd))$, and singlet fields
$(\Mm_{r})^{f g'}$, $(\Pl_{j})^{fg}$, and $(\Pr_{j})^{f'g'}$.
The superpotential is
\eqn\WDspsp{
W= \tr \tsD^{2(k+1)}
+\sum_{r=0}^{k-1} \Mm_{k-r-1}\; \ql\tsD^{2r+1}\qr
+\sum_{j=0}^{k} \left[\Pl_{k-j}\; \qr\tsD^{2j}\qr
+\Pr_{k-j}\; \ql\tsD^{2j}\ql\right].}
Note that the mesons of the $Sp(\ncl)$ group couple to the dual
quarks of $Sp(\ncrd)$; this sort of transformation on the flavor
indices will be seen repeatedly.

Under perturbation by a mass term, $W=\tr X^{2(k+1)} + m\tr X^2$,
the gauge group breaks to a product of decoupled defining models
\eqn\MPspsp{Sp(\ncl-\ncr+p_0)\times Sp(p_0)
\times Sp(p_1)\times\ldots\times Sp(p_k)}
(for $\ncl\geq\ncr$) with $\sum_{\ell=0}^{k}p_\ell =\ncr$.
The first (second) factor has $\nfl$ ($\nfr$) fundamental
flavors, while the others have $\nfl+\nfr$ flavors.
The magnetic theory flows to the dual of this product.

When the gauge coupling of one $Sp$ factor is much
larger than that of the other, the former can be dualized as
in \refs{\sem,\intpou} with the latter carried along
as a weakly coupled spectator.
For example, for $\nfr+\ncl=\ncr+2$ the $Sp(\ncr)$ factor
confines \intpou, and the $Sp(N_c)$ theory below the confinement
scale is a model of \kispso\ (see sect. \SQspasym) with
an antisymmetric tensor,
$\nfl+\nfr$ flavors and $W=X^{k+1}$.  In the magnetic theory
$Sp(\ncrd)$ confines similarly (since $\nfr+\ncld=\ncrd+2$)
leaving a theory with $Sp(\tilde N_c)=Sp[k(\nfl+\nfr-2)-\ncl]$,
which is the dual \kispso\ of the confined
electric model.\footnote{*}{In addition, both the electric
and magnetic theory have extra singlets
and a non-perturbative term in the superpotential which are
preserved by the duality.}
With $\nfr+\ncl>\ncr+2$ both theories again flow to models
studied in \kispso, with part of the flavor group weakly
gauged and with a superpotential.  The preservation of
duality in this way is essential
for the consistency of the renormalization group flow given
arbitrary gauge couplings.  The details of the general case
are given in sect. \Zspsp.3.

Additional discussion of this model is presented in sect. \Zspsp.

\subsec{$SO(\ncl)\times SO(\ncr)$}
\subseclab{\SQsoso}

The gauge group is $SO(\ncl)\times SO(\ncr)$; the field $X$ is
in the ${\bf (\ncl,\ncr)}$, and
the $\nfl$ ($\nfr$) fields $Q^f$ ($ Q'^{g'}$) are in the vector
representation of $SO(\ncl)\ ( SO(\ncr))$ .  The superpotential
\eqn\Wsoso{W= \tr \ts^{2(k+1)}}
truncates the chiral ring; the chiral mesons
are $(\Mm_{r})^{fg'}= Q^f\ts^{(2r+1)} Q'^{g'}$, $r=0\dots k-1$
$(\Pl_{j})^{fg}=Q^f\ts^{2j} Q^g$, and
$(\Pr_{j})^{f'g'}=Q'^{f'}\ts^{2j} Q'^{g'}$, $j=0\dots k$.
The $M$ and $M'$ operators are symmetric in their flavor
indices.

The dual theory has gauge group $SO(\ncld)\times SO(\ncrd)$, with
$\ncld = (k+1)(\nfl+\nfr+4)-\nfl-\ncr$ and
$\ncrd = (k+1)(\nfl+\nfr+4)-\nfr-\ncl$.
It has a field $Y$ in the ${\bf (\ncld,\ncrd)}$,
$\nfr$ ($\nfl$) fields $q_{g'}$ ($q'_f$) in the vector
representation of $SO(\ncld)\ (SO(\ncrd))$, and singlet fields
$(\Mm_{r})^{f g'}$, $(\Pl_{j})^{fg}$, and $(\Pr_{j})^{f'g'}$.
The superpotential is
\eqn\WDsoso{
W= \tr \tsD^{2(k+1)}
+\sum_{r=0}^{k-1} \Mm_{k-r-1}\; \ql\tsD^{2r+1}\qr
+\sum_{j=0}^{k} \left[\Pl_{k-j}\; \qr\tsD^{2j}\qr
+\Pr_{k-j}\; \ql\tsD^{2j}\ql\right].}
Note that the mesons of the $SO(\ncl)$ group couple to the dual
quarks of $SO(\ncrd)$.

Under perturbation by a mass term, $W=\tr X^{2(k+1)} + m\tr X^2$,
the gauge group breaks to a product of decoupled defining models
\eqn\MPsoso{
SO(\ncl-\ncr+p_0)\times SO(p_0)\times SO(p_1)\times\ldots\times
SO(p_k)}
(for $\ncl\geq\ncr$) with $\sum_{\ell=0}^{k}p_\ell =\ncr$.
The first (second) factor has $\nfl$ ($\nfr$) fields in
the vector representation, while the others have $\nfl+\nfr$
vectors.  The magnetic theory flows to the dual of this product.

As in sect. \SQspsp, if $\nfr+\ncl=\ncr-4$ the $SO(\ncr)$ factor
can confine \isson, leaving a theory of \kispso\ (see sect.
\SQsosym) with $SO(\ncl)$, a symmetric tensor, $\nfl+\nfr$
flavors and $W=X^{k+1}$.  In the magnetic theory $SO(\ncrd)$
similarly confines
(since $\nfr+\ncld=\ncrd-4$) leaving a theory with
$SO[k(\nfl+\nfr+4)-\ncl]$, which is the dual \kispso\ of the
confined electric model.  More general cases are discussed
in sect. \Zsoso.3.

Additional discussion of this model is presented in sect. \Zsoso.

\subsec{$SU(\ncl)\times SU(\ncr)$}
\subseclab{\SQsusu}

The gauge group is $SU(\ncl)\times SU(\ncr)$; the field $X$ is
in the ${\bf (\ncl,\ncr)}$, the field $\tilde X$ is
in the conjugate representation, and
the $\nfl$ ($\nfr$) flavors $\Ql^f,\Qlt^{\dot g}$
($ \Qr^{f'},\Qlt'^{\dot g'}$) are in the fundamental
representation of $SU(\ncl)\ ( SU(\ncr))$ .  The superpotential
\eqn\Wsusu{W= \tr(\ts\tst)^{k+1}}
truncates the chiral ring; the chiral mesons
are $(\Pl_{j})^{f\dot g} = \Ql^f (\tst\ts)^{j}\Qlt^{\dot g}$,
$(\Pr_{j})^{f'\dot g'} = \Qr^{f'} (\tst\ts)^{j}\Qlt'^{\dot g'}$,
$(\Mm_{r})^{fg'}=\Ql^f (\tst\ts)^{r-1}\tst\Qr^{g'}$ and
$(\Mmt_{r})^{\dot f\dot g'}=\Qlt^{\dot f}
\ts(\tst\ts)^{r-1}\Qlt'^{\dot g'}$, where $j=0\dots k$ and
$r=0\dots k-1$.

The dual theory has gauge group $SU(\ncld)\times SU(\ncrd)$, with
$\ncld = (k+1)(\nfl+\nfr)-\nfl-\ncr$ and
$\ncrd = (k+1)(\nfl+\nfl)-\nfr-\ncl$.
It has fields $Y$ and $\tilde Y$ in the ${\bf (\ncld,\ncrd)}$
and its conjugate, $\nfr$ ($\nfl$) flavors $\ql_{f'},\qlt_{\dot
g'}$ ($ \ql'_{f},\qlt'_{\dot g})$
in the fundamental
representation of $SU(\ncld)\ (SU(\ncrd))$, and singlet fields
$(\Pl_{j})^{f\dot g}$, $(\Pr_{j})^{f'\dot g'}$,
$(\Mm_{r})^{fg'}$, and $(\Mmt_{r})^{\dot f'\dot g'}$.
The superpotential is
\eqn\WDsusu{\eqalign{
W= \tr (\tsD\tstD)^{k+1}
+\sum_{j=0}^{k} \left[\Pl_{k-j} \qr (\tstD\tsD)^{j}\qrt
+\Pr_{k-j} \qlt (\tsD\tstD)^{j}\ql\right]\cr
+\sum_{r=0}^{k-1} \left[\Mm_{k-r-1} \ql \tstD(\tsD\tstD)^{r}\qr
+\Mmt_{k-r-1} \qrt (\tsD\tstD)^{r}\tsD\qlt\right].}}
Note that the mesons of the $SU(\ncl)$ group couple to the dual
quarks of $SU(\ncrd)$.

As in sects. \SQsuasym\ and \SQsusym, the three
vector-like $U(1)$ symmetries, counting $X$-number, $\Ql$-number
and $\Qr$-number, are mixed under duality.

Under perturbation by a mass term,
$W = \tr(X\tilde X)^{k+1} + m \tr X\tilde X$,
the gauge group breaks to a product of decoupled defining models
\eqn\MPsusu{SU(\ncl-\ncr+p_0)\times SU(p_0)
\times U(p_1)\times\ldots\times U(p_k)}
(for $\ncl\geq\ncr)$ with $\sum_{\ell=0}^{k}p_\ell =\ncr$.
The first (second) factor has $\nfl$ ($\nfr$) fundamental
flavors,  while the others have $\nfl+\nfr$ flavors.
The magnetic theory flows to the dual of this product.

As in sect. \SQspsp, if $\nfr+\ncl=\ncr+1$ the $SU(\ncr)$ factor
can confine \sem\ leaving a theory of \kutsch\ (see sect.
\SQsuadj) with $SU(\ncl)$, an adjoint tensor $\widehat X\sim
X\tilde X$, $\nfl+\nfr+1$ flavors and
$W=\tr \widehat X^{k+1} + B\widehat X\tilde B$,
where $B$, a baryon of $SU(\ncr)$, is in the fundamental
representation of $SU(\ncl)$.   In the magnetic theory
$SU(\ncrd)$ confines similarly (since $\nfr+\ncld=\ncrd+1$)
leaving a theory with $SU(\tilde N_c)=SU[k(\nfl+\nfr)-\ncl+1]$
and a similar superpotential;
this is dual \refs{\kut,\Ahsonyank,\kutsch} to the confined
electric model. More general cases are discussed in section
\Zsusu.4.

The model also has a set of flat directions in which an operator
$B_n\equiv X^n Q^{\ncl-n}Q'^{\ncr-n}$, with gauge indices
contracted using two
epsilon tensors, gets an expectation value.  The gauge group is
broken to $SU(n)$ with an adjoint field $\tilde X$ and
$\nfl+\nfr$ flavors,
along with some additional singlets;
the infrared superpotential is $W=\tr\tilde X^{k+1}$.  This
is the theory considered in \kut\ and \kutsch\ and discussed in
sect. \SQsuadj.  Under duality the operator $B_n$ is mapped to
$Y^{k(\nfl+\nfr)-n} q^{\nfr+n-\ncr} q'^{\nfl+n-\ncl}$.
Its expectation value causes
the magnetic theory to flow to the dual expected from \kutsch.

Additional discussion of this model is presented in sect. \Zsusu.

\subsec{$SO(\ncl)\times Sp(\ncr)$}
\subseclab{\SQsosp}

The gauge group is $SO(\ncl)\times Sp(\ncr)$; the field $X$ is
in the ${\bf (\ncl,2\ncr)}$,
the $\nfl$ fields $\Ql^f$ are in the vector
representation of $SO(\ncl)$, and the $\nfrp$ fields
$\Qr^{g'}$ are in the fundamental representation of
$Sp(\ncr)$.  The $Sp(\ncr)$ global anomaly of
\ref\wittenanom{E. Witten, \pl{117}{1982}{324}.} requires
$\nfrp+\ncl$ to be even.  This model is chiral in that mass
terms cannot be written for all the matter fields.  The
superpotential
\eqn\Wsosp{W= \tr \ts^{4(k+1)}}
truncates the chiral ring; the chiral mesons
are $(\Mm_{r})^{fg'}= Q^f\ts^{(2r+1)} Q'^{g'}$, $r=0\dots 2k$
$(\Pl_{j})^{fg}=Q^f\ts^{2j} Q^g$, and
$(\Pr_{j})^{f'g'}=Q'^{f'}\ts^{2j} Q'^{g'}$, $j=0\dots 2k+1$.
The $\Pl_{j}$ operators are (anti-)symmetric in their flavor
indices for (odd) even $j$; the reverse is true for the $\Pr_{j}$
operators.

The dual theory has gauge group $SO(\ncld)\times Sp(\ncrd)$, with
$\ncld = 2(k+1)(\nfl+\nfrp)-\nfl-2\ncr$ and
$2\ncrd =2(k+1)(\nfl+\nfrp)-\nfrp-\ncl$.  (Note that the
anomaly cancellation in the electric
$Sp(\ncr)$ theory ensures that $\ncrd$ is an integer.)
It has a field $Y$ in the ${\bf (\ncld,2\ncrd)}$,
$\nfrp$ ($\nfl$) fields $q_{g'}$ ($q'_{f}$) in the vector
(fundamental) representation of $SO(\ncld)\ (Sp(\ncrd))$, and
singlet fields
$(\Mm_{r})^{f g'}$, $(\Pl_{j})^{fg}$, and $(\Pr_{j})^{f'g'}$.
(Note that $\ncld+\nfl$ is even, ensuring that $Sp(\ncrd)$ is not
anomalous.)  The superpotential is
\eqn\WDsosp{\eqalign{
W= &  \tr\tsD^{4(k+1)}
+\sum_{r=0}^{2k} \Mm_{2k-r}\; \ql\tsD^{2r+1}\qr\cr
& +\sum_{j=0}^{2k+1} \left[\Pl_{2k-j+1}\; \qr\tsD^{2j}\qr
+\Pr_{2k-j+1}\; \ql\tsD^{2j}\ql\right].}}
Notice the mesons of the $SO(\ncl)$ group couple to the
dual quarks of the $Sp(\ncrd)$ group.

When the theory is perturbed by taking
$W = \tr X^{4(k+1)} + \lambda \tr X^4$ (no mass term for $X$
can be written), the gauge group breaks to a product of  models
\eqn\MPsosp{SO(\ncl-2\ncr+2p_0)\times Sp(p_0)
\times U(p_1)\times\ldots\times U(p_k)}
 with $2p_0+\sum_{\ell=1}^{k}p_\ell =2\ncr$.  (Here we
consider $\ncl\geq 2\ncr$; the other case is similar.)
The first and second factor form a model of the same type as the
original, but with $k=0$; the unitary factors are defining models
with $\nfl+\nfrp$ flavors.
The magnetic theory flows to the dual of this product.

As in sect. \SQspsp, if $\nfl+2\ncr=\ncl-4$ the $SO(\ncl)$ factor
confines \isson\ leaving a theory of \rlmsspso\ (see sect.
\SQspadj)
with $Sp(\ncr)$, an adjoint tensor $\widehat X\sim X^2$,
$\nfl+\nfrp$ fundamentals and
$W=\tr \widehat X^{2(k+1)}$.  In the magnetic theory
$SO(\ncrd)$ confines similarly (since $\nfrp+\ncl=\ncrd-4$)
 leaving a theory
with $Sp[\half(2k+1)(\nfl+\nfrp)-\ncr-2]$ which is dual
\rlmsspso\ to the confined electric model.  Similarly, if
$\ncl+\nfrp=2(\ncr+2)$, the $Sp(N_c')$ and $Sp(\tilde N_c')$
factors confine; the confined theories are those of
sect. \SQsoadj\ and satisfy the duality of \rlmsspso.

Additional discussion of this model is presented in sect. \Zsosp.

\subsec{$SU(M)\times SO(N)$ with a symmetric tensor of $SU(M)$}
\subseclab{\SQsusos}

The gauge group is $SU(M)\times SO(N)$; the field $X$ is
in the  ${\bf (M,N)}$, the
field $\tilde X$ is a symmetric tensor in the
${\bf (\overline{\half M(M+1)},1)}$,
the $m_f (\tilde m_f)$ fields $\Ql^f$ $(\Qlt^{\dot g})$ are in
the (anti)fundamental representation of $SU(M)$, and the $n_f$
fields
$S^t$ are in the vector representation of
$SO(N)$.  For $SU(M)$ to be non-anomalous
requires that $m_f+N=\tilde m_f+M+4$.
The superpotential
\eqn\Wsusos{W= \tr(\ts\tst\ts)^{(k+1)}}
truncates the chiral ring; the chiral mesons
are $(M_j)^{f\dot g}\equiv Q^f(\tst\ts ^2)^{j} \tilde Q^{\dot
g}$, $(P_j)^{fg}\equiv Q^f (\tst\ts ^2)^{j}\tst Q^g$,
$(N_j)^{tu}\equiv S^t(\ts\tst\ts)^{j}S^u$, $(\tilde R_j)^{\dot
g t}\equiv  \tilde Q^{\dot g} (\ts^2\tst)^{j}\ts S^t$, $j=0\dots
k$, $(\tilde P_r)^{\dot f\dot g}\equiv
\tilde Q^{\dot f}\ts ^2(\tst\ts ^2)^{r}\tilde Q^{\dot g}$
and $(R_r)^{ft}\equiv Q^f (\tst\ts ^2)^{r}\tst\ts S^{t}$,
$r=0\dots k-1$. The $N$, $P$ and $\tilde P$ operators are
symmetric in their flavor indices.

The dual theory has gauge group
$SU(\tilde M)\times SO(\tilde N)$, with
$\tilde M= (k+1)(m_f+\tilde m_f +n_f+4)-m_f-N$ and
$\tilde N= (k+1)(m_f+\tilde m_f+n_f+4)-n_f-M$. It has a field
$Y$ in the ${\bf (\tilde M,\tilde N)}$, a field $\tilde Y$ in
the ${\bf (\overline{\half \tilde M(\tilde M+1)},1)}$,
$n_f$ ($\tilde m_f$) fields $q_t$ ($\tilde q_{\dot g}$) in the
(anti)fundamental representation of $SU(\tilde M)$,
$m_f$  fields $s_f$  in the vector
representation of $SO(\tilde N)$, and singlet fields
$(M_j)^{f\dot g},(P_j)^{fg},(\tilde P_r)^{\dot f\dot
g},(N_j)^{tu}, (R_r)^{ft}$ and $(\tilde R_j)^{\dot g t}$.  All
gauge anomalies  automatically vanish.
The superpotential is
\eqn\WDsusos{W= \tr(\tsD\tstD\tsD)^{(k+1)} +
{\rm singlet\ meson\ couplings}.}
We omit the meson couplings for brevity; these terms are
analogous to those of all other models and are determined by the
flavor symmetries (see sect. \Zsusos).  We note that the mesons
of the $SO(N)$ group couple to the dual quarks of the $SU(\tilde
M)$ group while the antiquark mesons of $SU(M)$ couple to the
antiquarks of $SU(\tilde M)$.

If the $SU(M)$ factor confines in the electric theory,
then the $SU(\tilde M)$ factor confines in the magnetic theory;
in the infrared the models are electric and magnetic duals under
the duality of \kispso\ (see sect. \SQsosym.)
If the $SO(N)$ factor confines, then
$SO(\tilde N)$ confines also, and the infrared duality is that
of sect. \SQsusym.

Additional discussion of this model is presented in sect.
\Zsusos.

\subsec{$SU(M)\times SO(N)\times SO(N')$}

\subseclab{\SQsusoso}

The gauge group is $SU(M)\times SO(N)\times SO(N')$; the field
$X$ is in the  ${\bf (M,N,1)}$, the
field $\tilde X$ is in the ${\bf (\overline M,1,N')}$,
the $m_f(\tilde m_f)$ fields $\Ql^f$ $(\Qlt^{\dot g})$ are in the
(anti)fundamental representation of $SU(M)$, and the $n_f (n'_f)$
fields $S^t (S'^t)$ are in the vector representation of
$SO(N)$ $(SO(N'))$.  Anomaly cancellation requires that
$m_f+N=\tilde m_f + N'$.  The superpotential which truncates the
chiral ring is
\eqn\Wsusoso{W= \tr(\ts\tst)^{2(k+1)}.}
The chiral mesons  are $M_j\equiv Q(\tst\ts)^{2j} \tilde Q$,
$P_j\equiv Q (\tst\ts)^{2j}\tst ^2 Q$,
$\tilde P_j\equiv
\tilde Q\ts^2(\tst\ts)^{2j}\tilde Q$,
$N_j\equiv S(\tst\ts)^{2j}S$,
$N'_j\equiv S'(\tst\ts)^{2j}S'$,
$L_j\equiv S\ts\tst(\tst\ts)^{2j}S'$,
$\tilde R_j\equiv \tilde Q (\ts\tst)^{2j}\ts S$,
$R'_j\equiv Q (\tst\ts)^{2j}\tst S'$,
$R_r\equiv Q (\tst\ts)^{2r}\tst ^2\ts S$ and
$\tilde R'_r\equiv
\tilde Q (\ts\tst)^{2r}\ts ^2\tst S'$, with  $j=0\dots
k$ and $r=0\dots k-1$.
The $N$, $N'$, $P$ and $\tilde P$ operators are symmetric in
their flavor indices.

The dual theory has gauge group
$SU(\tilde M)\times SO(\tilde N)\times SO(\tilde N')$, with
$\tilde M= (k+1)(m_f+\tilde m_f +n_f+n'_f+4)-m_f-N$,
$\tilde N= (k+1)(m_f+\tilde m_f+n_f+n'_f+4)-n_f-M$ and
$\tilde N'= (k+1)(m_f+\tilde m_f+n_f+4)-n'_f-M$.
It has a field $Y$ in the ${\bf (\tilde M,\tilde N,1)}$,
a field $\tilde Y$ in the ${\bf (\overline{\tilde M},1,\tilde
N')}$, $n_f$ ($n'_f$) fields $q_t$ ($\tilde q_{t'}$) in the
(anti)fundamental representation of $SU(\tilde M)$,
$m_f$ ($\tilde m_f$) fields $s_f$ ($s'_{\dot g}$) in the vector
representation of $SO(\tilde N) (SO(\tilde N'))$, and singlet
fields $M_j,P_j,\tilde P_j,N_j, N'_j,L_j,R_r,\tilde R_j,
R'_j$ and $\tilde R'_r$.  All gauge anomalies
automatically vanish.  The superpotential is
\eqn\WDsusoso{W= \tr(\tsD\tstD)^{2(k+1)}
                 + {\rm singlet\ meson\ couplings}.}
We omit the meson couplings for brevity; these terms are
analogous to those of all other models
and are determined by the flavor symmetries (see sect. \Zsusoso).
We note that the mesons of the $SU(M)$ group couple to the
dual quarks of the $SO(\tilde N)\times SO(\tilde N')$ group.

If the $SU(M)$ factor confines in the electric theory,
then the $SU(\tilde M)$ factor confines in the magnetic theory;
in the infrared the models are electric and magnetic duals under
the duality of sect. \SQsoso.  If the $SO(N')$ factor confines,
then $SO(\tilde N')$
confines also, and the infrared duality is that of sect.
\SQsusos.

Additional discussion of this model is presented in sect.
\Zsusoso.

\subsec{$SU(M)\times Sp(N)$ with an antisymmetric tensor of
$SU(M)$}
\subseclab{\SQsuspa}

The gauge group is $SU(M)\times Sp(N)$; the field $X$ is
in the  ${\bf (M,2N)}$, the
field $\tilde X$ is an antisymmetric tensor in the
${\bf (\overline{\half M(M-1)},1)}$,
the $m_f (\tilde m_f)$ fields $\Ql^f$ $(\Qlt^{\dot g})$ are in
the (anti)fundamental representation of $SU(M)$, and the $n_f$
fields $S^t$ are in the fundamental representation of
$Sp(N)$.  Anomaly cancellation requires that $m_f+2N=\tilde
m_f+M-4$ and that $n_f+M$ be even.  The superpotential
\eqn\Wsuspa{W= \tr(\ts\tst\ts)^{(k+1)}}
truncates the chiral ring; the chiral mesons
are $(M_j)^{f\dot g}\equiv Q^f(\tst\ts ^2)^{j} \tilde Q^{\dot
g}$, $(P_j)^{fg}\equiv Q^f (\tst\ts ^2)^{j}\tst Q^g$,
$(N_j)^{tu}\equiv S^t(\ts\tst\ts)^{j}S^u$,
$(\tilde R_j)^{\dot g t}\equiv
\tilde Q^{\dot g} (\ts ^2\tst)^{j}\ts S^t$, $j=0\dots k$,
$(\tilde P_r)^{\dot f\dot g}\equiv
\tilde Q^{\dot f}\ts ^2(\tst\ts ^2)^{r}\tilde Q^{\dot g}$
and $(R_r)^{ft}\equiv Q^f (\tst\ts ^2)^{r}\tst\ts S^{t}$,
$r=0\dots k-1$.
The $N$, $P$ and $\tilde P$ operators are antisymmetric in their
flavor indices.

The dual theory has gauge group
$SU(\tilde M)\times Sp(\tilde N)$, with
$\tilde M= (k+1)(m_f+\tilde m_f +n_f-4)-m_f-2N$ and
$2 \tilde N= (k+1)(m_f+\tilde m_f+n_f-4)-n_f-M$.  (Anomaly
cancellation in the electric theory ensures that $m_f+\tilde
m_f+n_f$ is even.)
It has a field $Y$ in the ${\bf (\tilde M, 2\tilde N)}$, a field
$\tilde Y$ in the ${\bf (\overline{\half \tilde M(\tilde
M-1)},1)}$,
$n_f$ ($\tilde m_f$) fields $q_t$ ($\tilde q_{\dot g}$) in the
(anti)fundamental representation of $SU(\tilde M)$,
$m_f$  fields $s_f$  in the fundamental
representation of $Sp(\tilde N)$, and singlet fields
$(M_j)^{f\dot g},(P_j)^{fg},(\tilde P_r)^{\dot f\dot
g},(N_j)^{tu},
(R_r)^{ft}$ and $(\tilde R_j)^{\dot g t}$.  All gauge anomalies
automatically vanish.
The superpotential is
\eqn\WDsuspa{W= \tr(\tsD\tstD\tsD)^{(k+1)} +
{\rm singlet\ meson\ couplings}.}
We omit the meson couplings for brevity; these terms are
analogous
to those of all other models and are determined by the flavor
symmetries (see sect. \Zsuspa).
We note that the mesons of the $Sp(N)$ group couple to the
dual quarks of the $SU(\tilde M)$ group while the antiquark
mesons of $SU(M)$ couple to the antiquarks of $SU(\tilde M)$.

If the $SU(M)$ factor confines in the electric theory,
then the $SU(\tilde M)$ factor confines in the magnetic theory;
in the infrared the models are electric and magnetic duals under
the duality of \kispso\ (see sect. \SQspasym.)
If the $Sp(N)$ factor confines, then
the $Sp(\tilde N)$ confines also, and the infrared duality is
that of sect. \SQsuasym.

Additional discussion of this model is presented in sect.
\Zsuspa.

\subsec{$SU(M)\times Sp(N)\times Sp(N')$}
\subseclab{\SQsuspsp}

The gauge group is $SU(M)\times Sp(N)\times Sp(N')$; the field
$X$ is in the  ${\bf (M,2N,1)}$, the
field $\tilde X$ is in the ${\bf (\overline M,1,2N')}$,
the $m_f (\tilde m_f)$ fields $\Ql^f$ $(\Qlt^{\dot g})$ are in
the (anti)fundamental representation of $SU(M)$, and the $n_f
(n'_f)$ fields  $S^t (S'^{t'})$ are in the fundamental
representation of $Sp(N)(Sp(N'))$.  Anomaly cancellation
requires that $m_f+2N=\tilde m_f + 2N'$ and that $n_f+M$ and
$n_f'+M$ be even.

The superpotential
\eqn\Wsuspsp{W= \tr(\ts\tst)^{2(k+1)}}
truncates the chiral ring; the chiral mesons
are $M_j\equiv Q(\tst\ts)^{2j} \tilde Q$,
$P_j\equiv Q (\tst\ts)^{2j}\tst ^2 Q$,
$\tilde P_j\equiv
\tilde Q\ts^2(\tst\ts)^{2j}\tilde Q$,
$N_j\equiv S(\tst\ts)^{2j}S$,
$N'_j\equiv S'(\tst\ts)^{2j}S'$,
$L_j\equiv S\ts\tst(\tst\ts)^{2j}S'$,
$\tilde R_j\equiv \tilde Q (\ts\tst)^{2j}\ts S$,
$R'_j\equiv Q (\tst\ts)^{2j}\tst S'$, $j=0\dots k$,
$R_r\equiv Q (\tst\ts)^{2r}\tst ^2\ts S$ and
$\tilde R'_r\equiv
\tilde Q (\ts\tst)^{2r}\ts ^2\tst S'^{t'}$,
$r=0\dots k-1$.
The $N$, $N'$, $P$ and $\tilde P$ operators are antisymmetric in
their flavor indices.

The dual theory has gauge group
$SU(\tilde M)\times Sp(\tilde N)\times Sp(\tilde N')$, with
$\tilde M= (k+1)(m_f+\tilde m_f +n_f+n'_f-4)-m_f-2N$,
$2 \tilde N= (k+1)(m_f+\tilde m_f+n_f+n'_f-4)-n_f-M$ and
$2 \tilde N'= (k+1)(m_f+\tilde m_f+n_f-4)-n'_f-M$.  (Anomaly
cancellation in the electric theory ensures that $m_f+\tilde
m_f+n_f+n'_f$
is even.) It has a field $Y$ in the ${\bf (\tilde M,
2\tilde N,1)}$,
a field $\tilde Y$ in the ${\bf (\overline{\tilde M},1,
2\tilde N')}$,
$n_f$ ($n'_f$) fields $q_t$ ($\tilde q_{t'}$) in the
(anti)fundamental
representation of $SU(\tilde M)$,
$m_f$ ($\tilde m_f$) fields $s_f$ ($s'_{\dot g}$) in the
fundamental representation of $Sp(\tilde N) (Sp(\tilde N'))$, and
singlet fields $M_j,P_j,\tilde P_j,N_j, N'_j,L_j,R_r,\tilde R_j,
R'_j$ and $\tilde R'_r$.  All gauge anomalies
automatically vanish.  The superpotential is
\eqn\WDsuspsp{W= \tr(\tsD\tstD)^{2(k+1)}
                 + {\rm singlet\ meson\ couplings}.}
We omit the meson couplings for brevity; these terms are
analogous
to those of all other models and are determined by the flavor
symmetries (see sect. \Zsuspsp).
We note that the mesons of the $SU(M)$ group couple to the
dual quarks of the $Sp(\tilde N)\times Sp(\tilde N')$ group.

If the $SU(M)$ factor confines in the electric theory,
then the $SU(\tilde M)$ factor confines in the magnetic theory;
in the infrared the models are electric and magnetic duals under
the duality of sect. \SQspsp.  If the $Sp(N')$ factor confines,
then the $Sp(\tilde N')$
confines also, and the infrared duality is that of sect.
\SQsuspa.

Additional discussion of this model is presented in sect.
\Zsuspsp.

\subsec{$SU(M)\times Sp(N)$ with a symmetric tensor of $SU(M)$}
\subseclab{\SQsusps}

The gauge group is $SU(M)\times Sp(N)$; the field $X$ is
in the ${\bf (M,2N)}$, the
field $\tilde X$ is a symmetric tensor in the
${\bf (\overline{\half M(M+1)},1)}$,
the $m_f (\tilde m_f)$ fields $\Ql^f$ $(\Qlt^{\dot g})$ are in
the (anti)fundamental representation of $SU(M)$, and the $n_f$
fields $S^t$ are in the fundamental representation of
$Sp(N)$.  Anomaly cancellation requires that $m_f+2N=\tilde
m_f+M+4$ and that $n_f+M$ be even.  The superpotential
\eqn\Wsusps{W= \tr(\ts\tst\ts)^{2(k+1)}}
truncates the chiral ring; the chiral mesons
are $(M_j)^{f\dot g}\equiv Q^f(\tst\ts ^2)^{j} \tilde Q^{\dot
g}$, $(P_j)^{fg}\equiv Q^f (\tst\ts ^2)^{j}\tst Q^g$,
$(N_j)^{tu}\equiv S^t(\ts\tst\ts)^{j}S^u$,
$(\tilde R_j)^{\dot g t}\equiv
\tilde Q^{\dot g} (\ts ^2\tst)^{j}\ts S^t$, $j=0\dots 2k+1$,
$(\tilde P_r)^{\dot f\dot g}\equiv
\tilde Q^{\dot f}\ts ^2(\tst\ts ^2)^{r}\tilde Q^{\dot g}$
and $(R_r)^{ft}\equiv Q^f (\tst\ts ^2)^{r}\tst\ts S^{t}$,
$r=0\dots 2k$.  The $P_j$ ($N_j$ and $\tilde P_j$) operators are
symmetric in their flavor
indices for even (odd) $j$ and antisymmetric in their flavor
indices for odd (even) $j$.

The dual theory has gauge group
$SU(\tilde M)\times Sp(\tilde N)$, with
$\tilde M=2(k+1)(m_f+\tilde m_f+n_f)-m_f-2N$ and
$2\tilde N=2(k+1)(m_f+\tilde m_f+n_f)-n_f-M$.  (Anomaly
cancellation in the electric theory
ensures that $m_f+\tilde m_f+n_f$ is even.)
It has a field $Y$ in the ${\bf (\tilde M, 2\tilde N)}$, a field
$\tilde Y$ in the ${\bf (\overline{\half \tilde M(\tilde
M+1)},1)}$,
$n_f$ ($\tilde m_f$) fields $q_t$ ($\tilde q_{\dot g}$) in the
(anti)fundamental representation of $SU(\tilde M)$,
$m_f$  fields $s_f$  in the fundamental
representation of $Sp(\tilde N)$, and singlet fields
$(M_j)^{f\dot g},(P_j)^{fg},(\tilde P_r)^{\dot f\dot
g},(N_j)^{tu},
(R_r)^{ft}$ and $(\tilde R_j)^{\dot g t}$.  All gauge anomalies
automatically vanish.
The superpotential is
\eqn\WDsusps{W= \tr(\tsD\tstD\tsD)^{2(k+1)} +
{\rm singlet\ meson\ couplings}.}
We omit the meson couplings for brevity; these terms are
analogous to those of all other models and are determined by the
flavor symmetries (see sect. \Zsusps).
We note that the mesons of the $Sp(N)$ group couple to the
dual quarks of the $SU(\tilde M)$ group while the antiquark
mesons of $SU(M)$ couple to the antiquarks of $SU(\tilde M)$.

If the $SU(M)$ factor confines in the electric theory,
then the $SU(\tilde M)$ factor confines in the magnetic theory;
in the infrared the models are electric and magnetic duals under
the duality of \rlmsspso\ (see sect. \SQspadj.)
If the $Sp(N)$ factor confines, then
the $Sp(\tilde N)$ confines also, and the infrared duality is
that of sect. \SQsusa.

Additional discussion of this model is presented in sect.
\Zsusps.

\subsec{$SU(M)\times SO(N)$ with an antisymmetric tensor of
$SU(M)$}
\subseclab{\SQsusoa}

The gauge group is $SU(M)\times SO(N)$; the field $X$ is
in the  ${\bf (M,N)}$, the
field $\tilde X$ is an antisymmetric tensor in the
${\bf (\overline{\half M(M-1)},1)}$,
the $m_f (\tilde m_f)$ fields $\Ql^f$ $(\Qlt^{\dot g})$ are in
the (anti)fundamental representation of $SU(M)$, and the $n_f$
fields $S^t$ are in the vector representation of
$SO(N)$.  Anomaly cancellation requires that $m_f+N=\tilde
m_f+M-4$.  The superpotential
\eqn\Wsusoa{W= \tr(\ts\tst\ts)^{2(k+1)}}
truncates the chiral ring; the chiral mesons
are $(M_j)^{f\dot g}\equiv Q^f(\tst\ts ^2)^{j} \tilde Q^{\dot
g}$, $(P_j)^{fg}\equiv Q^f (\tst\ts ^2)^{j}\tst Q^g$,
$(N_j)^{tu}\equiv S^t(\ts\tst\ts)^{j}S^u$,
$(\tilde R_j)^{\dot g t}\equiv
\tilde Q^{\dot g} (\ts ^2\tst)^{j}\ts S^t$, $j=0\dots 2k+1$,
$(\tilde P_r)^{\dot f\dot g}\equiv
\tilde Q^{\dot f}\ts ^2(\tst\ts ^2)^{r}\tilde Q^{\dot g}$
and $(R_r)^{ft}\equiv Q^f (\tst\ts ^2)^{r}\tst\ts S^{t}$,
$r=0\dots 2k$.  The $P_j$ ($N_j$ and $\tilde P_j$) operators are
antisymmetric in their flavor
indices for even (odd) $j$ and symmetric in their flavor
indices for odd (even) $j$.

The dual theory has gauge group
$SU(\tilde M)\times SO(\tilde N)$, with
$\tilde M=2(k+1)(m_f+\tilde m_f+n_f+n'_f)-m_f-N$,
and $\tilde N=2(k+1)(m_f+\tilde m_f+n_f)-n_f-M$.
It has a field $Y$ in the ${\bf (\tilde M,\tilde N)}$, a field
$\tilde Y$ in the ${\bf (\overline{\half \tilde M(\tilde
M-1)},1)}$,
$n_f$ ($\tilde m_f$) fields $q_t$ ($\tilde q_{\dot g}$) in the
(anti)fundamental representation of $SU(\tilde M)$,
$m_f$  fields $s_f$  in the vector
representation of $SO(\tilde N)$, and singlet fields
$(M_j)^{f\dot g},(P_j)^{fg},(\tilde P_r)^{\dot f\dot
g},(N_j)^{tu},
(R_r)^{ft}$ and $(\tilde R_j)^{\dot g t}$.  All gauge anomalies
automatically vanish.
The superpotential is
\eqn\WDsusoa{W= \tr(\tsD\tstD\tsD)^{2(k+1)} +
{\rm singlet\ meson\ couplings}.}
We omit the meson couplings for brevity; these terms are
analogous
to those of all other models and are determined by the flavor
symmetries (see sect. \Zsusoa).
We note that the mesons of the $SO(N)$ group couple to the
dual quarks of the $SU(\tilde M)$ group while the antiquark
mesons of $SU(M)$ couple to the antiquarks of $SU(\tilde M)$.

If the $SU(M)$ factor confines in the electric theory,
then the $SU(\tilde M)$ factor confines in the magnetic theory;
in the infrared the models are electric and magnetic duals under
the duality of \rlmsspso\ (see sect. \SQsoadj.)
If the $SO(N)$ factor confines, then
$SO(\tilde N)$ confines also, and the infrared duality is that
of sect. \SQsusa.

Additional discussion of this model is presented in sect.
\Zsusoa.

\subsec{$SU(M)\times Sp(N)\times SO(N')$}
\subseclab{\SQsuspso}

The gauge group is $SU(M)\times Sp(N)\times SO(N')$; the field
$X$ is
in the  ${\bf (M,2N,1)}$, the
field $\tilde X$ is in the ${\bf (\overline M,1,N')}$,
the $m_f (\tilde m_f)$ fields $\Ql^f$ $(\Qlt^{\dot g})$ are in
the (anti)fundamental representation of $SU(M)$, and the $n_f
(n'_f)$ fields  $S^t (S'^t)$ are in the fundamental (vector)
representation of
$Sp(N)(SO(N'))$.  Anomaly cancellation requires that
$m_f+2N=\tilde m_f + N'$ and that $n_f+M$ be even.  The
superpotential
\eqn\Wsuspso{W= \tr(\ts\tst)^{4(k+1)}}
truncates the chiral ring; the chiral mesons
are $M_j\equiv Q(\tst\ts)^{2j} \tilde Q$,
$P_j\equiv Q (\tst\ts)^{2j}\tst ^2 Q$,
$\tilde P_j\equiv
\tilde Q\ts^2(\tst\ts)^{2j}\tilde Q$,
$N_j\equiv S(\tst\ts)^{2j}S$,
$N'_j\equiv S'(\tst\ts)^{2j}S'$,
$L_j\equiv S\ts\tst(\tst\ts)^{2j}S'$,
$\tilde R_j\equiv \tilde Q (\ts\tst)^{2j}\ts S$,
$R'_j\equiv Q (\tst\ts)^{2j}\tst S'$, $j=0\dots 2k+1$,
$R_r\equiv Q (\tst\ts)^{2r}\tst ^2\ts S$ and
$\tilde R'_r\equiv
\tilde Q (\ts\tst)^{2r}\ts ^2\tst S'$,
$r=0\dots 2k$.
The $N'$ and $P$  ($N$ and $\tilde P$) operators are
symmetric in their flavor
indices for even (odd) $j$ and antisymmetric in their flavor
indices for odd (even) $j$.

The dual theory has gauge group
$SU(\tilde M)\times Sp(\tilde N)\times
SO(\tilde N')$, with
$\tilde M=2(k+1)(m_f+\tilde m_f+n_f+n'_f)-m_f-2N$,
$2\tilde N=2(k+1)(m_f+\tilde m_f+n_f+n'_f)-n_f-M$,
and $\tilde N'=2(k+1)(m_f+\tilde m_f+n_f+n'_f)-n'_f-M$.
It has a field $Y$ in the ${\bf (\tilde M,2\tilde N,1)}$,
a field $\tilde Y$ in the ${\bf (\overline{\tilde M},1,\tilde
N')}$,
$n_f$ ($n'_f$) fields $q_t$ ($\tilde q_{t'}$) in the
(anti)fundamental
representation of $SU(\tilde M)$,
$m_f$ ($\tilde m_f$) fields $s_f$ ($s'_{\dot g}$) in the
fundamental
(vector) representation of $Sp(\tilde N) (SO(\tilde N'))$, and
singlet fields $M_j,P_j,\tilde P_j,N_j,
N'_j,L_j,R_r,\tilde R_j,
R'_j$ and $\tilde R'_r$.  All gauge anomalies
automatically vanish: $2\tilde{N}+n_f=\tilde N'+n'_f$ and
$\tilde M+m_f$ is even.  The superpotential is
\eqn\WDsuspso{W= \tr(\tsD\tstD)^{4(k+1)}
                 + {\rm singlet\ meson\ couplings}.}
We omit the meson couplings for brevity; these terms are
analogous
to those of all other models and are determined by the flavor
symmetries (see sect. \Zsuspso).
We note that the mesons of the $SU(M)$ group couple to the
dual quarks of the $Sp(\tilde N)\times SO(\tilde N')$ group.

If the $SU(M)$ factor confines in the electric theory,
then the $SU(\tilde M)$ factor confines in the magnetic theory;
in the infrared the models are electric and magnetic duals under
the duality of sect. \SQsosp.  If the $SO(N')$ factor confines,
then the $SO(\tilde N')$
confines also, and the infrared duality is that of sect.
\SQsusps.
Confinement of the $Sp$ groups leads to the duality  of sect.
\SQsusoa.

Additional discussion of this model is presented in sect.
\Zsuspso.

\newsec{$SU(N_c)$ with an antisymmetric flavor and $N_f$
fundamental
flavors}
\seclab{\Ssuasym}

In this and all later sections we provide details on the models
which were summarized in sect. 2.

We consider the model of sect. \SQsuasym.  This theory has an
anomaly free $SU(N_f)_L\times SU(N_f)_R\times U(1)_X\times
U(1)_B\times U(1)_R$ global symmetry with matter transforming as
\thicksize=1pt
\vskip12pt
\begintable
\tstrut  | $SU(\ncl)$ | $SU(\nfl)_L$ | $SU(\nfl)_R$ | $U(1)_X$
|
$U(1)_B$ | $U(1)_R$ \crthick
$\Ql$ | ${\bf \ncl}$ | ${\bf \nfl}$ | ${\bf 1}$ | 0 |
${1\over\ncl}$ |
$1-{\ncl+2k\over (k+1)\nfl}$\cr
$\Qlt$ | ${\bf \overline\ncl}$ | ${\bf 1}$ | ${\bf \nfl}$ | 0 |
$-{1\over\ncl}$ | $1-{\ncl+2k\over (k+1)\nfl}$\cr
$\ts$ | ${\bf asym}$ | ${\bf 1}$ | ${\bf 1}$ | $1$ |
${2\over\ncl}$ | ${1\over k+1}$\cr
$\tst$ | ${\bf \overline{asym}}$ | ${\bf 1}$ | ${\bf 1}$ | $-1$
|$-{2\over\ncl}$ | ${1\over k+1}$
\endtable
\noindent
The theory \Wsuasym\ also has a discrete
$\ZZ_{2\nf (k+1)}$ symmetry generated by
\eqn\adissyme{\eqalign{
\ts,\ \tst &\rightarrow \alpha ^{N_f}\ts,\ \tst\cr
\Ql,\ \Qlt&\rightarrow \alpha ^{-(N_c-2)}\Ql,\ \Qlt,\cr}}
with $\alpha =e^{2\pi i/2N_f(k+1)}$.

\subsec{Duality}
\subseclab{\Dsuasym}

The dual theory is $SU(\ncld)$, where $\ncld=(2k+1)\nfl-4k-\ncl$,
as described in sect. \SQsuasym.
Taking the singlets $\Pl_{j}$, $\Mm_{r}$, and $\Mmt_{r}$ to
transform as the mesons of the
electric theory, the dual theory \WDsuasym\ has the global
$SU(N_f)_L\times SU(N_f)_R\times U(1)_X\times U(1)_B\times
U(1)_R$ symmetry with the matter transforming as:
\thicksize=1pt
\vskip12pt
\begintable
\tstrut  | $SU(\ncld)$ | $SU(\nfl)_L$ | $SU(\nfl)_R$ | $U(1)_X$
| $U(1)_B$ | $U(1)_R$ \crthick
$\ql$ | ${\bf \ncld}$ | ${\bf \overline\nfl}$ | ${\bf 1}$ |
${k(\nfl-2)\over\ncld}$ | ${1\over\ncld}$ |
$1-{\ncld+2k\over (k+1)\nfl}$\cr
$\qlt$ | ${\bf \overline\ncld}$ | ${\bf 1}$ | ${\bf
\overline\nfl}$ |
$-{k(\nfl-2)\over\ncld}$ |
$-{1\over\ncld}$ | $1-{\ncld+2k\over (k+1)\nfl}$\cr
$\tsD$ | ${\bf asym}$ | ${\bf 1}$ | ${\bf 1}$ |
${\ncl-\nfl\over\ncld}$ | ${2\over\ncld}$ | ${1\over k+1}$\cr
$\tstD$ | ${\bf \overline{asym}}$ | ${\bf 1}$ | ${\bf 1}$ |
$-{\ncl-\nfl\over\ncld}$ | $-{2\over\ncld}$ | ${1\over k+1}$\cr
$\Pl_{j}$ | ${\bf 1}$ | ${\bf \nfl}$ | ${\bf \nfl}$ | 0 | 0 |
${\ncld-\ncl+(2j+1)\nfl\over\nfl(k+1)}$\cr
$\Mm_{r}$ | ${\bf 1}$ | ${\bf asym}$ | ${\bf 1}$ | $-1$ | 0 |
${\ncld-\ncl+2\nfl(r+1)\over\nfl(k+1)}$\cr
$\Mmt_{r}$ | ${\bf 1}$ | ${\bf 1}$ |
${\bf {asym}}$ | $1$ | 0 |
${\ncld-\ncl+2\nfl(r+1)\over\nfl(k+1)}$
\endtable
\noindent
These are anomaly free in the dual gauge theory (which is our
first check on the duality).

Taking the gauge singlets of the dual theory to transform as the
mesons of the electric theory under the discrete symmetry
\adissyme, the
dual superpotential respects the symmetry provided that
the other fields transform as
\eqn\adissymm{\eqalign{\tsD,\ \tstD&\rightarrow \CC^{\nfl}\;
\alpha^{\nfl}\; \tsD,\ \tstD\cr
\ql,\qlt &\rightarrow \CC^{\nfl}\; e^{-2\pi i(2/\nfl)}
\alpha^{-(\ncld-2)}\; \ql,\ \qlt,\cr}}
where $\CC$ is charge conjugation.   The transformation
\adissymm\
is indeed non-anomalous in the dual gauge theory: it is not
violated by instantons in the dual gauge group.
Note the mixing of the discrete symmetry with charge conjugation
and with the $\ZZ_{\nfl}$ centers of the $SU(\nfl)_L\times
SU(\nfl)_R$ flavor symmetries.

At $\ev{X}=\ev{\tilde X}=\ev{Q}=\ev{\tilde Q}=0$
the full flavor symmetry is unbroken and the 't Hooft anomalies
computed with the fermions of the electric theory must match
those computed with the fermions of the magnetic theory.  These
conditions are indeed satisfied; in both the electric and
 magnetic theories, we find:
\eqn\asthoofti{\eqalign{U(1)_R\qquad &-{N_c(N_c+3k)\over
k+1}-1\cr
U(1)_R^3\qquad &N_c^2-1-{2N_c(N_c+2k)^3\over N_f^2(k+1)^3}-
N_c(N_c-1){k^3\over (k+1)^3}\cr
SU(N_f)^3\qquad &N_cd_3(N_f)\cr
SU(N_f)^2U(1)_R\qquad &-{N_c(N_c+2k)\over (k+1)N_f}d_2(N_f)\cr
SU(N_f)^2U(1)_X\qquad &0\cr
SU(N_f)^2U(1)_B\qquad &d_2(N_f)\cr
U(1)_X^2U(1)_R\qquad &-N_c(N_c-1){k\over k+1}\cr
U(1)_B^2U(1)_R\qquad &-2{(2k+1)\over k+1}\cr
U(1)_XU(1)_BU(1)_R\qquad &-2(N_c-1){k\over k+1},}}
where $d_2(N_f)$ and $d_3(N_f)$ are the quadratic and cubic
$SU(N_f)$ Casimirs of the fundamental representation.

The electric theory has a variety of baryon-like operators, with
gauge indices contracted with an $\epsilon$-tensor.  Under the
duality transformation, these operators are mapped in a
non-trivial
manner to baryon-like operators of the magnetic gauge theory.
An example of this mapping is
\eqn\abmapi{\ts^{n}\Ql^{\ncl-2n}\rightarrow \tsD^{k(\nfl-2)-n}
\ql^{\nfl+2n-\ncl},}
where we have suppressed the flavor indices.  Note that this map
is consistent with all of the global symmetries, including the
discrete symmetries mentioned above (the charge conjugation $\CC$
in \adissymm\ is needed for the map to respect the discrete
symmetry). We will see in the next
section that deforming by such an operator in the electric theory
leads to the correct physics in the dual theory.

\subsec{Deformations: superpotential}
\subseclab{\DFSsuasym}

Consider deforming \Wsuasym\ to include lower order terms:
\eqn\WXDsuasym{W_{pert}=\sum_{\ell=0}^k\lk{\ell}\;
\tr(\ts\tst)^{\ell+1},}
The theory has multiple vacua with $\ev{\Ql}=0$ and
$\ev{\ts\tst}\neq 0$ satisfying
$\partial W/\partial\ts=\partial W/\partial\tst=0$.
For  $N_c=2n_c+1$ the D-flat directions
with $\ev{\Ql}=0$ are
\eqn\Noddfd{\ev{\ts}=\ev{\tst}=
\pmatrix{x_1\sigma _2&\ &\ &\ \cr\ &\ &\ &\ \cr\ &\
&x_{n_c}\sigma _2&\ \cr\ &\ &\ &0\cr},}
generically breaking $SU(2n_c+1)$ to $Sp(1)^{n_c}$.
For $N_c=2n_c$ the D-flat directions with $\ev{\Ql}=0$ are
\eqn\Nevenfd{\ev{\ts}=\pmatrix{x_1\sigma _2&\ &\ \cr\ &\  &\ \cr\
&\ &x_n\sigma _2\cr}\qquad \ev{\tst}
=\pmatrix{\tilde x_1\sigma _2&\ &\ \cr\ &\ \ &\ \cr\ &\
&\tilde x_n\sigma _2\cr}\quad\hbox{with}\quad |x_i|^2-|\tilde
x_i|^2={\rm const.},}
again generically breaking $SU(2n_c)$ to $Sp(1)^{n_c}$.  The
theory with superpotential \WXDsuasym\ has multiple vacua with
expectation values of the
form \Noddfd\ or \Nevenfd\ with $x_i$ and $\tilde x_i$ satisfying
$W'(x_i\tilde x_i)x_i=W'(x_i\tilde x_i)\tilde x_i=0$.  The
equation $W'(z)=0$ has $k$ solutions $z_l$ which are generically
distinct.  In addition, there is the solution $x_i=\tilde x_i=0$.
Let $\vl_{0}$ be the number of eigenvalues with $x_i=\tilde
x_i=0$ and $\vl_{l}$ be the number with $x_i\tilde x_i$ equal to
$z_\ell$, with
$\sum_{\ell=0}^k\vl_{\ell}=n_c$ for $\nc=2n_c+\om$ with
$\om\equiv 0\ (1)$  for $\nc$ even (odd).  In such a vacuum
the gauge group is broken by $\ev{\ts\tst}$ as:
\eqn\BGsuasym{SU(\nc)\rightarrow SU(2\vl_0+\om)\times
Sp(\vl_{1})\times Sp(\vl_{2})\times \ldots\times Sp(\vl_{k}).}
In the generic situation $\ts$ and $\tst$ are massive in each
vacuum and can be integrated out.  The $SU(2\vl_0+\om)$ gauge
group has $\nf$ flavors in the fundamental and each $Sp(\vl_l)$
factor has $2\nf$ matter fields in the  $2\vl_l$-dimensional
fundamental representation.  There is a stable vacuum (in this
theory as well as that with superpotential \Wsusym) provided
$\nf\geq 2\vl_0+\om$ \ref\ads{I. Affleck, M. Dine, and N.
Seiberg, \np{241}{1984}{493}; \np{256}{1985}{557}.} and
$\nf>\vl_\ell$, $\ell>0$ \intpou.

An interesting example of the deformation \WXDsuasym\ is a mass
term $m\ts\tst$. In any one vacuum, the unbroken symmetry will
be of the form \BGsuasym.   In the magnetic theory the analysis
is much the same and the gauge group is broken to
\eqn\BGsuasymD{SU(\ncld)\rightarrow SU(2\vlt_0+\tilde\om)\times
Sp(\vlt_1)\times\ldots\times Sp(\vlt_k).}
The fields $\tsD$ and $\tstD$ are massive in each factor and can
be integrated out. The duality mapping is
$2\vlt_0+\tilde\om=N_f-(2\vl_0+\om)$,
$\vlt_\ell=\nf-2-\vl_\ell$ for $\ell>0$; in the $SU$ factor the
duality is that of \sem\ and in each $Sp$ factor it is that
of \refs{\sem, \intpou}.

Next we consider deforming the theory by giving a mass to one
flavor of the electric quarks $\Ql$.
The low energy theory is an $Sp(\nc)$ theory with matter $\ts$
and superpotential \Wsuasym\ with one fewer flavor, $\hat
N_f=\nf-1$; having fewer flavors, it is more strongly coupled in
the infrared. In the dual theory this perturbation corresponds
to adding a term
$m(\Pl_{0})^{\nf,\nf}$ to \WDsuasym.  The
$\Pl_{j}$ equations of motion imply that the vacua of this theory
satisfy
\eqn\speom{\ql_{\nf}(\tstD\tsD)^{\ell-1}\qlt_{\nf}=
-m\delta_{\ell,k+1};\;\;\ell=1,\ldots, k+1}
which, along with some additional conditions, give expectation
values proportional to:
\eqn\spmv{\eqalign{ \ev{\ql_{\nf}}_c=&\;\delta_{c,1};\cr
\ev{\qlt_{\nf}}_c=&\;\delta_{c,2k+1};\cr
\tsD_{c,d}=\tstD^{c+1,d+1}=
&\cases{\delta_{c+1,d}&$c=2r-1;\ r=1,\cdots, k$\cr
-\delta_{d+1, c}&$d=2r-1;\ r=1,\cdots, k$\cr
0& otherwise.\cr}\cr}}
These expectation values break the magnetic
$SU((2k+1)\nf-4k-\nc)$
gauge group to $SU((2k+1)(\nf-1)-4k-\nc)$ with $\nf-1$ remaining
light flavors. The low energy magnetic theory is at weaker
coupling and is the dual of the low energy electric theory.

We can also consider perturbing the electric theory
by the other meson operators.  The corresponding perturbation in
the dual theory yields a low energy theory which is dual to the
perturbed electric theory.

\subsec{Deformations: flat directions}
\subseclab{\DFFsuasym}

The electric theory can also be deformed by giving fields
expectation values along flat directions.  Such a perturbation
generally breaks the gauge group and takes the theory to weaker
coupling in the infrared.
The corresponding perturbation in the dual theory should lead to
an infrared theory which is dual to the infrared electric theory.
We will only mention an especially interesting flat direction:
that along which the baryon operator $B_n=\ts^n\Ql^{\nc-2n}$ gets
an expectation value.
In terms of the elementary fields, this flat direction is
$\ev{\Ql^{f,c}}=\sqrt{2}a\delta^{f,c}$ for $c\leq\nc-2n$ and zero
otherwise, $\ev{\ts^{cd}}=a(\delta^{c,d+1}-\delta^{d,c+1})$ for
$c,d>\nc-2n$ and zero otherwise, with all other expectation
values zero. Along this flat direction, which is not lifted by
the superpotential \Wsuasym, the gauge group is broken to
$Sp(n)$. In the low energy theory the matter field $\tst$ yields an
antisymmetric $Sp(n)$
tensor $\hat\ts$ and $N_c-2n$ fields in the $2n$-dimensional
fundamental representation of $Sp(n)$.  The $\Qlt$ and the
$N_f-N_c+2n$ flavors of $\Ql$ not
entering in $B_n$ yield additional matter fields in the
fundamental representation of $Sp(n)$, bringing the total number
to $2N_f$.  In addition, there are a variety of singlets.   The
low energy theory has a superpotential inherited from \Wsuasym,
$W=\tr \widehat \ts^{k+1}$,
where indices are contracted using $\ev{\ts}$.  This low energy
theory is that considered in \kispso, along with some additional
singlets.

In the dual theory the operator $B_n$ labeling the above flat
direction is mapped as in \abmapi\ to $b_{n}=
\tsD^{k(N_f-2)-n}\ql^{N_f+2n-N_c}$,
which gets an expectation value along the flat direction
$\ev{\ql_{f,c}}=\sqrt{2}a\delta_{f,c}$ for
$c> N_c-2n$ and zero otherwise,
$\ev{\tsD_{cd}}=a(\delta _{c,d+1}-\delta
_{c+1,d})$ for $ c,d\leq k(N_f-2)-n$ and
zero otherwise.  The dual $SU(\ncld)$ gauge group is broken to
$Sp(\tilde n)$, with $\tilde n=k(N_f-2)-n$. The low energy
$Sp(\tilde n)$ theory has $2N_f$ matter fields in the fundamental
representation, with $N_f-N_c+2n$ coming from $\tstD$ and the
remainder coming from $\qlt$ and the $N_c-2n$ fields $\ql$ not
entering in $b_{n}$. In addition, there is an antisymmetric
tensor  coming from $\tstD$, along with some singlets, some of
which are
eliminated by \WDsuasym. The low energy magnetic theory gets
a superpotential from \WDsuasym, such that it is precisely that
shown in \kispso\ to
be the dual of the above low energy electric theory.
The $Sp(n)$ duality discussed in \kispso\ is thus inherited
{}from the $SU(\nc)$ model discussed here.

\newsec{$SU(\ncl)$ with a symmetric flavor and $\nfl$ fundamental
flavors}
\seclab{\Ssusym}

We now consider the theory described in sect. \SQsusym. It is
very
similar to the model of the previous section.
The theory has an anomaly-free
$SU(N_f)_L\times SU(N_f)_R\times U(1)_\ts\times U(1)_B\times
U(1)_R$
global symmetry with matter in the representations
\thicksize=1pt
\vskip12pt
\begintable
\tstrut  | $SU(\ncl)$ | $SU(\nfl)_L$ | $SU(\nfl)_R$ | $U(1)_X$|
$U(1)_B$ | $U(1)_R$ \crthick
$\Ql$ | ${\bf \ncl}$ | ${\bf \nfl}$ | ${\bf 1}$ | 0 |
${1\over\ncl}$ |
$1-{\ncl-2k\over (k+1)\nfl}$\cr
$\Qlt$ | ${\bf \overline\ncl}$ | ${\bf 1}$ | ${\bf \nfl}$ | 0 |
$-{1\over\ncl}$ | $1-{\ncl-2k\over (k+1)\nfl}$\cr
$\ts$ | ${\bf sym}$ | ${\bf 1}$ | ${\bf 1}$ | $1$ |
${2\over\ncl}$ | ${1\over k+1}$\cr
$\tst$ | ${\bf \overline{sym}}$ | ${\bf 1}$ | ${\bf 1}$ | $-1$|
$-{2\over\ncl}$ | ${1\over k+1}$
\endtable
\noindent
In addition the theory \Wsusym\ has a discrete $\ZZ_{2\nf(k+1)}$
symmetry generated by
\eqn\sdissyme{\eqalign{
\ts,\ \tst&\rightarrow \alpha^{\nfl}\ts,\ \tst\cr
\Ql,\ \Qlt &\rightarrow \alpha^{-(\nc+2)}\Ql,\ \Qlt,}}
where $\alpha=e^{2\pi i/2\nfl(k+1)}$.

\subsec{Duality}
\subseclab{\Dsusym}

The dual theory is $SU(\ncld)$, where $\ncld=(2k+1)N_f+4k-\ncl$,
as described in sect. \SQsusym.
Taking the singlets $\Pl_{j}$, $\Mm_{r}$, and $\Mmt_{r}$ to
transform as in the electric theory, the dual theory \WDsusu\ has
the $SU(N_f)_L\times SU(N_f)_R\times U(1)_X\times U(1)_B\times
U(1)_R$ global flavor symmetry with the matter transforming as:
\thicksize=1pt
\vskip12pt
\begintable
\tstrut  | $SU(\ncld)$ | $SU(\nfl)_L$ | $SU(\nfl)_R$ | $U(1)_X$|
$U(1)_B$ | $U(1)_R$ \crthick
$\ql$ | ${\bf \ncld}$ | ${\bf \overline\nfl}$ | ${\bf 1}$ |
${k(\nfl+2)\over\ncld}$ | ${1\over\ncld}$ |
$1-{\ncld-2k\over (k+1)\nfl}$\cr
$\qlt$ | ${\bf \overline\ncld}$ | ${\bf 1}$ | ${\bf
\overline\nfl}$ |
$-{k(\nfl+2)\over\ncld}$ |
$-{1\over\ncld}$ | $1-{\ncld-2k\over (k+1)\nfl}$\cr
$\tsD$ | ${\bf sym}$ | ${\bf 1}$ | ${\bf 1}$ |
${\ncl-\nfl\over\ncld}$ | ${2\over\ncld}$ | ${1\over k+1}$\cr
$\tstD$ | ${\bf \overline{sym}}$ | ${\bf 1}$ | ${\bf 1}$ |
$-{\ncl-\nfl\over\ncld}$ | $-{2\over\ncld}$ | ${1\over k+1}$\cr
$\Pl_{j}$ | ${\bf 1}$ | ${\bf \nfl}$ | ${\bf \nfl}$ | 0 | 0 |
${\ncld-\ncl+(2j+1)\nfl\over\nfl(k+1)}$\cr
$\Mm_{r}$ | ${\bf 1}$ | ${\bf sym}$ | ${\bf 1}$ | $-1$ | 0 |
${\ncld-\ncl+2\nfl(r+1)\over\nfl(k+1)}$\cr
$\Mmt_{r}$ | ${\bf 1}$ | ${\bf 1}$ |
${\bf {sym}}$ | $1$ | 0 |
${\ncld-\ncl+2\nfl(r+1)\over\nfl(k+1)}$
\endtable
\noindent
It is a check on the duality that these are anomaly free in the
dual gauge theory.  Similarly, the
discrete symmetry \sdissyme\ is a symmetry of the dual theory
with the singlets transforming as the mesons of the electric
theory provided the other fields transform as
\eqn\sdissymm{\eqalign{\tsD,\ \tstD&\rightarrow
\alpha^{\nfl}\tsD,\ \tstD\cr
\ql,\ \qlt &\rightarrow
e^{2\pi i(2/N_f)}\alpha^{-(\ncld+2)}\ql,\ \qlt,}}
which is indeed a non-anomalous discrete symmetry of the dual
gauge theory.  (As in \adissymm\ and \refs{\sem, \isson}, a factor
of charge conjugation may also be needed in \sdissymm\ in order for
the baryons to transform properly under duality.)

We have verified that the 't Hooft anomaly matching conditions
are satisfied, providing a highly non-trivial check on the
duality.

The electric theory has a variety of baryon-like operators, with
gauge indices contracted with  $\epsilon$-tensors.  Under the
duality transformation, these operators are mapped in a
non-trivial
manner to baryon-like operators of the magnetic gauge theory.
An example of this mapping is
\eqn\sbmapi{
\ts^{n}\Ql^{\ncl-n}\Ql^{\ncl-n}\rightarrow \tsD^{k(2\nfl+4)-n}
\ql^{\nfl+n-\ncl}\ql^{\nfl+n-\ncl},}
contracted with two $\epsilon$-tensors, where we have suppressed
all indices.  Note that this map is consistent
with all of the global symmetries, including the discrete
symmetries mentioned above. (Because it involves two epsilon
tensors it is insensitive to the possible $\CC$ in \sdissymm).
We will see in the next section that
deforming by such an operator in the electric theory leads to the
correct physics in the dual theory.

\subsec{Deformations: superpotential}
\subseclab{\DFSsusym}

Consider deforming \Wsusym\ to include lower order terms:
\eqn\WXDsusym{W_{pert}=
\sum_{\ell=0}^k\lk{\ell}\;\tr(\ts\tst)^{\ell+1},}
The theory has multiple vacua with $\ev{\Ql}=0$ and
$\ev{\ts\tst}\neq 0$ satisfying $\partial W/\partial\ts=\partial
W/\partial\tst=0$.  The D-flat directions with $\ev{\Ql}=0$ are
\eqn\snoqfd{\ev{\ts}=\pmatrix{x_1&\ &\ \cr\ &\ &\ \cr\ &\
&x_n\cr}\qquad \ev{\tst}
=\pmatrix{\tilde x_1&\ &\ \cr\ &\ &\ \cr\ &\
&\tilde x_n\cr}\quad\hbox{with}\quad |x_i|^2-|\tilde
x_i|^2={\rm const.},}
generically completely breaking $SU(\ncl)$.  The theory with
superpotential \WXDsusym\ has multiple vacua with expectation
values
of the form \snoqfd\ with $x_i$ and $\tilde x_i$ satisfying
$W'(x_i\tilde x_i)x_i=W'(x_i\tilde x_i)\tilde x_i=0$.  The
equation $W'(z)=0$ has $k$ solutions $z_\ell$ which are
generically distinct.
There is also the solution $x_i=\tilde x_i=0$.
Let $\vl_0$ be the number of
eigenvalues with $x_i=\tilde x_i=0$ and let
$\vl_\ell$ be the number of $x_i\tilde x_i$ equal to $z_\ell$,
with $\sum_{\ell=0}^k\vl_\ell=\ncl$.  In such a vacuum
the gauge group is broken by $\ev{\ts\tst}$ as:
\eqn\BGsusym{SU(\ncl)\rightarrow SU(\vl_0)\times
SO(\vl_1)\times SO(\vl_2)\times \cdots
\times SO(\vl_k).} In the generic situation $\ts$ and $\tst$ are
massive in each vacuum
and can be integrated out.  The $SU(\vl_0)$ factor has $\nfl$
fundamental flavors and each $SO(\vl_\ell)$
factor has $2N_f$ matter fields in the $\vl_\ell$-dimensional
vector representation.  The vacuum is stable provided
$\nfl\geq\vl_0$ \ads\ and $2\nf\geq\vl_\ell-4$ for every
$\ell=1,\ldots, k$ \isson.

An interesting example of the deformation \WXDsusym\ is a mass
term $m\ts\tst$. In any one vacuum, the unbroken symmetry will
be of the form \BGsusym.   In the magnetic theory the analysis
is much the same and the gauge group is broken to
\eqn\BGsuasymD{SU(\ncld)\rightarrow SU(\vlt_0)\times
SO(\vlt_1)\times SO(\vlt_2)\times\ldots\times SO(\vlt_k).}
The fields $\tsD$ and $\tstD$ are massive in each factor and can
be integrated\ out. The duality mapping is $\vlt_0=N_f-\vl_0$,
$\vlt_\ell=2\nf+4-\vl_\ell$ for $\ell>0$; in the $SU$ factor the
duality is that of \sem\ and in each $SO$ factor it is that
of \refs{\sem , \isson}.

Next consider deforming the electric theory by adding a mass for
the $N_f$-th quark flavor.  In the magnetic theory the term
$m(M_0)^{N_f,N_f}$ is added to \WDsusym.  The vacuum has
\eqn\smpv{\eqalign{\ql_{\nf}(\tstD\tsD)^{j-1}\qlt_{\nf}&=-m\delta
_{j,k+1};\ j=1,\ldots, k+1\cr
\ql_{N_f} (\tstD\tsD)^{r-1}\tstD\ql_{N_f}&=0\cr
\qlt_{N_f}\tsD(\tstD\tsD)^{r-1}\qlt_{N_f}&=0,\cr}}
which give expectation values $\ev{\ql_{N_f}}$,
$\ev{\qlt_{N_f}}$,
$\ev{\tsD}$ and $\ev{\tstD}$, breaking $SU((2k+1)\nf+4k-\ncl)$
to $SU((2k+1)(\nf-1)+4k-\ncl)$ with $N_f-1$ flavors.
The low energy magnetic theory is the
dual of the low energy electric theory.

\subsec{Deformations: flat directions}
\subseclab{\DFFsusym}

Consider the flat direction along which
the operator $B_n=\ts^n\Ql^{\ncl-n}\Ql^{\ncl-n}$ gets an
expectation
value. In terms of the elementary fields this flat direction is
$\ev{\Ql^{f,c}}=\sqrt{2}a\delta^{f,c}$ for $c\leq\ncl-n$ and zero
otherwise, $\ev{\ts^{cd}}=a\delta^{c,d}$ for $c>\ncl-n$
and zero otherwise, with all other expectation values zero.
Along this flat direction, which is not lifted by the
superpotential
\Wsusym, the gauge group is broken to $SO(n)$.
In the low energy theory there are $2N_f$ matter fields in the
$n$ dimensional vector representation of $SO(n)$, with some
coming from $\tst$ and others from the $\Ql$ and $\Qlt$ not
entering in $B_n$. There is also a symmetric $SO(n)$
tensor $\widehat\ts$ from $\tst$, along with a variety of
singlets. The low energy theory has a superpotential inherited
{}from \Wsusym,
$W=\Tr \widehat \ts^{k+1}$,
where indices are contracted using $\ev{\ts}$.  Up to some
additional
singlets, this low energy theory is that considered in \kispso.

In the dual theory the operator $B_n$ labeling the above flat
direction is mapped as in \sbmapi\ to $b_n =
\tsD^{k(2N_f+4)-n}\ql^{N_f+n-\ncl}
\ql^{N_f+n-\ncl}$, which gets an expectation value along the flat
direction $\ev{\ql_{f,{ c}}}=\sqrt{2}a\delta_{f,{ c}}$ for
$ c> \ncl-n$ and zero otherwise,
$\ev{\tsD_{c d}}=a\delta_{c,d}$ for
$c\leq k(2N_f+4)-n$ and zero
otherwise.  When this operator gets an expectation value the
dual $SU(\ncld)$ gauge
theory is therefore broken to $SO(\tilde n)$, with
$\tilde n=k(2N_f+4)-n$.
This low energy $SO(\tilde n)$ theory has $2N_f$
matter fields in the $\tilde n$-dimensional vector
representation, with some coming from $\tstD$ and others
{}from $\qlt$ and the $\ql$ not entering in $b_n$.
There is also a symmetric tensor coming from $\tstD$, and some
singlets, some of which are eliminated by \WDsusym.
The low energy magnetic theory gets
a superpotential from \WDsusym, such that it is precisely that
shown in \kispso\ to be the dual of the above low energy $SO(n)$
electric theory. The $SO(n)$ duality considered in \kispso\ is
thus inherited from the $SU(\nc)$ model discussed here.

\newsec{$SU(\ncl)$ with an antisymmetric tensor and a symmetric
tensor}
\seclab{\Ssusa}

We now consider the theory described in sect. \SQsusa. This is
a  chiral theory; anomaly cancellation requires $m_f-\tilde
m_f=8$. The theory has an anomaly-free $SU(m_f)\times
SU(\tilde m_f)\times U(1)_\ts\times U(1)_B\times U(1)_R$
global symmetry with matter in the representations
\thicksize=1pt
\vskip12pt
\begintable
\tstrut  | $SU(\ncl)$ | $SU(m_f)$ | $SU(\tilde m_f)$ | $U(1)_X$|
$U(1)_B$ | $U(1)_R$ \crthick
$\Ql$ | ${\bf \ncl}$ | ${\bf m_f}$ | ${\bf 1}$ |
$-(2k+1)+{2(4k+3)\over m_f}$ | ${1\over\ncl}$ |
$1-{\ncl+2(4k+3)\over 2(k+1)m_f}$\cr
$\Qlt$ | ${\bf \overline\ncl}$ | ${\bf 1}$ | ${\bf \tilde m_f}$|
$2k+1+{2(4k+3)\over \tilde m_f}$| $-{1\over\ncl}$ |
$1-{\ncl-2(4k+3)\over 2(k+1)\tilde m_f}$\cr
$\ts$ | ${\bf asym}$ | ${\bf 1}$ | ${\bf 1}$ | $1$ |
${2\over\ncl}$ | ${1\over 2(k+1)}$\cr
$\tst$ | ${\bf \overline{sym}}$ | ${\bf 1}$ | ${\bf 1}$ | $-1$|
$-{2\over\ncl}$ | ${1\over 2(k+1)}$
\endtable
\noindent
The dimensions of chiral operators at an interacting
fixed point are determined by the $\hat R$ charge appearing in
the same multiplet with the energy momentum tensor; however, in
this case $U(1)_{\hat R}$ may differ from $U(1)_R$ in the above
table by a linear combination of  $U(1)_X$ and $U(1)_B$.
The charge $U(1)_R$ thus determines dimensions only of operators
neutral under $U(1)_X$ and $U(1)_B$ (for example, terms in
the superpotential.)  These statements will be true in all our
chiral models.

In addition the theory \Wsusa\ has a discrete
$\ZZ_{2( m_f+\tilde m_f)(k+1)}$ symmetry generated by
\eqn\sdissase{\eqalign{
\ts,\ \tst&\rightarrow \alpha^{(m_f+\tilde m_f)/2}\ts,\ \tst\cr
\Ql,\ \Qlt&\rightarrow \alpha^{-N_c} \Ql,\ \Qlt,\cr
}}
where $\alpha=e^{2\pi i/2 (m_f+\tilde m_f)(k+1)}$.

\subsec{Duality}
\subseclab{\Dsusa}

The dual theory is $SU(\ncld)$, where
$\ncld=\half(4k+3)(m_f+\tilde m_f)-\ncl$, as
described in sect. \SQsusa.
Taking the singlets $\Pl_{j}$, $\Mm_{r}$, and $\Mmt_{r}$ to
transform as in the electric theory, the dual theory \WDsusa\ has
the $SU(m_f)\times SU(\tilde m_f)\times U(1)_X\times U(1)_B\times
U(1)_R$ global flavor symmetry with the other matter transforming
as:
\thicksize=1pt
\vskip12pt
\begintable
\tstrut  | $SU(\ncld)$ | $SU(m_f)$ | $SU(\tilde m_f)$ | $U(1)_X$|
$U(1)_B$ | $U(1)_R$ \crthick
$\ql$ | ${\bf \ncld}$ | ${\bf \overline m_f}$ | ${\bf 1}$ |
$2k+1 - {2(4k+3)\over m_f}$| ${1\over\ncld}$ |
$1-{\ncld+2(4k+3)\over 2(k+1)m_f}$\cr
$\qlt$ | ${\bf \overline\ncld}$ | ${\bf 1}$ |
${\bf \overline{\tilde m_f}}$ |
$-(2k+1) - {2(4k+3)\over \tilde m_f} $ |
$-{1\over\ncld}$ | $1-{\ncld-2(4k+3)\over 2(k+1)\tilde m_f}$\cr
$\tsD$ | ${\bf asym}$ | ${\bf 1}$ | ${\bf 1}$ |
$-1$ | ${2\over\ncld}$ | ${1\over 2(k+1)}$\cr
$\tstD$ | ${\bf \overline{sym}}$ | ${\bf 1}$ | ${\bf 1}$ |
$1$ | $-{2\over\ncld}$ | ${1\over 2(k+1)}$
\endtable
\noindent
It is a check on the duality that these are anomaly free in the
dual gauge theory.

Similarly, the
discrete symmetry \sdissase\ is a symmetry of the dual theory,
with the singlets transforming as the mesons of the electric
theory, provided the other fields transform as
\eqn\sdissasm{\eqalign{
\tsD &\rightarrow e^{2\pi i B_\tsD p}
        \alpha^{(m_f+\tilde{m}_f)/2}\tsD,\cr
\tstD &\rightarrow e^{-2\pi i B_{\tsD} p}
        \alpha^{(m_f+\tilde{m}_f)/2}\tstD,\cr
\ql &\rightarrow e^{2\pi i B_\ql p} \alpha^{-\ncd}\ql,\cr
\qlt &\rightarrow e^{-2\pi i B_{\ql} p} \alpha^{-\ncd}\qlt.\cr
}}
Here, $p={1\over 4}\left[
m_f+\tilde{m}_f+2{m_f+\tilde{m}_f+4\ncd\over
(k+1)(m_f+\tilde{m}_f)}\right]$, and $B_Y, B_{\ql}$ are the $U(1)_B$
charges of $\tsD$ and $\ql$. This is indeed a non-anomalous
discrete symmetry of the dual gauge theory.

We have
verified that the 't Hooft anomaly matching conditions are
satisfied, providing a highly non-trivial check on the duality.

The electric theory has a variety of baryon-like operators, with
gauge indices contracted with  $\epsilon$-tensors.  Under the
duality transformation, these operators are mapped in a
non-trivial manner to baryon-like operators of the magnetic gauge
theory. Two examples of this mapping are
\eqn\sabmapi{\eqalign{
\ts^{n}\Ql^{\ncl-2n}&\rightarrow
   \tsD^{(k+1/2)(m_f+\tilde m_f)-n-2}\ql^{m_f+2n-\ncl}\cr
\tst^{n}\Qlt^{\ncl-n}\Qlt^{\ncl-n}&\rightarrow
   \tstD^{(2k+1)k(m_f+\tilde m_f)-n+4} \qlt^{\tilde m_f+n-\ncl}
            \qlt^{\tilde m_f+n-\ncl}
,\cr}}
contracted with $\epsilon$-tensors, where we have suppressed all
indices.  Note that this map is consistent
with all of the global symmetries, including the discrete
symmetries mentioned above. We will see in the next
section that deforming by such an operator in the electric theory
leads to the correct physics in the dual theory.

\subsec{Deformations: superpotential}
\subseclab{\DFSsusa}

Consider deforming \Wsusa\ to include lower order terms:
\eqn\WXDsusa{W_{pert}=
\sum_{\ell=0}^k\lk{\ell}\;\tr(\ts\tst)^{2(\ell+1)},}
The theory has multiple vacua with $\ev{\Ql}=0$ and
$\ev{\ts\tst}\neq 0$
satisfying $\partial W/\partial\ts=\partial W/\partial\tst=0$.
The D-flat directions with $\ev{\Ql}=0$ require both $X$ and
$\tilde X$ to be $2\times 2$ block diagonal
\eqn\sanoqfd{\ev{\ts}=\pmatrix{x_1\sigma_2&\ &\ \cr\ &\ &\ \cr\
&\ &x_n\sigma_2\cr}\qquad \ev{\tst}
=\pmatrix{\tilde x_1\bigone&\ &\ \cr\ &\ &\ \cr\ &\
&\tilde x_n\bigone\cr}\quad\hbox{with}\quad |x_i|^2-|\tilde
x_i|^2={\rm const.},}
generically breaking $SU(\ncl)$ to $U(1)^{{\nc\over2}}$.
(If $\nc$ is odd then there is a zero in the lower right
corner of $X$ and $\tilde X$.)  The theory with
superpotential \WXDsusa\ has multiple vacua with expectation
values of the form \sanoqfd\ with $x_i$ and $\tilde x_i$
satisfying $W'(x_i\tilde x_i)x_i=W'(x_i\tilde x_i)\tilde x_i=0$.
The equation $W'(z)=0$ has $k$ solutions $z_\ell$ which are
generically distinct. There is also the solution $x_i=\tilde
x_i=0$.  Let $\vl_0$ be the number of
eigenvalues $x_i=\tilde x_i=0$ and let
$\vl_\ell$ be the number of $x_i\tilde x_i$ equal to $z_\ell$;
then $2(\vl_0+\sum_{\ell=0}^k\vl_\ell)+\eta=\ncl$, where
$\eta=0$ for even $\nc$ and $\eta=1$ for odd $\nc$.  In such a vacuum
the gauge group is broken by $\ev{\ts\tst}$ to
\eqn\BGsusa{SU(\ncl)\rightarrow SU(2\vl_0+\eta)\times
U(\vl_1)\times U(\vl_2)\times \cdots
\times U(\vl_k).}
The first group factor is a model of the type studied in this
section, with $k=0$; it has both a symmetric and an antisymmetric
tensor and $m_f(\tilde m_f)$ fields in the (anti)fundamental
representation.  In the remaining factors $\ts$ and $\tst$ are
generically massive and can be integrated out.  Each
$U(\vl_\ell)$
factor ($\ell>0)$ is a defining model with $m_f+\tilde m_f$
flavors of
fields in the fundamental and antifundamental representation.

In the magnetic theory the analysis is
much the same and the gauge group is broken to
\eqn\BGsusaD{SU(\ncld)\rightarrow SU(2\vlt_0+\tilde\eta)\times
U(\vlt_1)\times U(\vlt_2)\times\ldots\times U(\vlt_k).}
The duality mapping is
$2\vlt_0+\tilde\eta={3\over 2}(m_f+\tilde m_f)-(2\vl_0+\eta)$,
$\vlt_\ell=m_f+\tilde m_f-\vl_\ell$ for $\ell>0$; in the
first factor the duality is that of this section and in
the other factors it is that of \refs{\sem}.

Next consider deforming the electric theory by adding a mass term
$W_{tree}=mQ^{m_f}\tilde Q^{\tilde m_f}$.  In the magnetic theory
the term
$m(M_0)^{m_f,\tilde m_f}$ is added to \WDsusym.  The vacuum has
\eqn\smpv{\eqalign{\ql_{m_f}(\tstD\tsD)^{j}\qlt_{\tilde
m_f}&=-m\delta
_{j,2k+1}\cr
\ql_{m_f} (\tstD\tsD)^{r}\tstD\ql_{m_f}&=0\cr
\qlt_{\tilde m_f}\tsD(\tstD\tsD)^{r}\qlt_{\tilde m_f}&=0,\cr}}
which give expectation values $\ev{\ql_{m_f}}$, $\ev{\qlt_{\tilde
m_f}}$,
$\ev{\tsD}$ and $\ev{\tstD}$, breaking $SU(\half
(4k+3)(m_f+\tilde m_f)-\ncl)$ to
$SU(\half (4k+3)(m_f-1+\tilde m_f-1)-\ncl)$ with $m_f-1$ and
$\tilde m_f-1$ fields.  The low energy magnetic theory is the
dual of the low energy electric theory.

\subsec{Deformations: flat directions}
\subseclab{\DFFsusa}

The flat directions in this model are very similar to those
of the previous two models.  We therefore limit our discussion
to a general sketch.

Consider the flat direction along which the baryon
operator $B_n=\ts^n\Ql^{\nc-2n}$ gets an expectation value.
Along this flat direction the gauge group is broken to $Sp(n)$.
The low energy theory has a
symmetric (adjoint) $Sp(n)$ tensor $\widehat{\tst}$,
$2N_f$ fields in the $2n$-dimensional
fundamental representation of $Sp(n)$, some additional singlets,
and the superpotential $W=\tr \widehat {\tst}^{2(k+1)}$.
In the dual theory the operator $B_n$ is mapped as
in \abmapi\ to $\tsD^{(k+1/2)(m_f+\tilde m_f)-n-2}\ql^{m_f+2n-\ncl}$,
whose expectation value breaks the dual $SU(\ncld)$
gauge group to $Sp(\tilde n)$, with $\tilde n=(2k+1)N_f-n-2$.
The low energy $Sp(\tilde n)$ theory has $2N_f$ matter fields in
the fundamental representation, a symmetric tensor
coming from $\tstD$, and some singlets. This theory gets
a superpotential from \WDsusa, such that it is precisely that
shown in \rlmsspso\ to
be the dual of the above low energy electric theory (see section
\SQspadj.)

There is another flat direction along which the baryon
operator $\tilde B_n=\tst^n\Qlt^{\ncl-n}\Qlt^{\ncl-n}$ gets
an expectation value. Along this flat direction the gauge group
is broken to $SO(n)$.  The low energy theory has an
antisymmetric (adjoint) $SO(n)$ tensor $\hat{\ts}$,
$N_f$ fields in the $n$-dimensional
vector representation of $SO(n)$, some
additional singlet fields, and the superpotential
$W=\tr \widehat {\ts}^{2(k+1)}$. In the dual theory the
operator $\tilde B_n$ is mapped as in \abmapi\ to
$\tstD^{(2k+1)(m_f+\tilde m_f)-n+4}
\qlt^{\tilde m_f+n-\ncl}\qlt^{\tilde m_f+n-\ncl}$,
whose expectation value breaks the dual $SU(\ncld)$ gauge group
to $SO(\tilde n)$, with $\tilde n=(2k+1)N_f-n+4$.
The low energy $SO(\tilde n)$ theory has $N_f$ matter fields in
the vector representation, an antisymmetric tensor
coming from $\tsD$, and some singlets.  This theory gets
a superpotential from \WDsusa, such that it is precisely that
shown in \rlmsspso\ to
be the dual of the above low energy electric theory (see section
\SQsoadj.) Thus, the $Sp(n)$ and $SO(n)$ duality considered in
\rlmsspso\ is inherited from the $SU(\nc)$ model discussed here.

\newsec{$Sp(\ncl)\times Sp(\ncr)$}
\seclab{\Sspsp}

We consider the theory discussed in sect. \SQspsp.
The electric theory \Wspsp\ has
an $SU(2\nfl)\times SU(2\nfr)\times U(1)_R$ flavor symmetry
with matter transforming as:
\thicksize=1pt
\vskip12pt
\begintable
\tstrut  | $Sp(\ncl)$ | $Sp(\ncr)$ | $SU(2\nfl)$ | $SU(2\nfr)$|
$U(1)_R$ \crthick
$\Ql$ | ${\bf 2\ncl}$ | ${\bf 1}$ | ${\bf 2\nfl}$ | ${\bf 1}$ |
$R_\Ql=1-{(\ncl+1)(k+1)-k\ncr\over\nfl(k+1)}$\cr
$\Qr$ | ${\bf 1}$ | ${\bf 2\ncr}$ | ${\bf 1}$ | ${\bf 2\nfr}$ |
$R_\Qr=1-{(\ncr+1)(k+1)-k\ncl\over\nfr(k+1)}$\cr
$\ts$ | ${\bf 2\ncl}$ | ${\bf 2\ncr}$ | ${\bf 1}$ | ${\bf 1}$ |
$R_\ts={1\over k+1}$
\endtable
\noindent
In addition, the theory has a discrete $\ZZ_{2\nfl\nfr(k+1)}$
symmetry generated by
\eqn\spspdsyme{
\ts\rightarrow \alpha ^{N_fN_f'}\ts,\quad Q\rightarrow \alpha
^{-N_c'N_f'}Q,\quad Q'\rightarrow \alpha ^{-N_cN_f}Q',}
with $\alpha=e^{2\pi i/2\nfl\nfr(k+1)}$.

\subsec{The dual model}
\subseclab{\Dspsp}

The dual theory is an $Sp(\ncld)\times Sp(\ncrd)$ gauge theory
with $\ncld=(k~+~1)(\nfl+\nfr-2)-\nfl-\ncr$
and $\ncrd=(k~+~1)(\nfl+\nfr-2)-\nfr-\ncl$ as described in sect.
\SQspsp.  Taking the singlets
$\Mm_{r}$, $\Pl_{j}$ and $\Pr_{j}$ to transform as in the
electric theory, the magnetic theory \WDspsp\ has a
$SU(2\nfl)\times SU(2\nfr)\times U(1)_R$
flavor symmetry with matter fields transforming as:
\thicksize=1pt
\vskip12pt
\begintable
\tstrut  | $Sp(\ncld)$ | $Sp(\ncrd)$ | $SU(2\nfl)$ | $SU(2\nfr)$|
$U(1)_R$ \crthick
$\ql$ | ${\bf 2\ncld}$ | ${\bf 1}$ | ${\bf 1}$ |
${\bf \overline{2\nfr}}$ |
$1-{(\ncld+1)(k+1)-k\ncrd\over\nfr(k+1)}$\cr
$\qr$ | ${\bf 1}$ | ${\bf 2\ncrd}$ | ${\bf \overline{2\nfl}}$ |
${\bf 1}$ |
$1-{(\ncrd+1)(k+1)-k\ncld\over\nf(k+1)}$\cr
$\tsD$ | ${\bf 2\ncld}$ | ${\bf 2\ncrd}$ | ${\bf 1}$ | ${\bf 1}$|
${1\over k+1}$\cr
$\Mm_{r}$ | ${\bf 1}$ | ${\bf 1}$ | ${\bf 2\nfl}$ | ${\bf
2\nfr}$| $R_\Ql+R_\Qr+(2r+1)R_\ts$\cr
$\Pl_{j}$ | ${\bf 1}$ | ${\bf 1}$ | ${\bf asym}$ | ${\bf 1}$ |
$2R_\Ql+2jR_\ts$\cr
$\Pr_{j}$ | ${\bf 1}$ | ${\bf 1}$ |
${\bf 1}$ | ${\bf asym}$ |
$2R_\Qr+2jR_\ts$
\endtable
\noindent
These are indeed anomaly free in the dual theory.
Similarly, the discrete $\ZZ_{2\nfl\nfr(k+1)}$ symmetry
\spspdsyme\ of
the electric theory is a symmetry of the dual theory with the
singlets transforming as the associated mesons of the electric
theory provided
that the other fields transform as
\eqn\spspdsymm{\tsD\rightarrow\alpha^{\nfl\nfr}\tsD,
\quad\ql\rightarrow
e^{2\pi i(\nfl-2)/2\nfr}\alpha^{-\ncrd\nfl}\ql,
\quad\qr\rightarrow
e^{2\pi i(\nfr-2)/2\nfl}\alpha^{-\ncld\nfr}\qr, }
which is indeed non-anomalous in the dual gauge theory. There is a
freedom to modify \spspdsymm\ to give an additional minus sign to both
$q$ and $q'$; this additional transformation is a symmetry of the
dual theory which is free to mix with the
$\ZZ_{2N_fN_f'(k+1)}$ discrete
symmetry.

As another check on the duality we verify that that
't Hooft anomaly matching conditions are satisfied.  With both
the electric and the magnetic fermions we find
\eqn\spspthoofti{\eqalign{
U(1)_R\qquad &\ncl(2\ncl+1)+\ncr(2\ncr+1)-4\ncl\ncr{k\over
k+1}\cr
    & \ -4\ncl{(k+1)(\ncl+1)-k\ncr\over (k+1)}
           -4\ncr{(k+1)(\ncr+1)-k\ncl\over (k+1)}\cr
U(1)_R^3\qquad &\ncl(2\ncl+1)+\ncr(2\ncr+1)
        -4\ncl\ncr\left[{k\over k+1}\right]^3\cr
    -&{4\ncl\over \nfl^2}
            \left[{(k+1)(\ncl+1)-k\ncr\over (k+1)}\right]^3
     -{4\ncr\over \nfr^2}
             \left[{(k+1)(\ncr+1)-k\ncl\over (k+1)}\right]^3\cr
SU(2\nfl)^3\qquad &2\ncl d_3(2\nfl)\cr
SU(2\nfl)^2U(1)_R\qquad &
     -2\ncl{(k+1)(\ncl+1)-k\ncr\over \nfl(k+1)}d_2(2\nfl)
.}}

\subsec{Deformations: superpotential}
\subseclab{\DFSspsp}
\def\stw{\sigma_2}

Consider perturbing the superpotential as:
\eqn\WDFspsp{W_{pert}=\sum_{n=0}^{k}\lk{n}\; \tr\ts^{2(n+1)}.}
In the D-flat vacua with $\ev{\Ql}=\ev{\Qr}=0$, we may write:
\eqn\vacsp{\ev{\ts}=\pmatrix{
x_1\stw&&&&&&&&&&\cr
&x_2\stw&&&&&&&&&\cr
&&.&&&&&&  \cr
&&&\; .&&&&&  \cr
&&&&\; .&&&&  \cr
&&&&&&&x_\ncr\stw&&&&  \cr
},}
where we have assumed $\ncl\geq\ncr$.
The vanishing of the F-terms then give
$x_i W'(x_i^2)=0$
which in general has $k$ distinct solutions for
$x$ plus the solution $x=0$.
Each eigenvalue $x_i$ of
$\ev{\ts}$ may take any of these values. Thus the general
form of the unbroken symmetry in a given vacuum labeled by
integers $\{\vl_0,\vl_1,\ldots,\vl_k\}$ is:
\eqn\spspbrk{
Sp(\ncl)\times Sp(\ncr)\rightarrow Sp(\ncl-\ncr+\vl_0)\times
Sp(\vl_0)
\times Sp(\vl_1)\times\ldots\times Sp(\vl_k)}
with $\sum_{\ell=0}^{k}p_\ell =\ncr$.
In the generic situation, $\ts$ is massive in each vacuum
and can be integrated out.  The $Sp(\ncl-\ncr+\vl_0)$ factor has
$\nfl$ fundamental flavors ($2\nfl$ fields); the $Sp(\vl_0)$
factor has $\nfr$ fundamental flavors; and each $Sp(\vl_\ell)$
factor has $\nfl+\nfr$ flavors in the fundamental
representation.  The vacuum is stable provided
$\nfl>\ncl-\ncr+\vl_0$, $\nfr>\vl_0$, and $\nfl+\nfr>\vl_\ell$
\intpou\ for every $\ell=1,\dots, k$.

In particular, consider the special case where we add just a
mass term for $\ts$ to the superpotential
\Wspsp. The electric gauge group is broken as described in
eq. \spspbrk.  In the dual theory, $\tsD$ is also massive.
\def\wl{q}
For $\ncrd\leq\ncld$, the magnetic gauge group, in a vacuum
labeled by integers $\{\wl_0,\wl_1,\ldots,\wl_k\}$, is broken to
\eqn\spspmbrk{
Sp(\ncld)\times Sp(\ncrd)\rightarrow
Sp(\ncld-\ncrd+\wl_0)\times Sp(\wl_0)\times
Sp(\wl_1)\times\ldots\times
Sp(\wl_k)}
The first factor has $\nfr$ flavors, the second $\nfl$ flavors
and each of the remaining factors has $\nfl+\nfr$ flavors. The
duality between \spspbrk\ and \spspmbrk\ is
$q_0=N_f-(N_c-N_c'+p_0)-2$,
$q_{\ell\geq 1}=N_f+N_f'-p_\ell-2$, the duality of
\refs{\sem,\intpou}.
The situation is similar for $\ncrd>\ncld$.

Next, we consider perturbing the theory \Wspsp\ by a mass term
$W=mQ^{2\nfl-1}Q^{2\nfl}$
for one flavor of the $\Ql$ fields. In the infrared limit of the
electric theory the number of flavors
$\nfl$ will be reduced by one. In the magnetic theory, adding
$(M_0)^{2\nfl-1,2\nfl}$ to the superpotential, the vacua satisfy
\eqn\spspqmd{q_{2\nfl-1}Y^{2j}q_{2\nfl}=-m\delta _{j,k}}
which, along with some other conditions, give expectation values
proportional to
\eqn\spspqmdv{\eqalign{ q_{2\nfl-1,c}=&\delta_{c,1};\cr
q_{2\nfl,c}=
&\delta_{c,2(k+1)};\cr
Y_{c,d'}=&\cases{\delta_{c,d'+1}&$d'=1\dots 2k$;\cr
0& otherwise.\cr}\cr}}
These expectation values break $Sp(\ncld)\times Sp(\ncrd)$ to
$Sp(\ncrd-(k+1))\times Sp(\ncld-k)$, exactly corresponding to
reducing $\nfl$ by one.  The low energy magnetic theory is the
dual of the low energy electric theory.

\subsec{``Confining'' or ``dualizing'' one gauge group}
\subseclab{\Cspsp}
\def\nch{\widehat N_c'}

Consider the limit where the gauge coupling for $Sp(\ncl)$ on the
electric side is much weaker than that of $Sp(\ncr)$.  Turning
off $Sp(\ncl)$ entirely leaves $Sp(\ncr)$ with
$\nfl^{eff}(\ncr)=\ncl+\nfr$  flavors.  With $N_f'$ chosen so
that $N_f^{eff}=N_c'+2$, this theory confines to give a theory
with mesons in the antisymmetric representation of the
$SU(2N_f^{eff})$
flavor symmetry with a pfaffian superpotential \intpou.  Upon
weakly gauging a $Sp(N_c)$ subgroup of the $SU(2N_f^{eff})$
flavor symmetry, the theory is $Sp(N_c)$ with an antisymmetric
tensor $\widehat X\sim X^2$, $2N_f'$ fundamentals $T\sim XQ'$,
$2N_f$  fundamentals $Q$ as in the original theory, and singlets
$U\sim Q'Q'$.
In addition to the superpotential $W\sim \tr\widehat X^{k+1}$
inherited from \Wspsp, there is a pfaffian superpotential
\intpou\
involving $\hat X$, $T$, and $U$.  This is a model of the type
described in sect. \SQspasym.

The corresponding limit in the dual theory is when $Sp(\tilde
N_c)$ is
weakly gauged.  Turning off $Sp(\tilde N_c)$ leaves the
$Sp(\tilde N_c')$ theory with
$N_f^{eff}=\tilde N_c+N_f=\tilde N'_c+2$ flavors and thus, as
in the electric side, it confines.  With $Sp(\tilde N_c)$ weakly
gauged, the dual theory is $Sp(\tilde N_c)$, where $\tilde N_c=
k(N_f+N_f'-2)-N_c$,
with an antisymmetric tensor $\widehat Y\sim
YY$, $2N_f$ fundamentals $t\sim Yq'$, the $2N_f'$ original
fundamentals
$q$, and singlets $u\sim q'q'$.  In addition to the
superpotential
inherited from \WDspsp, there is a
pfaffian superpotential involving $\hat
Y$, $t$, and $u$.  Up to the singlets and the pfaffian
superpotential,
the duality between the above $Sp(N_c)$ theory and the dual
$Sp(\tilde N_c)$ theory is the duality of \kispso\ discussed in
sect. \SQspasym.  It is possible to show that the additional
singlets and pfaffian superpotential preserve the
duality.

More generally, we can consider the electric theory with
$Sp(N_c)$
weakly gauged for arbitrary $N_f'$.  The $Sp(N_c')$ theory can
be given a dual description as in \refs{\sem , \intpou} in terms
of an $Sp(\widehat N_c')$ gauge theory with $\widehat
N_c'=\ncl+\nfr-2-\ncr$.
Weakly gauging $Sp(\ncl)$, the theory becomes an
$Sp(\ncl)\times Sp(\widehat N_c')$ gauge theory
with matter in the representations
\eqn\coggei{\eqalign{\hat\ts&\qquad({\bf \ncl,1;1,1})\cr
T&\qquad ({\bf \ncl,1;1,2\nfr})\cr
U&\qquad ({\bf 1,1;1,\nfr(2\nfr-1)})\cr
V&\qquad ({\bf 1,\widehat N_c';1,\overline{2\nfr}})\cr
Z&\qquad ({\bf \ncl,\widehat N_c';1,1})\cr
Q&\qquad ({\bf \ncl,1;2\nfl,1})}}
of the $Sp(\ncl)\times Sp(\nch)\times SU(2\nfl)\times SU(2\nfr)$
gauge and flavor group with a superpotential
\eqn\coggew{W=\tr \hat \ts^{k+1}+UVV+\hat XZZ+TVZ.}
As above, $\hat \ts\sim\ts^2$, $T\sim\ts\Qr$, $U\sim\Qr\Qr$.  The
mesons  $\Pl_{j}$, $\Pr_{j}$, and $\Mm_{r}$ of
the original theory reduce in part to mesons
$\widehat\Pl_{j}=\widehat
\Ql\widehat\ts^{j}\widehat\Ql$, where $\widehat\Ql$ is either
$\Ql$ or $T$ in the theory \coggei, along with
the meson $U$ and an extra meson $\Pl_{k}=\Ql\widehat\ts^k\Ql$,
which is a redundant operator in the theory \coggew.

Likewise, consider the dual theory \WDspsp\ in the limit where
$Sp(\ncld)$ is turned off.  In this limit the $Sp(\ncrd)$
theory has the dual $Sp(\nch)$, with $\nch=\ncl+\nfr-2-\ncr$ as
above, and upon weakly gauging $Sp(\ncld)$, the theory can be
described by a $Sp(\ncld)\times Sp(\nch)$ gauge theory,  with
matter in the representations
\eqn\coggmi{\eqalign{\hat\tsD&\qquad({\bf \ncld,1,1,1})\cr
t&\qquad ({\bf \ncld,1,\overline{2\nfl},1})\cr
u&\qquad ({\bf 1,1,\overline{\nfl(2\nfl-1)},1})\cr
v&\qquad ({\bf 1,\nch,2\nfl,1})\cr
z&\qquad ({\bf \ncld,\nch,1,1})\cr
\ql&\qquad ({\bf \ncld,1,1,\overline{2\nfr}})}}
of the $Sp(\ncld)\times Sp(\nch)\times SU(2\nfl)\times SU(2\nfr)$
gauge and flavor group and a superpotential
\eqn\coggmw{W=\tr\hat\tsD^{k+1}+uvv+\hat\tsD
zz+tvz+M_ku+Uq\tsD^kq+
\sum _{j=1}^k\widehat M_{j}\widehat q\widehat \tsD^{k-j}\widehat
q,} with the additional terms those in \WDspsp\ and $\widehat q$
referring to $t$ or $\ql$.  Note that the $u$ equations of motion
eliminate the redundant $M_k$ and that the $U$ equation of motion
sets the redundant $q\tsD^kq$ to zero.  Similarly, the $M_k$
equation of motion sets $u$ to zero.

It follows from the
$Sp(\ncl)\times Sp(\ncr)\leftrightarrow Sp(\ncld)\times
Sp(\ncrd)$  duality of this section that the above
$Sp(\ncl)\times Sp(\nch)$ theory with matter \coggei\ and
superpotential \coggew\ is dual to the above $Sp(\ncld)\times
Sp(\nch)$ theory with matter \coggmi\ and superpotential \coggmw
. This duality again is a consequence of that considered in
\kispso.  Consider starting from $Sp(\ncl)$ with an antisymmetric
traceless tensor, a superpotential, and
$2(\nfl+\nfr+\nch)$ fields in the fundamental as in sect.
\SQspasym.  From \kispso\ the dual theory is
$Sp(k(\nfl+\nfr+\nch-2)-\ncl)$.  Adding the third term in the
electric superpotential \coggew\ causes
the magnetic theory to be broken to
$Sp(k(\nfl+\nfr-2)+\nch-\ncl)=Sp(\ncld)$, and its superpotential
picks up the third term in \coggmw.  The duality
found above then follows upon adding $Sp(\ncl)$ singlets $U$ and
$V$ to the electric theory and $Sp(\ncld)$ singlets $u$ and $v$
to the magnetic theory, along with the remaining terms in
\coggew\ and \coggmw, and upon gauging a particular $Sp(\nch)$
subgroup of the flavor symmetry in both the electric and magnetic
theories.

\newsec{$SO(\ncl)\times SO(\ncr)$}
\seclab{\Ssoso}

We consider the theory described in sect. \SQsoso.
The flavor group of these models is $SU(\nfl)\times SU(\nfr)
\times U(1)_R$; the matter fields transform as:
\thicksize=1pt
\vskip12pt
\begintable
\tstrut  | $SO(\ncl)$ | $SO(\ncr)$ | $SU(\nfl)$ | $SU(\nfr)$ |
$U(1)_R$ \crthick
$\Ql$ | ${\bf \ncl}$ | ${\bf 1}$ | ${\bf \nfl}$ | ${\bf 1}$ |
$R_{\Ql}=1-{(\ncl-2)(k+1)-k\ncr\over\nfl(k+1)}$\cr
$\Qr$ | ${\bf 1}$ | ${\bf \ncr}$ | ${\bf 1}$ | ${\bf \nfr}$ |
$R_{\Qr}=1-{(\ncr-2)(k+1)-k\ncl\over\nfr(k+1)}$\cr
$\ts$ | ${\bf \ncl}$ | ${\bf \ncr}$ | ${\bf 1}$ | ${\bf 1}$ |
$R_\ts={1\over k+1}$
\endtable
\noindent
The theory has a discrete $\ZZ_{2\nfl\nfr(k+1)}$
symmetry
generated by
\eqn\sosodsyme{
\ts\rightarrow \alpha ^{N_f N_f'}\ts,\quad Q\rightarrow \alpha
^{-N_c' N_f'}Q,\quad Q'\rightarrow \alpha ^{-N_c N_f}Q',}
with $\alpha=e^{2\pi i/2\nfl\nfr(k+1)}$.  (There are also
$\ZZ_{2N_f}$ and $\ZZ_{2N_f'}$ symmetries acting on $Q$ and $Q'$
respectively.)

\subsec{The dual model}
\subseclab{\Dsoso}

The dual magnetic theory is an $SO(\ncld)\times SO(\ncrd)$ gauge
theory with $\ncld=(k~+~1)(\nfl+\nfr+4)-\nfl-\ncr$ and
$\ncrd=(k~+~1)(\nfl+\nfr+4)-\nfr-\ncl$,
as described in sect. \SQsoso.
Taking $\Pl_{j}$, $\Pr_{j}$, and $\Mm_{r}$ to transform as in
the electric theory, the theory \WDsoso\ has an
$SU(\nfl)\times SU(\nfr)\times U(1)_R$
flavor symmetry with matter fields transforming as:
\thicksize=1pt
\vskip12pt
\begintable
\tstrut  | $SO(\ncld)$ | $SO(\ncrd)$ | $SU(\nfl)$ | $SU(\nfr)$|
$U(1)_R$ \crthick
$\ql$ | ${\bf \ncld}$ | ${\bf 1}$ | ${\bf 1}$ |
${\bf \overline{\nfr}}$ |
$1-{(\ncld-2)(k+1)-k\ncrd\over\nfr(k+1)}$\cr
$\qr$ | ${\bf 1}$ | ${\bf \ncrd}$ | ${\bf \overline{\nfl}}$ |
${\bf 1}$ | $1-{(\ncrd-2)(k+1)-k\ncld\over\nfl(k+1)}$\cr
$\tsD$ | ${\bf \ncld}$ | ${\bf \ncrd}$ | ${\bf 1}$ | ${\bf 1}$|
${1\over k+1}$\cr
$\Mm_{r}$ | ${\bf 1}$ | ${\bf 1}$ | ${\bf \nfl}$ | ${\bf \nfr}$|
$R_\Ql+R_{\Qr}+(2r+1)R_\ts$\cr
$\Pl_{j}$ | ${\bf 1}$ | ${\bf 1}$ | ${\bf sym}$ | ${\bf 1}$ |
$2R_\Ql+2jR_\ts$\cr
$\Pr_{j}$ | ${\bf 1}$ | ${\bf 1}$ | ${\bf 1}$ | ${\bf sym}$ |
$2R_{\Qr}+2jR_\ts$
\endtable
\noindent
These are indeed anomaly free in the dual theory.
Similarly, the discrete $\ZZ_{2\nfl\nfr(k+1)}$ symmetry
\sosodsyme\ of
the electric theory is a symmetry of the dual theory with the
singlets transforming as the associated mesons of the electric
theory provided
that the other fields transform as
\eqn\sosodsymm{\tsD\rightarrow\alpha^{\nfl\nfr}\tsD,
\quad\ql\rightarrow
e^{2\pi i(\nfl+4)/\nfr}\alpha^{-\ncrd\nfl}\ql,
\quad\qr\rightarrow
e^{2\pi i(\nfr+4)/2\nfl}\alpha^{-\ncld\nfr}\qr,}
which is indeed non-anomalous in the dual gauge theory.
There is a freedom to modify \sosodsymm\ to give an extra
minus sign to
both $q$ and $q'$; this operation is a symmetry of
the dual theory which can mix with the
$\ZZ_{2N_fN_f'(k+1)}$ symmetry.

We have verified that the 't Hooft anomaly matching conditions
are indeed satisfied.

\subsec{Deformations}
\subseclab{\DFsoso}

We consider the case $\ncl\geq\ncr$.
Consider deforming \Wsoso\ to include lower order terms:
\eqn\WDFsoso{W_{pert}=\sum_{n=0}^k \lk{n}\tr\ts^{2(n+1)}.}
The theory has vacua with $\ev{\Ql}=\ev{\Qr}=0$ for which the
D-terms give
\eqn\sosofd{\ev{\ts}=\pmatrix{
x_1&&&&&&&&\cr
&x_2&&&&&&&\cr
&&.&&&&&&  \cr
&&&\; .&&&&&  \cr
&&&&\; .&&&&  \cr
&&&&&x_\ncr&&&&}}
generically breaking the gauge group to $SO(\ncl-\ncr)$.
The theory with
superpotential \WDFsoso\ has multiple vacua with expectation
values of the form \sosofd\ with $x_i$  satisfying $x_i
W'(x_i^2)=0$
which generically has $k$ distinct non-zero solutions $z_\ell$;
zero is also a solution.   Let $\vl_0$ be the number of
eigenvalues with $x_i=0$ and let
$\vl_\ell$ be the number of $x_i=z_\ell$, with $\sum
_{\ell=0}^k p_\ell=\ncr$.  In such a vacuum
the gauge group is broken by $\ev{\ts}$ to
\eqn\sosohiggs{SO(\ncl)\times SO(\ncr)\rightarrow
SO(\ncl-\ncr+\vl_0)
\times SO(\vl_0)\times SO(\vl_1)\times \ldots\times SO(\vl_k).}
In the generic situation $\ts$ is massive in each vacuum
and can be integrated out.  The $SO(\ncl-\ncr+\vl_0)$ factor has
$\nfl$ vector representations ; the $SO(\vl_0)$ factor has $\nfr$
vector representations; and each $SO(\vl_\ell)$
factor has $\nfl+\nfr$ matter fields in the vector
representation.  The vacuum is stable provided
$\nfl\geq\ncl-\ncr+\vl_0-4$, $\nfr\geq\vl_0-4$, and
$\nfl+\nfr\geq \vl_\ell-4$ \intpou\ for every
$\ell=1,\ldots, k$.

A similar analysis in the dual theory leads to a dual version of
\sosohiggs\ with the duality in each factor that of
\refs{\sem, \isson}.

We now consider adding a mass term $mQ^{\nfl}\cdot Q^{\nfl}$,
giving a mass to a quark flavor in the electric theory.  In the
magnetic theory
the added term to the superpotential gives expectation values
\eqn\sosomqm{q_{\nfl}\tsD^{2j}q_{\nfl}=-m\delta _{j,k}}
which, along with the other $F$ terms and the $D$ terms,
give expectation values $\ev{q_{\nfl}}$ and $\ev{\tsD}$ which
break $SO(\ncld)\times SO(\ncrd)$
to $SO(\ncld-(k+1))\times SO(\ncrd-k)$.  For example, for
$k=1$ the expectation values are $(q_{\nfl})_c=\delta
_{c,1}+i\delta_{c,2}$ and $\tsD_{c,d'}=(\delta _{c,1}-i\delta
_{c,2})\delta_{d'1}$,
which reduces $\ncl$ by two and $\ncr$ by one.  The low energy
magnetic theory is the dual of the low energy electric theory.

\subsec{Dualizing one gauge group}

Consider the limit where the gauge coupling for $SO(\ncl)$ on the
electric side is much weaker than that of $SO(\ncr)$.  Turning
off $SO(\ncl)$ entirely leaves $SO(\ncr)$ with
$\nfl^{eff}(\ncr)=\ncl+\nfr$ fields in the vector representation.
If $N_f'$ is chosen so that
$N_f^{eff}=N_c'-4$, this theory confines to give a theory with
mesons in the symmetric representation of the $SU(N_f^{eff})$
flavor symmetry \isson.  When the $SO(N_c)$ subgroup of
$SU(N_f^{eff})$ is weakly gauged, the theory has a symmetric
tensor $\widehat X\sim X^2$ of $SO(N_c)$, $N_f'$ vectors $T\sim
XQ'$, $N_f$ vectors $Q$ as in the original theory, and singlets
$U\sim Q'Q'$. This is a theory of \kispso\ (sect. \SQsosym).

The corresponding limit in the dual theory is when $SO(\tilde
N_c)$ is weakly gauged.  Turning off $SO(\tilde N_c)$ leaves the
$SO(\tilde N'_c)$ theory with
$N_f^{eff}=\tilde N_c+N_f=\tilde N'_c-4$ vectors, and thus,
as on the electric side, it confines.  When
$SO(\tilde N_c)$ is weakly gauged, where $\tilde N_c=
k(N_f+N_f'+4)-N_c$,  the theory has
a symmetric tensor $\widehat Y\sim YY$ of $SO(\tilde N_c)$,
$N_f$ vectors $t\sim Yq'$, the $N_f'$ original vectors
$q$, and singlets $u\sim q'q'$.   Up to the singlets,
the duality between the above $SO(N_c)$ theory and the dual
$SO(\tilde N_c)$ theory is the duality of \kispso\ discussed in
sect. \SQsosym.  It is possible to show that the additional
singlets preserve the duality.

More generally, we can consider the electric theory with
$SO(N_c)$ weakly gauged for arbitrary $N_f'$.  The analysis is
very similar to the case studied in sect. \Cspsp, and
we refer the reader to that section for details. The main point
is that the electric and magnetic theories flow to
$SO(N_c)\times SO(\nch)$ and $SO(\tilde N_c)\times SO(\nch)$
respectively.  The matter content is such that the electric
model is an example studied in \kispso\ (sect. \SQsosym)
with extra singlets, a superpotential, and an $SO(\nch)$
subgroup of its flavor group gauged.  The magnetic theory
is the dual theory under the duality of sect. \SQsosym,
along with extra singlets, the dual superpotential, and
the same $SO(\nch)$ subgroup of its flavor group gauged.

\newsec{$SU(\ncl)\times SU(\ncr)$}
\seclab{\Ssusu}

This theory was outlined in sect. \SQsusu.  The theory \Wsusu\
has an $[SU(\nfl)\times SU(\nfr)]^2\times U(1)^3\times U(1)_R $
global flavor symmetry with the fields transforming as
\thicksize=1pt
\vskip12pt
\begintable
\tstrut  | $\Ql;\Qlt$ | $\Qr;\Qrt$ | $\ts;\tst$ \crthick
$SU(\ncl)$ | ${\bf\ncl;\overline\ncl}$ | ${\bf 1;1}$ |
${\bf\ncl;\overline\ncl}$ \cr
$SU(\ncr)$ | ${\bf 1;1}$ | ${\bf\ncr;\overline\ncr}$ |
${\bf\ncr;\overline\ncr}$ \cr
$SU(\nfl)_L$ | ${\bf\nfl;1}$ | ${\bf 1;1}$ | ${\bf 1;1}$ \cr
$SU(\nfr)_L$ | ${\bf 1;1}$ | ${\bf\nfr;1}$ | ${\bf 1;1}$\cr
$SU(\nfl)_R$ | ${\bf 1;\nfl}$ | ${\bf 1;1}$ | ${\bf 1;1}$\cr
$SU(\nfr)_R$ | ${\bf 1;1}$ | ${\bf 1;\nfr}$ | ${\bf 1;1}$\cr
$U(1)_B$ | $\pm {1\over\ncl}$ | $\pm {1\over\ncr}$ |
$\pm \left({1\over\ncl}+{1\over\ncr}\right)$\cr
$U(1)_{B'}$ | $\pm {1\over\ncl}$ | $\mp {1\over\ncr}$ |
$\pm \left({1\over\ncl}-{1\over\ncr}\right)$\cr
$U(1)_X$ | 0 | 0 | $\pm 1$\cr
$U(1)_R$ | $1+{(k\ncr-(k+1)\ncl)\over\nfl (k+1)}$ |
$1+{(k\ncl-(k+1)\ncr)\over\nfr (k+1)}$ |
${1\over k+1}$
\endtable
\noindent
We have chosen $U(1)_B$ and $U(1)_{B'}$
such that all mesons are invariant.

There is also a $\ZZ_{2\nfl\nfr(k+1)}$ discrete symmetry generated
by
\eqn\susudise{
X,\tilde X\rightarrow \alpha^{\nfl\nfr}X,\tilde X \qquad
\Ql,\Qlt\rightarrow\alpha^{-\ncr\nfr}\Ql,\Qlt\qquad
\Qr,\Qrt\rightarrow\alpha^{-\ncl\nfl}\Qr,\Qrt,}
with $\alpha =e^{2\pi i/2\nfl\nfr(k+1)}$.

\subsec{The dual theory}
\subseclab{\Dsusu}

The dual is an $SU(\ncld)\times SU(\ncrd)$ gauge theory with
$\ncld=(k+1)(\nfr+\nfl)-\nfl-\ncr$ and
$\ncrd=(k+1)(\nfr+\nfl)-\nfr-\ncl$,
as described in sect. \SQsusu.
Taking the singlets to transform as the mesons of the electric
theory,
the dual theory has the same flavor group with the various fields
transforming as:
\thicksize=1pt
\vskip12pt
\begintable
\tstrut  | $\ql;\qlt$ | $\qr;\qrt$ | $\tsD;\tstD$ \crthick
$SU(\ncld)$ | ${\bf\ncld;\overline\ncld}$ | ${\bf 1;1}$ |
${\bf\ncld;\overline\ncld}$ \cr
$SU(\ncrd)$ | ${\bf 1; 1}$ | ${\bf\ncrd;\overline\ncrd}$ |
${\bf \ncrd;\overline\ncrd}$ \cr
$SU(\nfl)_L$ | ${\bf 1;1}$ | ${\bf \overline\nfl; 1}$ | ${\bf
1;1}$ \cr
$SU(\nfl)_R$ | ${\bf 1;1}$ | ${\bf 1; \overline\nfl}$ | ${\bf
1;1}$ \cr
$SU(\nfr)_L$ | ${\bf \overline\nfr;1}$ | ${\bf 1; 1}$ | ${\bf
1;1}$\cr
$SU(\nfr)_R$ | ${\bf 1;\overline\nfr}$ | ${\bf 1; 1}$ | ${\bf
1;1}$\cr
$U(1)_B$ | $\pm {1\over\ncld}$ | $\pm {1\over\ncrd}$ |
 $\pm \left({1\over\ncld}+{1\over\ncrd}\right)$ \cr
$U(1)_{B'}$ | $\pm {1\over\ncld}$ | $\mp {1\over\ncrd}$ |
$\pm\left({1\over\ncld}-{1\over\ncrd}\right)$\cr
$U(1)_X$ | $\pm {k\nfl\over\ncld}$ | $\pm {k\nfr\over\ncrd}$
  | $\mp\left(1-{k\nfl\over\ncld}-{k\nfr\over\ncrd}\right)$\cr
$U(1)_R$ | $1+{(k\ncrd-(k+1)\ncld)\over\nfr (k+1)}$ |
$1+{(k\ncld-(k+1)\ncrd)\over\nfl (k+1)}$ | ${1\over k+1}$
\endtable
\noindent
which are indeed anomaly free in the dual theory.
Similarly, the discrete symmetry \susudise\ is a symmetry of the
dual theory with the singlets transforming as the corresponding
mesons of the electric theory provided it acts on the other
fields as
\eqn\susudism{\eqalign{
Y &\rightarrow e^{2\pi iB_Y\ncr/2}\ \ e^{2\pi iB'_Y\ncl/2}
\ \ \alpha ^{\nfl\nfr}\ Y \cr
\tilde Y&\rightarrow e^{-2\pi iB_Y\ncr/2}\ e^{-2\pi iB'_Y\ncl/2}
\ \alpha ^{\nfl\nfr}\ \tilde Y \cr
\ql&\rightarrow e^{2\pi iB_\ql\ncr/2}\ \ e^{2\pi iB'_\ql\ncl/2}
\ \ \alpha^{-\nfl\ncrd}\ e^{2\pi iN_f/2N_f'}\ \ql\cr
\qlt&\rightarrow e^{-2\pi iB_\ql\ncr/2}\ e^{-2\pi iB'_\ql\ncl/2}
\ \alpha^{-\nfl\ncrd}\ e^{2\pi iN_f/2N_f'}\ \qlt
\cr
\qr&\rightarrow e^{2\pi iB_{\qr}\ncr/2}\ \ e^{2\pi iB'_{\qr}\ncl/2}
\ \ \alpha^{-\nfr\ncld}\ e^{2\pi iN_f'/2N_f}\ \qr\cr
\qrt&\rightarrow e^{-2\pi iB_{\qr}\ncr/2}\ e^{-2\pi iB'_{\qr}\ncl/2}
\ \alpha^{-\nfr\ncld}\ e^{2\pi iN_f'/2N_f}\ \qrt
\cr}}
Here $B_Y$ and $B'_Y$ are the $U(1)_B$ and $U(1)_{B'}$
charges of $\tsD$, and similarly for the other fields. This is
indeed a non-anomalous discrete symmetry of the dual gauge theory.

We have verified that all of the
't Hooft anomaly matching conditions are indeed satisfied.

Under the duality the baryons of the electric theory,
\eqn\susubary{B\{n_j,m_r\}=\prod_{j=0}^k Q_{(j)}^{n_j}
\prod_{r=0}^{k-1} S_{(r)}^{m_r},}
with gauge indices contracted with an $\epsilon$-tensor of
$SU(\ncl)$, where $Q_{(j)}=(\ts\tst)^j \Ql$ and
$S_{(r)}=(\ts\tst)^r \ts\Qrt$, are mapped to the analogous baryons
\eqn\susubaryD{
b\{\tilde n_j,\tilde m_r\}=\prod_{j=0}^k {q'}_{(j)}^{\tilde n_j}
\prod_{r=0}^{k-1} {s'}_{(r)}^{\tilde m_r}
}
of the dual theory, where $q'_{(j)}=(\tsD\tstD)^j \qr$ and
$s'_{(r)}=(\tsD\tstD)^r \tsD\qlt$.  The mapping is
$B\{n_j,m_r\}\rightarrow b\{\tilde n_{j},\tilde m_r\}$,
where $\tilde n_j=\nfl-n_{k-j}$ and $\tilde m_r=\nfr-m_{k-1-r}$.
This transformation is consistent with all of the flavor symmetries.
There is a similar mapping for all other baryons and
antibaryons in the theory.

\subsec{Deformations: superpotential}
\subseclab{\DFSsusu}

Consider deforming the superpotential as:
\eqn\susudef{W_{pert}=\sum_{n=0}^k\lk{n}\tr (\ts\tst)^{n+1}.}
Let us assume that $\ncr\geq\ncl$; the general D-flat vacua are
of
the form
\eqn\vacsux{\ev{\ts}=\pmatrix{
x_1&&&&&&&&&\cr
&x_2&&&&&&&&\cr
&&.&&&&&  \cr
&&&\; .&&&&  \cr
&&&&\; .&&&  \cr
&&&&&x_\ncl&&  \cr
};\;
\ev{\tst}=\pmatrix{
\tilde x_1&&&&&&&&&\cr
&\tilde x_2&&&&&&&&\cr
&&.&&&&&  \cr
&&&\; .&&&&  \cr
&&&&\; .&&&  \cr
&&&&&\tilde x_\ncl&&  \cr}}
where $|x_i|=|\tilde x_i|$.
The equations of motion for $\ts$ and $\tst$ give
$x_iW'(x_i\tilde x_i)=\tilde x_iW'(x_i\tilde x_i)=0$
There are generically $k$ distinct solutions to these equations
$z_\ell=x_\ell\tilde x_\ell$,
plus the solution $x=\tilde x=0$.
A vacuum state is thus labeled by integers
$\{\vl_0,\vl_1,\ldots, \vl_k\}$, with $\sum_{j=0}^k \vl_j=\ncl$,
where $\vl_0$ is the number of eigenvalues which are zero.
The unbroken symmetry in such a vacuum is
\eqn\susubrk{SU(\ncl)\times SU(\ncr)\rightarrow
SU(\ncr-\ncl+\vl_0)\times SU(\vl_0)
\times SU(\vl_1)\ldots\times SU(\vl_k).}
In the generic situation $X$ and $\tilde X$ are massive in each
vacuum and can be integrated out.  In the
first two factors there are $\nfl$ and $\nfr$ flavors
of quarks respectively and in each of the other factors there are
$\nfl+\nfr$ flavors of quarks.
In particular, this is the case when adding a mass term
$W=m\ts\tst$ to \Wsusa.

In the dual theory there is a breaking similar to \susubrk\ and
the duality between the two theories is the duality of \sem\ in
each factor.

\subsec{Deformations: flat directions}
\subseclab{\DFFsusu}

Consider the flat direction along which
the operator $B_n=\ts^n\Ql^{\ncl-n}\Qr^{\ncr-n}$, with gauge indices
contracted with two $\epsilon$-tensors, gets an expectation
value. In terms of the elementary fields this flat direction is
$\ev{\Ql^{f,c}}=\sqrt{2}a\delta^{f,c}$ for $c\leq\ncl-n$ and zero
otherwise, $\ev{\ts^{cd}}=a\delta^{c,d}$ for $c>\ncl-n$
and zero otherwise, with all other expectation values zero.
Along this flat direction, which is not lifted by the
superpotential
\Wsusu, the gauge group is broken to $SU(n)$.
In the low energy theory there are $\nfl+\nfr$ matter fields in
the fundamental representation of $SU(n)$, with some coming from
$\tst$ and some from the quarks not entering in $B_n$. There
is also an adjoint $\widehat\ts$ from $\tst$. The
low energy theory has a superpotential inherited from \Wsusu ,
$W=\Tr \widehat \ts^{k+1}$,
where indices are contracted using $\ev{\ts}$.  Up to some
additional singlets, this low energy theory is that considered
in \kref.

In the dual theory the operator $B_n$ labeling the above flat
direction is mapped to $b_{n}=
\tsD^{k(\nfl+\nfr)-n}\qr^{\nfl-\ncl+n}\ql^{\nfr-\ncr+n}$.
When this operator gets an expectation value the
dual $SU(\ncld)\times SU(\ncrd)$ gauge
theory is broken to $SU(\tilde n)$, with
$\tilde n=k(\nfl+\nfr)-n$.
This low energy $SU(\tilde n)$ theory has $\nfl+\nfr$
matter fields in the fundamental
representation, with some coming from $\tstD$ and some
{}from the dual quarks not entering in $b_{n}$.
There is also an adjoint coming from $\tstD$, and some
singlets, some of which are eliminated by \WDsusu.
The low energy magnetic theory gets a superpotential
{}from \WDsusu, such that it is precisely that shown in \kref\ to
be the dual of the above low energy electric theory.
The $SU(n)$ duality discussed in \kref\ is thus inherited
{}from the model discussed here.

\subsec{Dualizing one gauge group}

Consider the limit where the gauge coupling for $SU(\ncl)$ on the
electric side is much weaker than that of $SU(\ncr)$.  Turning
off $SU(\ncl)$ entirely leaves $SU(\ncr)$ with
$\nfl^{eff}(\ncr)=\ncl+\nfr$ flavors in
the fundamental representation.  If $N_f'$ is chosen so that
$N_f^{eff}=N_c'+1$, this theory confines to give a theory with
mesons in the $(N_f^{eff},N_f^{eff})$ representation of the
$SU(N_f^{eff})\times SU(N_f^{eff})$
flavor symmetry \powerh.  When the diagonal $SU(N_c)$ subgroup
of $SU(N_f^{eff})\times SU(N_f^{eff})$ is weakly gauged, the
theory has an adjoint tensor $\widehat X\sim X^2$ of $SU(N_c)$,
$N_f'$ flavors $T\sim X \tilde Q'$, $\tilde T\sim \tilde X Q'$,
$N_f$ flavors $Q, \tilde Q$ as in the original theory, a flavor
$B,\tilde B$ from the baryons of $SU(\ncr)$, and
some singlets. The superpotential
is  $W=\tr\widehat X^{k+1} + B\widehat X\tilde B$ plus other
terms. This is a theory of \refs{\kut,\kutsch,\Ahsonyank}
(sect. \SQsuadj).

The corresponding limit in the dual theory is when $SU(\tilde
N_c)$ is weakly gauged.  Turning off $SU(\tilde N_c)$ leaves the
$SU(\tilde N'_c)$ theory with
$N_f^{eff}=\tilde N_c+N_f=\tilde N'_c+1$ flavors, and thus,
as on the electric side, it confines.  When
$SU(\tilde N_c)$ is weakly gauged, where $\tilde N_c=
k(N_f+N_f')+1-N_c$, the theory has
an adjoint tensor $\widehat Y\sim YY$ of $SU(\tilde N_c)$,
$N_f$ flavors $t\sim Y\tilde q'$, $\tilde t\sim \tilde Yq'$,
the $N_f'$ original flavors $q,\tilde q$,
a flavor $b, \tilde b$ from the baryons of $SU(\ncrd)$,
and some singlets. The superpotential
is  $W=\tr\widehat Y^{k+1} + b\widehat Y\tilde b$ plus other
terms. Up to the singlets and the terms in the superpotential
which we have not written,
the duality between the above $SU(N_c)$ theory and the dual
$SU(\tilde N_c)$ theory is the duality of
\refs{\kut,\kutsch,\Ahsonyank} discussed in
sect. \SQsuadj.  It is possible to show that the additional
singlets and superpotential preserve the duality.

More generally, we can consider the electric theory with
$SU(N_c)$
weakly gauged for arbitrary $N_f'$.  The analysis is
very similar to the case studied in sect. \Cspsp, and
we refer the reader to that section for details. The main point
is that the electric and magnetic theories flow to
$SU(N_c)\times SU(\nch)$ and $SU(\tilde N_c)\times SU(\nch)$
respectively.  The matter content is such that the electric
model is an example studied in \kutsch\ (sect. \SQsuadj)
with extra singlets, a superpotential, and an $SU(\nch)$
subgroup of its flavor group gauged.  The magnetic theory
is the dual theory under the duality of sect. \SQsuadj,
along with extra singlets, the dual superpotential, and
the same $SU(\nch)$ subgroup of its flavor group gauged.

\newsec{$SO(\ncl)\times Sp(\ncr)$}
\seclab{\Ssosp}
\def\nfrp{n'_f}

Consider the $SO(\ncl)\times Sp(\ncr)$ theories described in
sect. \SQsosp.   These are chiral theories in that the field $\ts$
can not be given a mass term, as there is no gauge invariant
$\tr \ts^2$; the  basic gauge invariants involving $\ts$ are
$\tr\ts^{4(j+1)}$.  Also, to cancel the global anomaly of the $Sp$
factor we must take $\ncl+\nfrp$ even.  This theory has an
anomaly-free $SU(\nfl)\times SU(\nfrp)\times U(1)_R$ flavor
symmetry with matter fields in the representations
\thicksize=1pt
\vskip12pt
\begintable
\tstrut  | $SO(\ncl)$ | $Sp(\ncr)$ | $SU(\nfl)$ | $SU(\nfrp)$ |
$U(1)_R$ \crthick
$\Ql$ | ${\bf \ncl}$ | ${\bf 1}$ | ${\bf \nfl}$ | ${\bf 1}$ |
$R_{\Ql}=1-{1\over\nfl}\left(\ncl-2-{\ncr(2k+1)\over
k+1}\right)$\cr
$\Qr$ | ${\bf 1}$ | ${\bf 2\ncr}$ | ${\bf 1}$ | ${\bf \nfrp}$ |
$R_{\Qr}=1-{1\over\nfrp}
         \left(2\ncr+2-{\ncl(2k+1)\over 2(k+1)}\right)$\cr
$\ts$ | ${\bf \ncl}$ | ${\bf 2\ncr}$ | ${\bf 1}$ | ${\bf 1}$ |
$R_{\ts}={1\over 2(k+1)}$
\endtable
\noindent
In addition, the theory \Wsosp\ has a $\ZZ_{4\nfl\nfrp(k+1)}$
discrete symmetry
generated by
\eqn\spsodsyme{\ts\rightarrow \alpha ^{\nfl\nfrp}\ts,\quad
\Ql\rightarrow \alpha^{-2\ncr\nfrp} \Ql,
\quad
\Qr\rightarrow \alpha^{-\ncl\nfl}\Qr,}
with $\alpha =e^{2\pi i/4\nfl\nfrp(k+1)}$, and a $\ZZ_{2\nfl}$
symmetry generated by
\eqn\spsodsymiie{\Ql\rightarrow \beta \Ql,}
where $\beta=e^{2\pi i/2\nfl}$.

\subsec{Dual theory}
\subseclab{\Dsosp}

The dual theory is $SO(\ncld)\times Sp(\ncrd)$, where
$\ncld=2(k+1)(\nfrp+\nfl)-\nfl-2\ncr$ and
$2\ncrd=2(k+1)(\nfrp+\nfl)-\nfrp-\ncl$, as described in
sect. \SQsosp.
Taking the singlets $\Mm_{r}$, $\Pl_{j}$, and $\Pr_{j}$ to
transform as
the mesons of
the electric theory,
the theory \WDsosp\ has a
$SU(\nfl)\times SU(\nfrp)\times U(1)_R$ global symmetry with
matter
in the representations
\thicksize=1pt
\vskip12pt
\begintable
\tstrut  | $SO(\ncld)$ | $Sp(\ncrd)$ | $SU(\nfl)$ | $SU(\nfrp)$|
$U(1)_R$ \crthick
$\ql$ | ${\bf \ncld}$ | ${\bf 1}$ | ${\bf 1}$ |
${\bf \overline{\nfrp}}$ |
$1-{1\over\nfl}\left(\ncld-2-{\ncrd(2k+1)\over k+1}\right)$\cr
$\qr$ | ${\bf 1}$ | ${\bf 2\ncrd}$ | ${\bf \overline{\nfl}}$ |
${\bf 1}$ |
$1-{1\over\nfrp}\left(2\ncrd+2-{\ncld(2k+1)\over
2(k+1)}\right)$\cr
$\tsD$ | ${\bf \ncld}$ | ${\bf 2\ncrd}$ | ${\bf 1}$ | ${\bf 1}$|
${1\over 2(k+1)}$\cr
$\Mm_{r}$ | ${\bf 1}$ | ${\bf 1}$ | ${\bf \nfl}$ | ${\bf \nfrp}$|
$R_\Ql+R_\Qr+(2r+1)R_{\ts}$\cr
\vctr{$\Pl_{j}$} | \vctr{${\bf 1}$} | \vctr{${\bf 1}$} |
$\eqalign{{\bf sym}, {\rm j\ even}\nr \kern-5.2em {\bf asym},
{\rm j\ odd}}$ |
\vctr{${\bf 1}$} |
\vctr{$2R_{\Ql}+2jR_\ts$}\cr
\vctr{$\Pr_{j}$} | \vctr{${\bf 1}$} | \vctr{${\bf 1}$} |
\vctr{${\bf 1}$} |
$\eqalign{{\bf asym}, {\rm j\ even}\nr \kern-5.1em {\bf sym},
{\rm j\ odd}}$ |
\vctr{$2R_{\Qr}+2jR_\ts$}
\endtable
\noindent
which are indeed anomaly free in the dual gauge theory.
Taking the singlets to transform as the mesons of the electric theory
under \spsodsyme, the superpotential \WDsosp\ is invariant under the
$\ZZ_{4N_fn_f'(k+1)}$ symmetry if the other fields transform as
\eqn\spsodsymm{\tsD\rightarrow \alpha ^{\nfl\nfrp}\tsD,
\quad
\ql\rightarrow e^{2\pi i\nfl/2\nfrp}
\alpha^{-2\ncrd\nfl}\ql,
\quad
\qr\rightarrow e^{2\pi i\nfrp/2\nfl}
\alpha^{-\nfrp\ncld}\qr.}
Similarly, \WDsosp\ is invariant under the $\ZZ_{2N_f}$
transformation on the singlets corresponding to
\spsodsymiie\ if the other fields transform as
\eqn\spsodsymiim{q'\rightarrow \beta ^{-1}q'.}
However, both \spsodsymm\ and \spsodsymiim\ are not quite symmetries of
the dual theory: under both of them the $Sp(\tilde N_c')$ instanton
't Hooft vertex picks up a minus sign.
(For odd $N_f$ this can be corrected
by combining the transformations \spsodsymm\ and \spsodsymiim\ with a
transformation which gives a minus sign to both $q$ and $q'$.)  Perhaps
this suggests that the symmetries \spsodsyme\ and \spsodsymiie\ of the
electric theory take the magnetic dual to a dyonic dual in which
the $Sp(\tilde N_c')$ theta angle is shifted by $\pi$ relative to that
of the magnetic dual.  Dyonic duals with theta angles shifted by $\pi$
appeared in \isson.

We have verified that all of the
't Hooft anomaly matching conditions are satisfied.

\subsec{Deformations}
\subseclab{\DFsosp}
\def\nclp{{n_{c}}}

Consider deforming \Wsosp\ to include lower order terms:
\eqn\WDFsosp{W_{pert}=\sum_{n=0}^k \lk{n}\tr\ts^{4(n+1)}.}
We consider the case $\ncl\geq 2N_c'$; the other case is very
similar.
The theory has vacua with $\ev{\Ql}=\ev{\Qr}=0$ for which the
D-terms give
\eqn\sospfd{\ev{\ts}=\pmatrix{
x_1\bigone_2&&&&\cr
&x_2\bigone_2&&&\cr
&&.&&  \cr
&&&\; .&  \cr
&&&&x_{N_c'}\bigone_2\cr
&&&&\cr
&&&&\cr}.}
A vacuum is labeled by integers
$\{\vl_0,\vl_1,\ldots,\vl_k\}$, with $\sum _{\ell=0}^k\vl
_\ell=N_c'$,
with the gauge group broken by $\ev{\ts}$ to
\eqn\sosphiggs{SO(\ncl)\times Sp(\ncr)\rightarrow
SO(N_c-2N_c'+2\vl_0)\times Sp(\vl _0)
\times U(\vl_1)\times U(\vl_2)\times \ldots\times U(\vl_k).}
Because $X$ cannot be given a mass term,
it will not be massive in the $SO(2\vl_0+\om)\times
Sp(\ncr-\nclp+\vl_0)$
factor; this factor will have a field $\widehat X$ transforming
in the defining representation
of both the $SO$ and $Sp$ groups.  For example, consider the
effect of the superpotential $W = \tr X^{4(k+1)} + \lambda\tr X^4$.
The $SO(N_c-2N_c'+2\vl_0)\times Sp(\vl _0)$ factor in \sosphiggs\
is then a theory of the type considered in this section, with a
field $\widehat X$ in the defining representation of both gauge
groups and a superpotential $W\sim \widehat X^4$
(i.e. $k=0$), along with $N_f$ fields in the vector
representation of the $SO$ factor and $n_f'$ fields in the
fundamental representation of the $Sp$ factor.  In the $U(\vl _\ell)$
factors in \sosphiggs\ the field $X$ is
massive and can be integrated out, leaving a defining model; each
has $\nfl+n_f'$ flavors in the fundamental
representation.  The vacuum is stable provided
$\nfl\geq 2\vl_0+\om-4$, $\nfr >\ncr-\nclp+\vl_0$, and
$\nfl+\nfr\geq \vl_\ell$ for every
$\ell=1,\ldots, k$.

A similar analysis in the dual theory yields vacua labeled by
$\wl _\ell$, with $\sum _{\ell=0}^k\wl _\ell =2\tilde N_c'$, with
the gauge group broken as
\eqn\sosphiggsm{SO(\tilde N_c-2\tilde N_c'+2\wl_0)\times Sp(\wl
_0) \times U(\wl_1)\times U(\wl_2)\times \ldots\times U(\wl_k)}
with a matter field $\widehat Y$ and a superpotential
$W\sim \tr\widehat Y^4$ in the $SO\times Sp$ factor, and with
the matter field $Y$ massive and decoupled in all of the
$U(\wl _\ell)$ factors.  The duality mapping between
\sosphiggs\ and \sosphiggsm\ is
the $k=0$ case of the duality of this section for the $SO\times
Sp$ factor and the duality of \sem\ for the $U$ factors:
$2\wl_0=2N_f+n_f'-(N_c-2N_c'+2\vl _0)$ and, for $\ell>0$,
$\wl_{\ell}=N_f+n_f'-\vl _\ell$.

\subsec{Dualizing one gauge group}

Consider the limit where the gauge coupling for $SO(\ncl)$ on the
electric side is much weaker than that of $Sp(\ncr)$.  Turning
off $SO(\ncl)$ entirely leaves $Sp(\ncr)$ with
$n_f^{eff}(\ncr)=\ncl+\nfrp$ fields in the fundamental
representation. If $n_f'$ is chosen so that
$n_f^{eff}=2(N_c'+2)$, this theory confines to give a theory with
mesons in the antisymmetric representation of the $SU(n_f^{eff})$
flavor symmetry \intpou.  When the $SO(N_c)$ subgroup of
$SU(n_f^{eff})$ is weakly gauged, the theory has an antisymmetric
(adjoint) tensor $\widehat X\sim X^2$ of $SO(N_c)$, $n_f'$
vectors $T\sim XQ'$, $N_f$ vectors $Q$ as in the original theory,
and singlets $U\sim Q'Q'$. This is a theory of \rlmsspso\ (sect.
\SQsoadj).

The corresponding limit in the dual theory is when $SO(\tilde
N_c)$ is
weakly gauged.  Turning off $SO(\tilde N_c)$ leaves the
$Sp(\tilde N'_c)$ theory with
$n_f^{eff}=\tilde N_c+N_f=2(\tilde N'_c+2)$ fields in the
fundamental representation, and thus,
as on the electric side, it confines.  When
$SO(\tilde N_c)$ is weakly gauged, where $\tilde N_c=
k(N_f+N_f'+4)-N_c$,  the theory has an antisymmetric
(adjoint) tensor $\widehat Y\sim YY$ of $SO(\tilde N_c)$,
$N_f$ vectors $t\sim Yq'$, the $n_f'$ original vectors
$q$, and singlets $u\sim q'q'$.   Up to the singlets,
the duality between the above $SO(N_c)$ theory and the dual
$SO(\tilde N_c)$ theory is the duality of \rlmsspso\ discussed
in  sect. \SQsoadj.  It is possible to show that the additional
singlets preserve the duality.

More generally, we can consider the electric theory with
$SO(N_c)$ weakly gauged for arbitrary $n_f'$.  The analysis is
very similar to the case studied in sect. \Cspsp, and
we refer the reader to that section for details. The main point
is that the electric and magnetic theories flow to
$SO(N_c)\times Sp(\nch)$ and $SO(\tilde N_c)\times Sp(\nch)$
respectively.  The matter content is such that the electric
model is an example studied in \rlmsspso\ (sect. \SQsoadj)
with extra singlets, a superpotential, and an $Sp(\nch)$
subgroup of its flavor group gauged.  The magnetic theory
is the dual theory under the duality of sect. \SQsoadj,
along with extra singlets, the dual superpotential, and
the same $Sp(\nch)$ subgroup of its flavor group gauged.

If the gauge coupling for $Sp(\ncr)$ is much weaker than that of
$SO(\ncl)$, the analysis is very similar to the above with
appropriate replacements of $N_f$ with $n_f'$, ``antisymmetric''
with ``symmetric'', {\it etc.}
The low energy electric and magnetic theories flow to
$Sp(N'_c)\times SO(\hat\nc)$ and $Sp(\tilde N_c)\times
SO(\hat\nc)$
respectively; the matter content is such that the
models are those of sect. \SQspadj\ with an $SO(\hat\nc)$
subgroup of the flavor symmetry gauged.  They are
dual under the duality of \rlmsspso.

\newsec{$SU(M)\times SO(\nncl)$ with a symmetric tensor of
$SU(M)$}
\seclab{\Ssusos}

Consider the $SU(M)\times SO(\nncl)$ theories described
in sect. \SQsusos.  These models have an anomaly-free
$SU(\mfl)\times SU(\mfr)\times SU(\nflp)\times U(1)_R$ flavor
symmetry with matter fields in the representations
\thicksize=1pt
\vskip12pt
\begintable
\tstrut  | $SU(\mcl)$ | $SO(\nncl)$ |
$SU(\mfl)$ | $SU(\mfr)$ | $SU(\nflp)$  | $U(1)_R$ \crthick
$\Ql$ | ${\bf \mcl}$ | ${\bf 1}$ |
${\bf\mfl}$ | ${\bf 1}$ | ${\bf 1}$ |
$R_{\Ql}$\cr
$\Qr$ | ${\bf \overline\mcl}$ | ${\bf 1}$ |
${\bf 1}$ | ${\bf\mfr}$ | ${\bf 1}$ |
$R_{\Qr}$\cr
$\Sl$ | ${\bf 1}$ | ${\bf \nncl}$ |
${\bf 1}$ | ${\bf 1}$ | ${\bf \nflp}$ |
$R_{\Sl}$\cr
$\ts$ | ${\bf \mcl}$ | ${\bf \nncl}$ |
${\bf 1}$ | ${\bf 1}$ | ${\bf 1}$ |
$R_{\ts}={1\over 2(k+1)}$\cr
$\tst$ | ${\bf \overline{sym}}$ | ${\bf 1}$ |
${\bf 1}$ | ${\bf 1}$ | ${\bf 1}$ |
$R_{\ts}={1\over 2(k+1)}$
\endtable
\noindent
(There are two additional $U(1)$ symmetries
which we have omitted from this
table.)  This is a chiral theory; cancellation of gauge anomalies
requires $\mfl+2\nncl=\mfr+M+4$.  The $R$ charges can be taken to be
\eqn\Rsuspo{\eqalign{
R_Q &= 1 - {(k+1)(2[M-N]+\mfl+\nflp-\mfr)+N\over 2(k+1)m_f}, \cr
R_{\tilde Q} &=
 1 - {(k+1)(-\mfl-\nflp+\mfr)+2(M-2k)\over 2(k+1)\tilde m_f}, \cr
R_S &= 1 - {2(k+1)(N-M+2)+M\over 2(k+1)n_f}. \cr
 &\cr}}

\subsec{The dual theory}
\subseclab{\Dsusos}

The dual theory is $SU(\mcld)\times SO(\nncld)$
where
$$\eqalign{\mcld=&\; (k+1)(\nflp+\mfl+\mfr+4)-\mfl-\nncl,\cr
\nncld=&\; (k+1)(\nflp+\mfl+\mfr+4)-\nflp-\mcl.\cr}$$
Taking the singlets $M_{j},\dots, R_r$ to transform as
the mesons of  the electric theory,
the magnetic theory \WDsusos\ has charged matter in the
representations
\thicksize=1pt
\vskip12pt
\begintable
\tstrut  | $SU(\mcld)$ | $SO(\nncld)$ |
$SU(\mfl)$ | $SU(\mfr)$ | $SU(\nflp)$ |  $U(1)_R$ \crthick
$\ql$ | ${\bf \mcld}$ | ${\bf 1}$ |
${\bf1}$ | ${\bf 1}$ | ${\bf \overline\nflp}$ |
$R_{\ql}$\cr
$\qr$ | ${\bf \overline\mcld}$ | ${\bf 1}$ |
${\bf 1}$ | ${\bf \overline\mfr}$ | ${\bf 1}$ |
$R_{\qr}$\cr
$\sll$ | ${\bf 1}$ | ${\bf \nncld}$ |
${\bf \overline\mfl}$ | ${\bf 1}$ | ${\bf 1}$ |
$R_{\sll}$\cr
$\tsD$ | ${\bf \mcld}$ | ${\bf \nncld}$ |
${\bf 1}$ | ${\bf 1}$ | ${\bf 1}$ |
$R_{\tsD}={1\over 2(k+1)}$\cr
$\tstD$ | ${\bf \overline{sym}}$ | ${\bf 1}$ |
${\bf 1}$ | ${\bf 1}$ | ${\bf 1}$ |
$R_{\tstD}={1\over 2(k+1)}$
\endtable
\noindent
The formulas for the $R$ charges are the same as in the electric
theory
with $M,N$ replaced by $\tilde M,\tilde N$ and with
$m_f$ exchanged with $n_f$.

\newsec{$SU(\mcl)\times SO(\nncl)\times SO(\nncr)$}
\seclab{\Ssusoso}

Consider the $SU(M)\times SO(\nncl)\times SO(\nncr)$ theories
described  in sect. \SQsusoso.
These models have an anomaly-free
$SU(\mfl)\times SU(\mfr)\times SU(\nflp)\times SU(\nfrp)\times
U(1)_R$ flavor symmetry with matter fields in the representations
\thicksize=1pt
\vskip12pt
\begintable
\tstrut  | $SU(\mcl)$ | $SO(\nncl)$ | $SO(\nncr)$ |
$SU(\mfl)$ | $SU(\mfr)$ | $SU(\nflp)$ | $SU(\nfrp)$ | $U(1)_R$
\crthick
$\Ql$ | ${\bf \mcl}$ | ${\bf 1}$ | ${\bf 1}$ |
${\bf\mfl}$ | ${\bf 1}$ | ${\bf 1}$ | ${\bf 1}$ |
$R_{\Ql}$\cr
$\Qr$ | ${\bf \overline\mcl}$ | ${\bf 1}$ | ${\bf 1}$ |
${\bf 1}$ | ${\bf\mfr}$ | ${\bf 1}$ | ${\bf 1}$ |
$R_{\Qr}$\cr
$\Sl$ | ${\bf 1}$ | ${\bf \nncl}$ | ${\bf 1}$ |
${\bf 1}$ | ${\bf 1}$ | ${\bf \nflp}$ | ${\bf 1}$ |
$R_{\Sl}$\cr
$\Sr$ | ${\bf 1}$ | ${\bf 1}$ | ${\bf \nncr}$ |
${\bf 1}$ |${\bf 1}$ | ${\bf 1}$ |  ${\bf \nfrp}$ |
$R_{\Sr}$\cr
$\ts$ | ${\bf \mcl}$ | ${\bf \nncl}$ | ${\bf 1}$ |
${\bf 1}$ | ${\bf 1}$ | ${\bf 1}$ | ${\bf 1}$ |
$R_{\ts}={1\over 2(k+1)}$\cr
$\tst$ | ${\bf \overline\mcl}$ | ${\bf 1}$ | ${\bf \nncr}$ |
${\bf 1}$ | ${\bf 1}$ | ${\bf 1}$ | ${\bf 1}$ |
$R_{\tst}={1\over 2(k+1)}$
\endtable
\noindent
(There are two additional $U(1)$ symmetries
which we have omitted from this
table.)
This is a chiral theory; cancellation of gauge anomalies
requires  $\mfl+\nncl=\mfr+\nncr$.  The $R$ charges can be taken to be
\eqn\Rsusoso{\eqalign{
R_Q &= 1 - {(k+1)(2[M-N]+\mfl+\nflp-\mfr-\nfrp)+N\over
2(k+1)m_f}, \cr R_{\tilde Q} &= 1
 - {(k+1)(2[M-N']-\mfl-\nflp+\mfr+\nfrp)+N'\over 2(k+1)\tilde
m_f}, \cr
R_S &= 1 - {2(k+1)(N-M-2)+M\over 2(k+1)n_f}, \cr
R_{S'} &= 1 - {2(k+1)(N'-M-2)+M\over 2(k+1)n'_f}. \cr}}

\subsec{The dual theory}
\subseclab{\Dsusoso}

The dual theory is $SU(\mcld)\times SO(\nncld)\times SO(\nncrd)$
where
$$\eqalign{\mcld=&\;
(k+1)(\nflp+\nfrp+\mfl+\mfr+4)-\mfl-\nncl,\cr
\nncld=&\; (k+1)(\nflp+\nfrp+\mfl+\mfr+4)-\nflp-\mcl,\cr
\nncrd=&\; (k+1)(\nflp+\nfrp+\mfl+\mfr+4)-\nfrp-\mcl.}$$
Taking the singlets $M_{j},\dots,\tilde R'_r$ to transform as
the mesons of  the electric theory, the magnetic theory
\WDsusoso\
has charged matter in the representations
\thicksize=1pt
\vskip12pt
\begintable
\tstrut  | $SU(\mcld)$ | $SO(\nncld)$ | $SO(\nncrd)$ |
$SU(\mfl)$ | $SU(\mfr)$ | $SU(\nflp)$ | $SU(\nfrp)$ | $U(1)_R$
\crthick
$\ql$ | ${\bf \mcld}$ | ${\bf 1}$ | ${\bf 1}$ |
${\bf 1}$ | ${\bf 1}$ | ${\bf \overline\nflp}$ | ${\bf 1}$ |
$R_{\ql}$\cr
$\qr$ | ${\bf \overline\mcld}$ | ${\bf 1}$ | ${\bf 1}$ |
${\bf 1}$ | ${\bf 1}$ | ${\bf 1}$ | ${\bf \overline\nfrp}$ |
$R_{\qr}$\cr
$\sll$ | ${\bf 1}$ | ${\bf \nncld}$ | ${\bf 1}$ |
${\bf \overline\mfl}$ | ${\bf 1}$ | ${\bf 1}$ | ${\bf 1}$ |
$R_{\sll}$\cr
$\sr$ | ${\bf 1}$ | ${\bf 1}$ | ${\bf \nncrd}$ |
${\bf 1}$ |${\bf \overline\mfr}$ | ${\bf 1}$ |  ${\bf 1}$ |
$R_{\sr}$\cr
$\tsD$ | ${\bf \mcld}$ | ${\bf \nncld}$ | ${\bf 1}$ |
${\bf 1}$ | ${\bf 1}$ | ${\bf 1}$ | ${\bf 1}$ |
$R_{\tsD}={1\over 2(k+1)}$\cr
$\tstD$ | ${\bf \overline\mcld}$ | ${\bf 1}$ | ${\bf \nncrd}$|
${\bf 1}$ | ${\bf 1}$ | ${\bf 1}$ | ${\bf 1}$ |
$R_{\tstD}={1\over 2(k+1)}$
\endtable
\noindent
The formulas for the $R$ charges are the same as in the electric
theory
with $M,N,N'$ replaced by $\tilde M,\tilde N,\tilde N'$ and with
$m_f,\tilde m_f$ exchanged with $n_f,n'_f$.

\newsec{$SU(M)\times Sp(\nncl)$ with an antisymmetric tensor
of $SU(M)$}
\seclab{\Ssuspa}

Consider the $SU(M)\times Sp(\nncl)$ theories described
in sect. \SQsuspa.
These models have an anomaly-free
$SU(\mfl)\times SU(\mfr)\times SU(\nflp)\times U(1)_R$ flavor
symmetry with matter fields in the representations
\thicksize=1pt
\vskip12pt
\begintable
\tstrut  | $SU(\mcl)$ | $Sp(\nncl)$ |
$SU(\mfl)$ | $SU(\mfr)$ | $SU(\nflp)$  | $U(1)_R$ \crthick
$\Ql$ | ${\bf \mcl}$ | ${\bf 1}$ |
${\bf\mfl}$ | ${\bf 1}$ | ${\bf 1}$ |
$R_{\Ql}$\cr
$\Qr$ | ${\bf \overline\mcl}$ | ${\bf 1}$ |
${\bf 1}$ | ${\bf\mfr}$ | ${\bf 1}$ |
$R_{\Qr}$\cr
$\Sl$ | ${\bf 1}$ | ${\bf 2\nncl}$ |
${\bf 1}$ | ${\bf 1}$ | ${\bf \nflp}$ |
$R_{\Sl}$\cr
$\ts$ | ${\bf \mcl}$ | ${\bf 2\nncl}$ |
${\bf 1}$ | ${\bf 1}$ | ${\bf 1}$ |
$R_{\ts}={1\over 2(k+1)}$\cr
$\tst$ | ${\bf \overline{asym}}$ | ${\bf 1}$ |
${\bf 1}$ | ${\bf 1}$ | ${\bf 1}$ |
$R_{\ts}={1\over 2(k+1)}$
\endtable
\noindent
(There are two additional $U(1)$ symmetries
which we have omitted from this
table.)  This is a chiral theory; cancellation of gauge anomalies
requires $\mfl+2\nncl=\mfr+M-4$, and $\mcl+\nflp$ even.  The $R$
charges can be taken to be
\eqn\Rsuspa{\eqalign{
R_Q &= 1 - {(k+1)(2[M-2N]+\mfl+\nflp-\mfr)+2N\over 2(k+1)m_f},
\cr R_{\tilde Q} &=
 1 - {(k+1)(-\mfl-\nflp+\mfr)+2(M+2k)\over 2(k+1)\tilde m_f}, \cr
R_S &= 1 - {2(k+1)(2N-M+2)+M\over 2(k+1)n_f}. \cr
 &\cr}}

\subsec{The dual theory}
\subseclab{\Dsuspa}

The dual theory is $SU(\mcld)\times Sp(\nncld)$
where
$$\eqalign{\mcld=&\; (k+1)(\nflp+\mfl+\mfr-4)-\mfl-2\nncl,\cr
2\nncld=&\; (k+1)(\nflp+\mfl+\mfr-4)-\nflp-\mcl.\cr}$$
Taking the singlets $M_{j},\dots,\tilde R_r$ to transform as
the mesons of  the electric theory,
the magnetic theory \WDsuspa\ has charged matter in the
representations
\thicksize=1pt
\vskip12pt
\begintable
\tstrut  | $SU(\mcld)$ | $Sp(\nncld)$ |
$SU(\mfl)$ | $SU(\mfr)$ | $SU(\nflp)$ |  $U(1)_R$ \crthick
$\ql$ | ${\bf \mcld}$ | ${\bf 1}$ |
${\bf1}$ | ${\bf 1}$ | ${\bf \overline\nflp}$ |
$R_{\ql}$\cr
$\qr$ | ${\bf \overline\mcld}$ | ${\bf 1}$ |
${\bf 1}$ | ${\bf \overline\mfr}$ | ${\bf 1}$ |
$R_{\qr}$\cr
$\sll$ | ${\bf 1}$ | ${\bf 2\nncld}$ |
${\bf \overline\mfl}$ | ${\bf 1}$ | ${\bf 1}$ |
$R_{\sll}$\cr
$\tsD$ | ${\bf \mcld}$ | ${\bf 2\nncld}$ |
${\bf 1}$ | ${\bf 1}$ | ${\bf 1}$ |
$R_{\tsD}={1\over 2(k+1)}$\cr
$\tstD$ | ${\bf \overline{asym}}$ | ${\bf 1}$ |
${\bf 1}$ | ${\bf 1}$ | ${\bf 1}$ |
$R_{\tstD}={1\over 2(k+1)}$
\endtable
\noindent
The formulas for the $R$ charges are the same as in the electric
theory
with $M,N$ replaced by $\tilde M,\tilde N$ and with
$m_f$ exchanged with $n_f$.

\newsec{$SU(M)\times Sp(\nncl)\times Sp(\nncr)$}
\seclab{\Ssuspsp}

Consider the $SU(M)\times Sp(\nncl)\times Sp(\nncr)$ theories
described in sect.  \SQsuspsp.
These models have an anomaly-free
$SU(\mfl)\times SU(\mfr)\times SU(\nflp)\times SU(\nfrp)\times
U(1)_R$ flavor symmetry with matter fields in the representations
\thicksize=1pt
\vskip12pt
\begintable
\tstrut  | $SU(\mcl)$ | $Sp(\nncl)$ | $Sp(\nncr)$ |
$SU(\mfl)$ | $SU(\mfr)$ | $SU(\nflp)$ | $SU(\nfrp)$ | $U(1)_R$
\crthick
$\Ql$ | ${\bf \mcl}$ | ${\bf 1}$ | ${\bf 1}$ |
${\bf\mfl}$ | ${\bf 1}$ | ${\bf 1}$ | ${\bf 1}$ |
$R_{\Ql}$\cr
$\Qr$ | ${\bf \overline\mcl}$ | ${\bf 1}$ | ${\bf 1}$ |
${\bf 1}$ | ${\bf\mfr}$ | ${\bf 1}$ | ${\bf 1}$ |
$R_{\Qr}$\cr
$\Sl$ | ${\bf 1}$ | ${\bf 2\nncl}$ | ${\bf 1}$ |
${\bf 1}$ | ${\bf 1}$ | ${\bf \nflp}$ | ${\bf 1}$ |
$R_{\Sl}$\cr
$\Sr$ | ${\bf 1}$ | ${\bf 1}$ | ${\bf 2\nncr}$ |
${\bf 1}$ |${\bf 1}$ | ${\bf 1}$ |  ${\bf \nfrp}$ |
$R_{\Sr}$\cr
$\ts$ | ${\bf \mcl}$ | ${\bf 2\nncl}$ | ${\bf 1}$ |
${\bf 1}$ | ${\bf 1}$ | ${\bf 1}$ | ${\bf 1}$ |
$R_{\ts}={1\over 2(k+1)}$\cr
$\tst$ | ${\bf \overline\mcl}$ | ${\bf 1}$ | ${\bf 2\nncr}$ |
${\bf 1}$ | ${\bf 1}$ | ${\bf 1}$ | ${\bf 1}$ |
$R_{\ts}={1\over 2(k+1)}$
\endtable
\noindent
(There are two additional $U(1)$ symmetries
which we have omitted from this
table.)  This is a chiral theory; cancellation of gauge anomalies
requires $\mfl+2\nncl=\mfr+2\nncr$, and $\mcl+\nflp$,
$\mcl+\nfrp$
both even.  The $R$ charges can be taken to be
\eqn\Rsuspsp{\eqalign{
R_Q &= 1 - {(k+1)(2[M-2N]+\mfl+\nflp-\mfr-\nfrp)+2N\over
2(k+1)m_f}, \cr
R_{\tilde Q} &= 1 -
  {(k+1)(2[M-2N']-\mfl-\nflp+\mfr+\nfrp)+2N'\over 2(k+1)\tilde
m_f}, \cr
R_S &= 1 - {2(k+1)(2N-M+2)+M\over 2(k+1)n_f}, \cr
R_{S'} &= 1 - {2(k+1)(2N'-M+2)+M\over 2(k+1)n'_f}. \cr}}

\subsec{The dual theory}
\subseclab{\Dsuspsp}

The dual theory is $SU(\mcld)\times Sp(\nncld)\times Sp(\nncrd)$
where
$$\eqalign{\mcld=&\;
(k+1)(\nflp+\nfrp+\mfl+\mfr-4)-\mfl-2\nncl,\cr
2\nncld=&\; (k+1)(\nflp+\nfrp+\mfl+\mfr-4)-\nflp-\mcl,\cr
2\nncrd=&\; (k+1)(\nflp+\nfrp+\mfl+\mfr-4)-\nfrp-\mcl.}$$
Taking the singlets $M_{j},\dots,\tilde R'_r$ to transform as
the mesons of  the electric theory, the magnetic theory
\WDsuspsp\
has charged matter in the representations
\thicksize=1pt
\vskip12pt
\begintable
\tstrut  | $SU(\mcld)$ | $Sp(\nncld)$ | $Sp(\nncrd)$ |
$SU(\mfl)$ | $SU(\mfr)$ | $SU(\nflp)$ | $SU(\nfrp)$ | $U(1)_R$
\crthick
$\ql$ | ${\bf \mcld}$ | ${\bf 1}$ | ${\bf 1}$ |
${\bf1}$ | ${\bf 1}$ | ${\bf \overline\nflp}$ | ${\bf 1}$ |
$R_{\ql}$\cr
$\qr$ | ${\bf \overline\mcld}$ | ${\bf 1}$ | ${\bf 1}$ |
${\bf 1}$ | ${\bf 1}$ | ${\bf 1}$ | ${\bf \overline\nfrp}$ |
$R_{\qr}$\cr
$\sll$ | ${\bf 1}$ | ${\bf 2\nncld}$ | ${\bf 1}$ |
${\bf \overline\mfl}$ | ${\bf 1}$ | ${\bf 1}$ | ${\bf 1}$ |
$R_{\sll}$\cr
$\sr$ | ${\bf 1}$ | ${\bf 1}$ | ${\bf 2\nncrd}$ |
${\bf 1}$ |${\bf \overline\mfr}$ | ${\bf 1}$ |  ${\bf 1}$ |
$R_{\sr}$\cr
$\tsD$ | ${\bf \mcld}$ | ${\bf 2\nncld}$ | ${\bf 1}$ |
${\bf 1}$ | ${\bf 1}$ | ${\bf 1}$ | ${\bf 1}$ |
$R_{\tsD}={1\over 2(k+1)}$\cr
$\tstD$ | ${\bf \overline\mcld}$ | ${\bf 1}$ | ${\bf 2\nncrd}$|
${\bf 1}$ | ${\bf 1}$ | ${\bf 1}$ | ${\bf 1}$ |
$R_{\tstD}={1\over 2(k+1)}$
\endtable
\noindent
The formulas for the $R$ charges are the same as in the electric
theory
with $M,N,N'$ replaced by $\tilde M,\tilde N,\tilde N'$ and with
$m_f,\tilde m_f$ exchanged with $n_f,n'_f$.

\newsec{$SU(M)\times Sp(\nncl)$ with a symmetric tensor of
$SU(M)$}
\seclab{\Ssusps}

Consider the $SU(M)\times Sp(\nncl)$ theories described
in sect. \SQsusps.
These models have an anomaly-free
$SU(\mfl)\times SU(\mfr)\times SU(\nflp)\times U(1)_R$ flavor
symmetry with matter fields in the representations
\thicksize=1pt
\vskip12pt
\begintable
\tstrut  | $SU(\mcl)$ | $Sp(\nncl)$ |
$SU(\mfl)$ | $SU(\mfr)$ | $SU(\nflp)$  | $U(1)_R$ \crthick
$\Ql$ | ${\bf \mcl}$ | ${\bf 1}$ |
${\bf\mfl}$ | ${\bf 1}$ | ${\bf 1}$ |
$R_{\Ql}$\cr
$\Qr$ | ${\bf \overline\mcl}$ | ${\bf 1}$ |
${\bf 1}$ | ${\bf\mfr}$ | ${\bf 1}$ |
$R_{\Qr}$\cr
$\Sl$ | ${\bf 1}$ | ${\bf 2\nncl}$ |
${\bf 1}$ | ${\bf 1}$ | ${\bf \nflp}$ |
$R_{\Sl}$\cr
$\ts$ | ${\bf \mcl}$ | ${\bf 2\nncl}$ |
${\bf 1}$ | ${\bf 1}$ | ${\bf 1}$ |
$R_{\ts}={1\over 2(k+1)}$\cr
$\tst$ | ${\bf \overline{sym}}$ | ${\bf 1}$ |
${\bf 1}$ | ${\bf 1}$ | ${\bf 1}$ |
$R_{\ts}={1\over 2(k+1)}$
\endtable
\noindent
(There are two additional $U(1)$ symmetries
which we have omitted from this
table.)  This is a chiral theory; cancellation of gauge anomalies
requires $\mfl+2\nncl=\mfr+M+4$, and $\mcl+\nflp$ even.  The $R$
charges can be taken to be
\eqn\Rsusps{\eqalign{
R_Q &= 1 - {2(k+1)(2[M-2N]+\mfl+\nflp-\mfr)+2N\over 4(k+1)m_f}
,\cr R_{\tilde Q} &=
 1 - {2(k+1)(-\mfl-\nflp+\mfr)+2(M+2k)\over 4(k+1)\tilde m_f},
\cr R_S &= 1 - {4(k+1)(2N-M+2)+M\over 4(k+1)n_f}. \cr
 &\cr}}

\subsec{The dual theory}
\subseclab{\Dsusps}

The dual theory is $SU(\mcld)\times Sp(\nncld)$
where
$$\eqalign{\mcld=&\; 2(k+1)(\nflp+\mfl+\mfr)-\mfl-2\nncl,\cr
2\nncld=&\; 2(k+1)(\nflp+\mfl+\mfr)-\nflp-\mcl.\cr}$$
Taking the singlets $M_{j},\dots,\tilde R_r$ to transform as
the mesons of  the electric theory,
the magnetic theory \WDsusps\ has charged matter in the
representations
\thicksize=1pt
\vskip12pt
\begintable
\tstrut  | $SU(\mcld)$ | $Sp(\nncld)$ |
$SU(\mfl)$ | $SU(\mfr)$ | $SU(\nflp)$ |  $U(1)_R$ \crthick
$\ql$ | ${\bf \mcld}$ | ${\bf 1}$ |
${\bf1}$ | ${\bf 1}$ | ${\bf \overline\nflp}$ |
$R_{\ql}$\cr
$\qr$ | ${\bf \overline\mcld}$ | ${\bf 1}$ |
${\bf 1}$ | ${\bf \overline\mfr}$ | ${\bf 1}$ |
$R_{\qr}$\cr
$\sll$ | ${\bf 1}$ | ${\bf 2\nncld}$ |
${\bf \overline\mfl}$ | ${\bf 1}$ | ${\bf 1}$ |
$R_{\sll}$\cr
$\tsD$ | ${\bf \mcld}$ | ${\bf 2\nncld}$ |
${\bf 1}$ | ${\bf 1}$ | ${\bf 1}$ |
$R_{\tsD}={1\over 2(k+1)}$\cr
$\tstD$ | ${\bf \overline{sym}}$ | ${\bf 1}$ |
${\bf 1}$ | ${\bf 1}$ | ${\bf 1}$ |
$R_{\tstD}={1\over 2(k+1)}$
\endtable
\noindent
The formulas for the $R$ charges are the same as in the electric
theory
with $M,N$ replaced by $\tilde M,\tilde N$ and with
$m_f$ exchanged with $n_f$.

\newsec{$SU(M)\times SO(\nncl)$ with an antisymmetric tensor of
$SU(M)$}
\seclab{\Ssusoa}

Consider the $SU(M)\times SO(\nncl)$ theories described
in sect. \SQsusoa.
These models have an anomaly-free
$SU(\mfl)\times SU(\mfr)\times SU(\nflp)\times U(1)_R$ flavor
symmetry with matter fields in the representations
\thicksize=1pt
\vskip12pt
\begintable
\tstrut  | $SU(\mcl)$ | $SO(\nncl)$ |
$SU(\mfl)$ | $SU(\mfr)$ | $SU(\nflp)$  | $U(1)_R$ \crthick
$\Ql$ | ${\bf \mcl}$ | ${\bf 1}$ |
${\bf\mfl}$ | ${\bf 1}$ | ${\bf 1}$ |
$R_{\Ql}$\cr
$\Qr$ | ${\bf \overline\mcl}$ | ${\bf 1}$ |
${\bf 1}$ | ${\bf\mfr}$ | ${\bf 1}$ |
$R_{\Qr}$\cr
$\Sl$ | ${\bf 1}$ | ${\bf \nncl}$ |
${\bf 1}$ | ${\bf 1}$ | ${\bf \nflp}$ |
$R_{\Sl}$\cr
$\ts$ | ${\bf \mcl}$ | ${\bf \nncl}$ |
${\bf 1}$ | ${\bf 1}$ | ${\bf 1}$ |
$R_{\ts}={1\over 2(k+1)}$\cr
$\tst$ | ${\bf \overline{asym}}$ | ${\bf 1}$ |
${\bf 1}$ | ${\bf 1}$ | ${\bf 1}$ |
$R_{\ts}={1\over 2(k+1)}$
\endtable
\noindent
(There are two additional $U(1)$ symmetries
which we have omitted from this
table.)  This is a chiral theory; cancellation of gauge anomalies
requires $\mfl+\nncl=\mfr+M-4$.  The $R$ charges can be taken to be
\eqn\Rsusoa{\eqalign{
R_Q &= 1 - {2(k+1)(2[M-N]+\mfl+\nflp-\mfr)+N\over 4(k+1)m_f},\cr
R_{\tilde Q} &=
 1 - {2(k+1)(-\mfl-\nflp+\mfr)+2(M-2)\over 4(k+1)\tilde m_f},\cr
R_S &= 1 - {4(k+1)(N-M-2)+M\over 4(k+1)n_f}. \cr
 &\cr}}

\subsec{The dual theory}
\subseclab{\Dsusoa}

The dual theory is $SU(\mcld)\times SO(\nncld)$
where
$$\eqalign{\mcld=&\; 2(k+1)(\nflp+\mfl+\mfr)-\mfl-\nncl,\cr
\nncld=&\; 2(k+1)(\nflp+\mfl+\mfr)-\nflp-\mcl.\cr}$$
Taking the singlets $M_{j},\dots,\tilde R_r$ to transform as
the mesons of  the electric theory,
the magnetic theory \WDsusoa\ has charged matter in the
representations
\thicksize=1pt
\vskip12pt
\begintable
\tstrut  | $SU(\mcld)$ | $SO(\nncld)$ |
$SU(\mfl)$ | $SU(\mfr)$ | $SU(\nflp)$ |  $U(1)_R$ \crthick
$\ql$ | ${\bf \mcld}$ | ${\bf 1}$ |
${\bf 1}$ | ${\bf 1}$ | ${\bf \overline\nflp}$ |
$R_{\ql}$\cr
$\qr$ | ${\bf \overline\mcld}$ | ${\bf 1}$ |
${\bf 1}$ | ${\bf \overline\mfr}$ | ${\bf 1}$ |
$R_{\qr}$\cr
$\sll$ | ${\bf 1}$ | ${\bf \nncld}$ |
${\bf \overline\mfl}$ | ${\bf 1}$ | ${\bf 1}$ |
$R_{\sll}$\cr
$\tsD$ | ${\bf \mcld}$ | ${\bf \nncld}$ |
${\bf 1}$ | ${\bf 1}$ | ${\bf 1}$ |
$R_{\tsD}={1\over 2(k+1)}$\cr
$\tstD$ | ${\bf \overline{asym}}$ | ${\bf 1}$ |
${\bf 1}$ | ${\bf 1}$ | ${\bf 1}$ |
$R_{\tstD}={1\over 2(k+1)}$
\endtable
\noindent
The formulas for the $R$ charges are the same as in the electric
theory
with $M,N$ replaced by $\tilde M,\tilde N$ and with
$m_f$ exchanged with $n_f$.

\newsec{$SU(\mcl)\times Sp(\nncl)\times SO(\nncr)$}
\seclab{\Ssuspso}

Consider the $SU(M)\times Sp(\nncl)\times SO(\nncr)$ theories
described  in sect. \SQsuspso.  These models have an
anomaly-free
$SU(\mfl)\times SU(\mfr)\times SU(\nflp)\times SU(\nfrp)\times
U(1)_R$ flavor symmetry with matter fields in the representations
\thicksize=1pt
\vskip12pt
\begintable
\tstrut  | $SU(\mcl)$ | $Sp(\nncl)$ | $SO(\nncr)$ |
$SU(\mfl)$ | $SU(\mfr)$ | $SU(\nflp)$ | $SU(\nfrp)$ | $U(1)_R$
\crthick
$\Ql$ | ${\bf \mcl}$ | ${\bf 1}$ | ${\bf 1}$ |
${\bf\mfl}$ | ${\bf 1}$ | ${\bf 1}$ | ${\bf 1}$ |
$R_{\Ql}$\cr
$\Qr$ | ${\bf \overline\mcl}$ | ${\bf 1}$ | ${\bf 1}$ |
${\bf 1}$ | ${\bf\mfr}$ | ${\bf 1}$ | ${\bf 1}$ |
$R_{\Qr}$\cr
$\Sl$ | ${\bf 1}$ | ${\bf 2\nncl}$ | ${\bf 1}$ |
${\bf 1}$ | ${\bf 1}$ | ${\bf \nflp}$ | ${\bf 1}$ |
$R_{\Sl}$\cr
$\Sr$ | ${\bf 1}$ | ${\bf 1}$ | ${\bf \nncr}$ |
${\bf 1}$ |${\bf 1}$ | ${\bf 1}$ |  ${\bf \nfrp}$ |
$R_{\Sr}$\cr
$\ts$ | ${\bf \mcl}$ | ${\bf 2\nncl}$ | ${\bf 1}$ |
${\bf 1}$ | ${\bf 1}$ | ${\bf 1}$ | ${\bf 1}$ |
$R_{\ts}={1\over 4(k+1)}$\cr
$\tst$ | ${\bf \overline\mcl}$ | ${\bf 1}$ | ${\bf \nncr}$ |
${\bf 1}$ | ${\bf 1}$ | ${\bf 1}$ | ${\bf 1}$ |
$R_{\ts}={1\over 4(k+1)}$
\endtable
\noindent
(There are two additional $U(1)$ symmetries
 which we have omitted from this
table.)  This is a chiral theory; cancellation of gauge anomalies
requires $\mfl+2\nncl=\mfr+\nncr$ and $\mcl+\nflp$ be even.
The $R$ charges can be taken to be
\eqn\Rsuspso{\eqalign{
R_Q &= 1
  - {2(k+1)(2[M-2N]+\mfl+\nflp-\mfr-\nfrp-4)+2N\over 4(k+1)m_f}
,\cr
R_{\tilde Q} &= 1 -
 {2(k+1)(2[M-N']-\mfl-\nflp+\mfr+\nfrp+4)+N'\over 4(k+1)\tilde
m_f} ,\cr
R_S &= 1 - {4(k+1)(2N-M+2)+M\over 4(k+1)n_f} ,\cr
R_{S'} &= 1 - {4(k+1)(N'-M-2)+M\over 4(k+1)n'_f} .\cr}}

\subsec{The dual theory}
\subseclab{\Dsuspso}

The dual theory is $SU(\mcld)\times Sp(\nncld)\times SO(\nncrd)$
where
$$\eqalign{\mcld=&\; 2(k+1)(\nflp+\nfrp+\mfl+\mfr)-\mfr-\nncr,\cr
2\nncld=&\; 2(k+1)(\nflp+\nfrp+\mfl+\mfr)-\nflp-\mcl,\cr
\nncrd=&\; 2(k+1)(\nflp+\nfrp+\mfl+\mfr)-\nfrp-\mcl.}$$
Taking the singlets $M_{j},\dots,\tilde R'_r$ to transform as
the mesons of  the electric theory, the magnetic theory
\WDsuspso\
has charged matter in the representations
\thicksize=1pt
\vskip12pt
\begintable
\tstrut  | $SU(\mcld)$ | $Sp(\nncld)$ | $SO(\nncrd)$ |
$SU(\mfl)$ | $SU(\mfr)$ | $SU(\nflp)$ | $SU(\nfrp)$ | $U(1)_R$
\crthick
$\ql$ | ${\bf \mcld}$ | ${\bf 1}$ | ${\bf 1}$ |
${\bf 1}$ | ${\bf 1}$ | ${\bf \overline\nflp}$ | ${\bf 1}$ |
$R_{\ql}$\cr
$\qr$ | ${\bf \overline\mcld}$ | ${\bf 1}$ | ${\bf 1}$ |
${\bf 1}$ | ${\bf 1}$ | ${\bf 1}$ | ${\bf \overline\nfrp}$ |
$R_{\qr}$\cr
$\sll$ | ${\bf 1}$ | ${\bf 2\nncld}$ | ${\bf 1}$ |
${\bf \overline\mfl}$ | ${\bf 1}$ | ${\bf 1}$ | ${\bf 1}$ |
$R_{\sll}$\cr
$\sr$ | ${\bf 1}$ | ${\bf 1}$ | ${\bf \nncrd}$ |
${\bf 1}$ |${\bf \overline\mfr}$ | ${\bf 1}$ |  ${\bf 1}$ |
$R_{\sr}$\cr
$\tsD$ | ${\bf \mcld}$ | ${\bf 2\nncld}$ | ${\bf 1}$ |
${\bf 1}$ | ${\bf 1}$ | ${\bf 1}$ | ${\bf 1}$ |
$R_{\tsD}={1\over 4(k+1)}$\cr
$\tstD$ | ${\bf \overline\mcld}$ | ${\bf 1}$ | ${\bf \nncrd}$ |
${\bf 1}$ | ${\bf 1}$ | ${\bf 1}$ | ${\bf 1}$ |
$R_{\tstD}={1\over 4(k+1)}$
\endtable
\noindent
The formulas for the $R$ charges are the same as in the electric
theory
with $M,N,N'$ replaced by $\tilde M,\tilde N,\tilde N'$ and with
$m_f,\tilde m_f$ exchanged with $n_f,n'_f$.

\newsec{Conclusion}

We have presented many generalizations of previously known
examples
of duality in N=1 supersymmetry \refs{\sem,\kut-\rlmsspso}.
In particular we have discussed chiral
models which are very similar in appearance to vector-like ones.
The models are all interconnected under superpotential
perturbations,
symmetry breaking and confinement.  We have by no means
presented a complete list of models of this type; many more
remain to be explored.  Furthermore,
much work needs to be done to clarify certain aspects of
the physics of these theories, and to discover the duality in
similar  models without a superpotential. Still, we hope that
these new examples will help guide us to an understanding of
duality in supersymmetric field theory.

\centerline{{\bf Acknowledgments}}

We would like to thank N. Seiberg for discussions.  This work was
supported in part by DOE grant \#DE-FG05-90ER40559. R.G.L. thanks
the CERN Theory Division, and R.G.L. and M.J.S. thank the Aspen
Center for Physics, where part of this work was done.

\listrefs

\end